\documentclass[twocolumn]{emulateapj}
\usepackage{apjfonts}
\usepackage{graphicx}
\newcommand{\bnabla}{\mbox{\boldmath$\nabla$}}

\newcommand{\be}{\begin{equation}}
\newcommand{\ee}{\end{equation} }
\newcommand{\ba}{\begin{eqnarray}}
\newcommand{\ea}{\end{eqnarray}}
\newcommand{\nn}{\mbox{} \nonumber \\ \mbox{} }
\newcommand{\kB}{k_{\rm B}}
\shorttitle{Evolution of a Radially Magnetized Protoplanetary Disk}
\shortauthors{Russo \& Thompson}	
\slugcomment{Astrophysical Journal 2015, in press}
\begin{document}
\title{Constrained Evolution of a Radially Magnetized Protoplanetary Disk: \nn Implications for Planetary Migration}
\author{Matthew Russo}
\affil{Department of Physics, University of Toronto, 60 St. George St., Toronto, ON M5S 1A7, Canada.}
\author{Christopher Thompson}
\affil{Canadian Institute for Theoretical Astrophysics, 60 St. George St., Toronto, ON M5S 3H8, Canada.}

%%%%%%%%%%%%%%%%%%%%%%%%%%%%%%%%%%%%%%%%%%%%%%%%%%%%%%%%%%%%%%%%%%%%%%%%%%%%%%%%%%%
%%%%%%%%%%%%%%%%%%%%%%%%%%%%%%%%%%%%%%%%%%%%%%%%%%%%%%%%%%%%%%%%%%%%%%%%%%%%%%%%%%%
\begin{abstract}
We consider the inner $\sim$ AU of a protoplanetary disk (PPD), at a stage where angular momentum transport 
is driven by the mixing of a radial magnetic field into the disk from a T-Tauri wind.  Because the radial profile
of the imposed magnetic field is well constrained, a deterministic calculation of the disk mass flow becomes possible.  
The vertical disk profiles obtained in Paper I imply a stronger magnetization in the inner disk, faster accretion,
and a secular depletion of the disk material.  Inward transport of solids allows the disk to maintain a broad optical
absorption layer even when the grain abundance becomes too small to suppress its ionization.  Thus a PPD may show
a strong middle-to-near infrared spectral excess even while its mass profile departs radically
from the minimum-mass solar nebula.   The disk surface density is buffered at $\sim 30$ g cm$^{-2}$:  below this, X-rays
trigger strong enough magnetorotational turbulence at the midplane to loft
mm-cm sized particles high in the disk, followed by catastrophic fragmentation.  A sharp density gradient
bounds the inner depleted disk, and propagates outward to $\sim 1$--2 AU over
a few Myr.   Earth-mass planets migrate through the inner disk over a similar timescale, whereas 
the migration of Jupiters is limited by the supply of gas.  Gas-mediated migration must
stall outside 0.04 AU, where silicates are sublimated and the disk shifts to a much lower column.  
A transition disk emerges when the dust/gas ratio in the MRI-active layer falls below 
$X_d \sim 10^{-6}(a_d/\mu{\rm m})$, where $a_d$ is the grain size.
\end{abstract}

\keywords{accretion, accretion disks --- magnetic fields --- planets and satellites: formation --- protoplanetary disks --- turbulence}
%%%%%%%%%%%%%%%%%%%%%%%%%%%%%%%%%%%%%%%%%%%%%%%%%%%%%%%%%%%%%%%%%%%%%%%%%%%%%%%%%%%
%%%%%%%%%%%%%%%%%%%%%%%%%%%%%%%%%%%%%%%%%%%%%%%%%%%%%%%%%%%%%%%%%%%%%%%%%%%%%%%%%%%

\section{Introduction}

Our goal here is to develop a deterministic model for the redistribution of mass in the inner part of
a protoplanetary disk (PPD).  When considering accretion through a thin disk, a central problem is
the dependence of the specific torque (or the `viscosity') on distance from the accreting star.
This question remains unresolved in most contexts, but it is essential to understanding the 
presence of exoplanets in small orbits around sun-like stars.  

The planets discovered by {\it Kepler} probe the development of the natal disk over a broad range
of radius and time \citep{lissauer14}.  Our focus 
here is on angular momentum transport driven by magnetic stresses internal to the disk, including both 
the laminar Maxwell stress $B_RB_\phi/4\pi$, and the turbulent stress that is driven by the 
magnetorotational instability (MRI; \citealt{bb94,gammie96}).  

When the active column is determined only by external ionization, these internal magnetic stresses do not support 
a steady mass flow through the disk.  The sign of the change in the surface mass density $\Sigma_g$
depends on the distribution of the seed magnetic field.   Most MRI-based models assume a vertical seed field, 
whose radial distribution cannot yet be reliably calculated.  This prevents a deterministic calculation of
the disk evolution -- a shortcoming which extends to torque models invoking magnetized winds \citep{pudritz86,suzuki10,bs13}.

It has long been realized that radial magnetic fields could have a strong influence on angular momentum 
transport in centrifugally supported flows, driven by the persistent winding of the magnetic field.  
A magnetocentrifugal wind from the central star, interacting with the upper layers of the disk, is a natural 
source for such a seed radial field.  Indeed, the kinetic pressure of a T-Tauri wind can dominate the pressure  of 
any wind that might flow from the disk surface during the later stages of PPD evolution.  

This seed magnetic field is relatively strong in the inner disk, as is the induced torque.  As we demonstrate in this paper, this has the interesting consequence  that MRI-driven mass transfer {\it removes} material from
the inner disk.   A detailed description of the vertical structure of such a radially magnetized disk, 
including non-ideal MHD effects, can be found in Russo \& Thompson (2015; hereafter Paper I).

A PPD modeled with a uniform viscosity coefficient $\alpha$ has a gas surface density profile slightly flatter than
the one obtained by radial smoothing in the solar system, $\Sigma_g \propto R^{-1}$ (e.g. \citealt{dalessio98,menou04}).
A similar profile is obtained by averaging over the Kepler planetary systems, 
but with a normalization an order of magnitude higher \citep{chiang13}.  Formation of these exoplanets in situ 
then poses a significant conundrum:  either their assembly must be delayed long after the condensation of solids 
in the inner disk, or rapid migration must be suppressed (e.g. \citealt{ogihara15}).

A much different surface density profile $\Sigma_g(R)$ is obtained following an accretion outburst driven by 
runaway MRI activation of a heavy disk (e.g. \citealt{zhu09}):  one finds $\Sigma_g(R)$ growing with radius up to 
a peak at $\sim 2$--3 AU, as limited by the onset of gravitational instability.  (This is a plausible mechanism 
for FU Orionis-type outbursts, which may occur several times during the early formation of a PPD.)
Evolving this profile forward with the torque mechanism developed in Paper I, we find that that the PPD
does not regenerate an inner surface density cusp following its final accretion outburst.

The similarity between the mass profile deduced from the Kepler systems and 
a uniform-$\alpha$ disk may therefore be illusory.   An alternative explanation can be found in terms of the stability
of planetary architectures, with the mean separation between planets limited to a multiple of the planetary
Hill radius \citep{chambers96,smith09,funk10}.

Another longstanding puzzle involves the relatively brief interval over which PPDs appear
as `transition disks' with extended internal cavities in reprocessed stellar light \citep{strom89,williams11}.
Here we show that the evolution toward a low-$\Sigma_g$ profile can occur very rapidly, while
the flux of dust into the inner disk remains high enough to maintain an optical absorption layer.
The dust loading of material flowing inside $\sim 1$ AU can be determined from a competition between
inward advection and adhesion and settling.  Evolution to a transition disk does not then require a 
sudden or major structural change, merely a reduction in the dust loading of the accreting gas.  

Although a secular depletion of gas in a PPD can clearly inhibit planetary migration, the presence
of close-in exoplanets is also suggestive of a regulatory mechanism that maintains $\Sigma_g$ high enough
to permit some migration over a $\sim 1$--10 Myr timescale.  
As a first estimate, consider migration driven by linear Lindblad and corotation torques (`linear Type I':
\citealt{gt80,ward1991}).  A planet of mass $M_p$ and semi-major axis $a$ moves radially on a timescale
\ba\label{eq:tI}
t_I &\equiv& {a\over da/dt} \sim \Omega^{-1}\left({c_g\over \Omega R}\right)^2 {M_\star^2\over \Sigma_g R^2 M_p} \nn
    &\sim& 1\times 10^6 \,\left({\Sigma_g\over 30~{\rm g~cm^{-2}}}\right)^{-1} \left({M_p\over 10 M_\oplus}\right)^{-1}
\quad{\rm yr},
\ea
where $M_\star$ is the mass of the central star, $\Omega$ is the Keplerian angular frequency, and 
$c_g = (\kB T/\mu)^{1/2}$ is the isothermal sound speed in gas of mean molecular weight $\mu$.
Expression (\ref{eq:tI}) points to a surface density $\Sigma_g \sim 30$ g cm$^{-2}$ over $\sim 0.03-1$ AU.  
Although this torque model is greatly oversimplified, a migration timescale comparable to the above estimate
is supported by more detailed considerations of planet migration in our model disk, as presented in the 
concluding section.

We identify the following feedback mechanism.  A column $\sim 30$ g cm$^{-2}$ is comparable
to the column $\delta\Sigma_{g,\rm ion}$ that can be ionized by stellar X-rays.
Maintaining a moderate ionization level, which is essential to the
MRI, also depends on a strong depletion in $\mu$m-sized dust grains.  In a disk with $\Sigma_g$
initially exceeding $2\delta\Sigma_{g,\rm ion}$, most of the solid material settles close to the midplane, where
the vertical component of gravity nearly vanishes.   

As $\Sigma_g$ is reduced by mass transfer through the outer MRI-active layers of the disk, the ionization level
rises near the midplane, permitting the lofting of solid particles from the settled layer.  This creates an abundance of 
small grains through catastrophic fragmentation, because collisions typically first occur high in the 
disk, where the particle drift speed greatly exceeds the fragmentation speed.  Thus the penetration of 
X-rays to the midplane, accompanied by the excitation of MRI turbulence, leads to a strong feedback 
on the amplitude of the turbulence.  

This mechanism for regulating $\Sigma_g$ depends on the persistence of a midplane particle layer over
a period of Myrs.   Small particles could well be depleted by planetesimal formation but, in the 
absence of large planets, are easily resupplied by inward drift from the outer disk, mediated by azimuthal
gas drag. 

\subsection{Plan of the Paper}  

The model of a radially magnetized PPD developed in Paper I 
is reviewed in Section \ref{s:review}, and its radial structure is described in Section \ref{sec:glob}.  
The inflow of dust through the turbulent upper layer of a PPD, and the reprocessing of stellar light by this 
layer, is addressed in Section \ref{sec:dust}.  The feedback of midplane particles
on the disk viscosity results in a buffered sequence of MRI-active disk solutions, which are described
in Section \ref{sec:solid}; a much lower gas column is obtained in the absence of dust.

The evolution of a global disk model is tackled in Section \ref{sec:evo}:  a similar end result is obtained
for different initial disk profiles.  Characteristic timescales for planet migration are outlined in Section \ref{s:disc},
along with broader implications of this disk model.  The Appendix presents our calibration of a post-FU Ori outburst disk, at the stage where dust condenses out of the gas phase and MRI turbulence is truncated near the midplane.

Throughout this paper we will sometimes use a shorthand $X = X_n\times 10^n$, with quantity $X$ in c.g.s. units.

\section{Review of Radially Magnetized Disk Model}\label{s:review}

Vertical disk profiles are obtained (Paper I) by starting with a weakly magnetized PPD and a 
stellar wind, which forms a turbulent boundary layer with the upper disk.  
The radial magnetic field that is implanted into the boundary layer is gradually raised in strength 
to a maximum value $\sim 1$ G at $0.1$ AU, corresponding to a split-monopole profile reaching 
$\sim 10^2$ G at the stellar radius $R_\star \sim 2R_\odot$.  The stellar wind sweeps back
any poloidal magnetic field whose pressure is less than a fraction $\sim 10^{-2}(\alpha/0.1)$ 
of the disk thermal pressure;  here $\alpha$ is the Shakura-Sunyaev viscosity coefficient.

The implanted radial field is quickly sheared in the toroidal direction, at a rate
$-(3\Omega/2)B_R$, thereby triggering MRI activity.   The growth of the toroidal field is eventually
limited by transport into the deeper, weakly conducting layers of the disk.   
The magnetic pressure generated in the disk-wind boundary layer pushes the radial field downward, thereby exciting
a toroidal field and MRI activity at progressively greater depths.  
This process terminates in a hydromagnetically `dead' zone that extends 
to the midplane.  Vertical transport of $B_R$ and $B_\phi$ is mediated by 
(i) Ohmic drift; (ii) ambipolar drift; (iii) MHD turbulence that is generated by MRI; and (iv) turbulence 
driven by the wind shear in the boundary layer.

\begin{table}
\caption{Model Parameters} \label{tab:model} 
\begin{center}
\begin{tabular}{ |l|l| }
%  \hline
%  \multicolumn{2}{|c|}{Disk Model Parameters} \\
%  \hline
%  $\Sigma_{tot}$ & 200 g cm$^{-2}$ \\
%  $\eta$ & $9/7$ \\
%  $T_{\rm bl}$ & $5000\,\rm{K}$ \\
%  $\mu_{g,h}$ & $1.27~m_p$ \\
%  $\mu_{g,c}$ & $2.32~m_p$ \\
%  $a_t$ & $5$ \\[1ex]
  \hline
  \multicolumn{2}{|c|}{Stellar Model Parameters} \\
  \hline 
  $M_*$ & $M_\odot$ \\
  $R_*$ & $2~R_\odot$ \\
  $R_X$ & $10~R_\odot$ \\
  $L_*$ & $L_\odot$ \\
  $L_X$ & $2\times 10^{30}$ erg s$^{-1}$ \\
  $T_X$ & 1 keV \\[1ex]
  \hline
  \multicolumn{2}{|c|}{Wind Model Parameters} \\
  \hline
  $\dot{M}_w$ & $10^{-9}~M_\odot$ yr$^{-1}$ \\
  $V_w$ & 400 km s$^{-1}$ \\
  $B^w_R$ & $1~(R/\rm 0.1~AU)^{-2}$ G \\[1ex]
  \hline
  \multicolumn{2}{|c|}{Turbulence Model Parameters} \\
  \hline
  $\widetilde\alpha_{{\rm MRI},0}$ & $0.1$ \\
  $\alpha_{{\rm mix},0}$ & $1$ \\
  ${\rm Am}_{\rm crit}$ & $10$ \\
  $\Lambda_{\rm O, crit}$ & $100$ \\[1ex]
  \hline
\end{tabular}
\end{center}
\end{table}

The T-Tauri wind is launched from the stellar magnetosphere and flows radially at $V_w\sim400$ km s$^{-1}$.  
It transfers mass at a rate $\dot{M}_w\sim 10^{-9}$ $M_\odot$ yr$^{-1}$, which is consistent with $\sim 10\%$ of the 
accretion rate through our model disk.  Such a ratio of outflow to inflow is motivated by observation \citep{calvet1997}
and by the spin rates of T-Tauri stars \citep{matt05}.  

The nominal upper boundary of the flared disk is defined by a balance between thermal pressure and
the normal component of the wind ram pressure.  The thickness of the disk material that is overturned by the wind,
and into which the radial magnetic field is mixed, is limited by radiative cooling.  
We find $\delta\Sigma_g \sim 10^{-3}$ g cm$^{-2}$ in an upper layer of atomic H that is regulated to 
$\sim 5000$ K by electronic transitions.
The deeper layers of the disk are mainly composed of molecular H$_2$, and cool to a temperature 
$\sim (1-5)\times 10^2$ K that is determined largely by the absorption of stellar optical light on 
dust grains \citep{chiang1997}.  The accretion rate and disk optical depth are low enough
that the midplane temperature and scaleheight are similar to those of a passively irradiated disk, $h_g = 
c_g/\Omega\approx 0.023\left(R/AU\right)^{9/7}$ AU.  Both the atomic and molecular layers are therefore taken to be isothermal, with a smooth but rapid transition in temperature at the cooling column.

Ionization of the disk is dominated by thermal X-rays from the stellar corona.
Particular care must be taken with respect to the dependence of ionization rate on radius.
It is determined by an approximate solution to the radiative transfer
equation that is consistent with the Monte Carlo results of \cite{igea1999}, but drops off with radius more
rapidly than the analytic fit used by \cite{baigood2009} (and employed in several recent numerical simulations). 
We take as fiducial values an X-ray luminosity $L_X = 2\times 10^{30}$ erg s$^{-1}$ and an optically thin, thermal
bremsstrahlung spectrum with temperature $\kB T_X = 1$ keV.  Refractory solids are assumed to be depleted from
the gas, and the mass fraction of volatiles (C, N, O, Ne, and S) is reduced by a factor $10^{-2}$ compared with
solar abundance.  The abundance of free metal atoms with low ionization potential (e.g. Mg) is taken to be
$10^{-3}$ of the solar abundance.

MRI activity is maintained at $\delta\Sigma_g $ ranging up to $10$--30 g cm$^{-2}$, below which the instability
is quenched by Ohmic and ambipolar drift.  Here the radial transport of angular momentum
may still be maintained by a laminar Maxwell stress $B_RB_\phi/4\pi$ \citep{turner2008,okuzumi2011,mohanty2013,lesur14}.
Depending on the magnetization, this `undead' zone may extend to columns as high as $\delta\Sigma_g\approx 100$ g cm$^{-2}$. 

Our baseline disk profile is constructed by ignoring the adsorption of charged particles on the 
surfaces of $\mu$m-sized dust grains.  This is consistent with a mass fraction $X_d$ of grains of radius
$a_d$ smaller than $X_d/a_d \lesssim 10^{-4}$ $\mu$m$^{-1}$ near the base of the MRI-active layer, and 
implies a net depletion below $\sim 10^{-2}$ of the solar abundance.\footnote{Throughout this paper, 
$X$ and $x$ are used to denote mass and number fraction, respectively.} In this paper, we investigate the effect 
of an enhanced dust abundance on the ionization and torque profiles.

The following two sections review further details of our calculation of the magnetic field profile in the disk,
the rate of radial mass transfer, and the limitations imposed by non-ideal MHD effects on internal disk torques.

\subsection{Non-ideal MHD Effects and Turbulence}\label{s:magdiff}

We calculate the vertical profile of the background magnetic field $(B_R,B_\phi)$ and mass density 
$\rho$ by combining the steady form of the induction equation
\be \label{e:induction}
\frac{\partial{\bf B}}{\partial t}={\bnabla}\times\left({\bf v}\times{\bf B}\right)-\frac{4\pi}{c}
{\bnabla}\times\left(\eta{\bf J}+\eta_a{\bf J}_{\perp}\right) = 0
\ee
with the equation of magnetostatic equilibrium, 
\be
{\partial\over\partial z}\left(\rho c_g^2 + {B^2\over 8\pi}\right) = -\rho \Omega^2 z.
\ee
Here ${\bf J}=c\bnabla\times{\bf B}/4\pi$ is the current density.   In layers where the undular Newcomb-Parker
mode is excited \citep{newcomb1961}, the density gradient is set to give marginal stability  of this mode, 
$-d\rho/dz=\rho z/\gamma h_g^2$, with adiabatic index $\gamma=7/5$.  Then the scale height $h_g \sim c_g/\Omega$.

The Ohmic diffusivity (as derived from the diagonal component of the resistivity tensor) 
is rescaled to include the combined effects of MRI-driven turbulence and Kelvin-Helmholtz mixing near the boundary layer,
\be
\eta = \eta_{\rm O} + \eta_{\rm MRI} + \eta_{\rm mix};\quad \eta_{\rm O} \approx 235(T/\rm{K})^{1/2} x_e^{-1}\; {\rm cm^2~s^{-1}}.
\ee
Hall drift and ambipolar drift are parameterized by \citep{wardle2012}:
\ba\label{eq:etas}
\eta_{\rm H}  &\approx& 2\times 10^5 x_e^{-1}\left(\frac{B/{\rm G}}{\rho_{-10}}\right)\;{\rm cm^2~s^{-1}};\nn
\eta_a &\approx& 7 \times 10^{3} x_e^{-1}\left(\frac{B/{\rm G}}{\rho_{-10}}\right)^{2}\;{\rm cm^2~s^{-1}}.
\ea
Ohmic diffusion dominates in dense and weakly magnetized regions, while ambipolar drift is dominant in the opposite limit.

Hall drift turns out to play a significant role in sustaining a non-axisymmetric MRI in a strong toroidal magnetic field
(Paper I), but is of secondary importance for vertical transport of the mean magnetic field.  That is because the
vertical component of the Hall-drift speed $v_{{\rm H},z} = -J_z/en_e$ is proportional to the radial gradient of $B_\phi$.
Even though $\eta_{\rm H}  > \eta_a$ throughout the MRI-active layer, vertical Hall drift is suppressed by an additional
factor $\sim h_g/R$.  This is true as long as the vertical field $B_z = O(h_g/R) B_R$.

The Ohmic and ambipolar Elsasser numbers 
\be\label{eq:lam0}
\Lambda_{\rm O}\equiv\frac{v_{\rm A}^{2}}{\eta_{\rm O}\Omega}; \quad \quad {\rm Am}\equiv\frac{v_{\rm A}^{2}}{\eta_a\Omega}
\ee 
provide the most accurate local measure of the ability diffusion to limit the growth of MRI turbulence.  Here $v_{\rm A}=B/\sqrt{4\pi\rho}$ 
is the Alfv\'en speed.  When either Elsasser number falls below a critical value, fields can diffuse across a perturbation before it can develop into turbulence.  

We assume that the MRI is cut off at critical Elsasser numbers
\be
\Lambda_{\rm O, crit}=100;\quad\quad {\rm Am}_{\rm crit}=10.
\ee
Our choice of $\Lambda_{\rm O, crit}$ comes from three-dimensional shearing box studies, which 
typically invoke weaker starting fields \citep{sano2002,simon2009,baistone2011,flock2012}.
Different values of $\Lambda_{\rm O, crit}$ and ${\rm Am}_{\rm crit}$ change the magnitude of the radial mass 
flux through the disk by varying the active column, but have a weak effect on the radial dependence of the mass flux.

The choice of ${\rm Am}_{\rm crit}$ is motivated in part by a linear, non-axisymmetric stability analysis
of a toroidally magnetized disk that includes both Hall and ambipolar drift (Paper I).  This stability analysis
shows that the MRI is more easily sustained at high magnetization in the presence of a toroidal, as opposed to
vertical seed field.  The Poynting flux associated with a growing MRI mode is predominantly horizontal, in 
contrast with the case of a vertical seed field, where the Poynting flux is directed vertically.   On this basis,
vigorous MRI-driven turbulence is not suppressed in a background toroidal field by propagation out of the growth layer, 
as numerical experiments find it to be when the seed field is strong and vertical, $B_z^2/8\pi P \gtrsim 10^{-2}$
\citep{bs13}.

A high toroidal magnetization is sustained in the upper disk, $B_\phi^2/8\pi P \sim 0.1$--10, and so we 
treat the MRI-generated stress as a perturbation to the background field pressure.
The ambipolar diffusivity $\eta_a$ is taken to depend only on the background field $B^2 = B_R^2 + B_\phi^2$.  
A first estimate of the equilibrium
magnetization is obtained by balancing winding of the mean radial seed field $B_R$ against vertical transport across 
a scale height $c_g/\Omega$ with a diffusivity
\be\label{e:nut}
\eta_{\rm MRI} = \alpha_{\rm MRI}{c_g^2\over \Omega} =
\widetilde\alpha_{\rm MRI} \left({B_{\phi}^2\over 8\pi P}\right)^\delta {c_g^2\over \Omega}
\ee
where $B_\phi$ is the mean toroidal seed field generated by winding and $\widetilde\alpha_{\rm MRI}$ is constant.
One finds (Paper I)
\be\label{eq:mageq}
{B_\phi^2\over 8\pi P} = \left({3\over 2\widetilde\alpha_{\rm MRI}}\right)^{2/(1+2\delta)}
     \left({B_R^2\over 8\pi P}\right)^{1/(1+2\delta)}.
\ee

The main challenge here is to make an appropriate choice for the amplitude $\widetilde\alpha_{\rm MRI}$ and 
index $\delta$.  We find that ${\rm Am} \sim 100$--300 up to a column $\delta\Sigma_g \sim 10$--$30$ g cm$^{-2}$
below the surface, beyond which the free electron fraction $x_e$ drops precipitously.  (See Figure 1 in Paper I.)  This
is higher by a factor $\sim 10$--30 than in the simulations of \cite{lesur14}, \cite{bai2014}, and \cite{gressel15}
which impose a vertical seed field, and in which MRI turbulence is only marginally excited.  On the other hand,
the magnetization is high enough that the scaling $\delta = 1/2$ which is found for weak seed fields in direct
numerical simulations \citep{hgb95} cannot be trusted. We 
choose $\nu_{\rm MRI}$ to scale directly with the seed toroidal magnetic pressure, corresponding to $\delta = 1$:
\be \label{e:alpha}
\alpha_{\rm MRI}\equiv \widetilde\alpha_{{\rm MRI},0}
\frac{B_{\phi}^{2}}{8\pi \rho c_g^2}\left(\frac{\Lambda_{\rm O}}{\Lambda_{\rm O}+
\Lambda_{\rm O, crit}}\right)\left(\frac{{\rm Am}}{{\rm Am}+{\rm Am}_{\rm crit}}\right).
\ee
Here $\widetilde\alpha_{{\rm MRI},0} = 0.1$. The last two factors on the right-hand side of Equation (\ref{e:alpha})
implement the cutoff of the MRI-generated stress at low values of ${\rm Am}$ and $\Lambda_{\rm O}$.

The main results of this paper are not sensitive to this calibration of $\nu_{\rm MRI}$.  The rate of mass
transfer scales as $\sim \widetilde\alpha_{\rm MRI}^{1/3}$, and one obtains inside-out mass depletion for $\delta$
in the full range $0.5-1$, as can be seen from Figure 2 of Paper I.  We view our normalization
as conservative, given that simulations of the MRI with a weak seed field demonstrate enormous growth of the
Maxwell stress.  

The active layer is defined by the column $\delta\Sigma_{\rm act}$, measured below the top of the disk, where 
$\widetilde\alpha_{\rm MRI}$ is reduced by a factor 2 by Ohmic and ambipolar diffusion.   Typically 
$\Lambda_{\rm O}$ reaches its threshold value higher in the disk than does ${\rm Am}$, so that the quenching of the MRI is
mainly driven by Ohmic losses.  

The disk-wind boundary layer is strongly magnetized given the chosen wind parameters, but our enforcement of marginal
Newcomb-Parker stability causes the scale height to saturate at $h_g \sim c_g/\Omega$.
The turbulent diffusivity in the mixing layer is therefore taken to be
\be\label{eq:numix}
\eta_{\rm mix}=\alpha_{\rm mix} h_g\left(c^2_g +v_{\rm A}^2\right)^{1/2}.
\ee
We take $\alpha_{\rm mix}=1$ (with a sharp cutoff below the base of the boundary layer as given by Equation (48) 
of Paper I) representing the fact that the depth of the mixing layer is set by a balance between
Kelvin-Helmholtz-driven overturns and radiative cooling.  Given rapid mixing in the
boundary layer, we adopt the upper boundary condition $B_\phi/B_R = -3/2$.
The deeper profile of the disk, in particular its magnetization
and turbulent amplitude, turn out to be insensitive to the value of $\eta_{\rm mix}$ and to the boundary condition
on $B_\phi$.

\subsection{Angular Momentum Redistribution}

Turbulence in a thin PPD redistributes the background magnetic field $(B_R,B_\phi)$ primarily in the vertical direction,
but also transfers angular momentum in the radial direction.  These two effects are respectively encapsulated in 
a renormalized magnetic diffusivity and viscosity.   Here we simply equate these diffusivities, corresponding
to unit magnetic Prandtl number:
\be\label{eq:pran}
\nu_{\rm MRI} = \eta_{\rm MRI}; \quad\quad \nu_{\rm mix} = \eta_{\rm mix}.
\ee
This choice is supported by some numerical simulations of the MRI with an imposed vertical field  \citep{fromang2009,guan2009,lesur2009}. 

The radial flow of disk material is driven by a combination of turbulent stresses and the laminar Maxwell stress.
The time evolution of the disk can be calculated, following \cite{pringle1974}, from the equation of conservation of
angular momentum for a ring of radial thickness $\delta R$ and mass $\delta m = 2\pi R\Sigma_g \delta R$,
\be
{d\over dt}\left(\delta m\,\Omega R^2\right) = \delta m\cdot v_R{\partial(\Omega R^2)\over\partial R} 
= \delta R {\partial G\over\partial R}.
\ee
The total turbulent $+$ laminar magnetic couple acting at radius $R$ is
\ba
G(R) &=& R\cdot 2\pi R \int dz\left(\rho\nu R{\partial\Omega\over\partial R} + {B_RB_\phi\over 4\pi}\right)\nn
  &=& - 3\pi R^2\Omega \langle\nu\rangle \Sigma_g + {R^2\over 2}\int dz B_R B_\phi \nn
  &=& -3\pi R^2\int dz \left( \alpha_{\rm MRI} + \alpha_{\rm mix} + \alpha_{\rm lam}\right)P_g,
\ea
where
\be\label{eq:nuav}
\langle\nu\rangle = \Sigma_g^{-1} \int dz\left(\nu_{\rm MRI} + \nu_{\rm mix}\right) \rho.
\ee
and $\alpha_{\rm MRI}$, $\alpha_{\rm mix}$ are the viscosity coefficients defined in Equations (\ref{e:nut}), (\ref{eq:numix}).
The parameter
\be\label{eq:alphalam}
\alpha_{\rm lam} \equiv {|B_RB_\phi|\over 6\pi P_{g}}
\ee
exceeds $\alpha_{\rm MRI}$ below the active region, and may also exceed $\alpha_{\rm mix}$ in the disk-wind 
boundary layer where $B_\phi/B_R\lesssim 10$. 

The local mass transfer rate 
\be
\dot M(R) = -2\pi R v_R\Sigma_g = \dot M_{\rm turb} + \dot M_{\rm lam},
\ee
can be written in terms of local disk quantities when $\langle\nu\rangle \Sigma_g \propto R^{\gamma_\nu}$ 
and $\Omega^{-1}\int dz B_R B_\phi \propto R^{\gamma_B}$,
\ba \label{e:Mdot}
\dot{M}_{\rm turb}(R) &=& 3\pi(1+2\gamma_\nu)\langle\nu\rangle\Sigma_{g}; \nn
\dot{M}_{\rm lam}(R) &=& {1+2\gamma_B\over 2\Omega} \int dz B_RB_\phi.
\ea 
The gas surface density evolves with time unless $\gamma_\nu = 0$, $\gamma_B = 0$:
\be
{\partial\Sigma_g\over\partial t} = {1\over 2\pi R}{\partial\dot M\over \partial R}.
\ee

Linear winding of the radial magnetic field is sustained at lower values of $x_e$ than is MRI turbulence.  
An upper limit on the Lorentz force is obtained by requiring the Ohmic contribution to the vertical electric
field to be smaller than the contribution from the disk rotation,
\be 
{4\pi\eta\over c^2}|J_{z}|< {\Omega R\over c}|B_{R}|.
\ee
The torque density is given by 
\be 
\frac{R}{c}J_{z}B_{R}={\partial\over\partial R}\left(R{B_RB_\phi\over 4\pi}\right),
\ee
so the laminar stress can be no larger than
\be 
-{B_R B_\phi\over 4\pi} \lesssim {\Omega R^2\over \eta} {B_R^2\over 4\pi}.
\ee 
Here $\eta \sim \max(\eta_{\rm O},\eta_a)$ as both Ohmic and ambipolar diffusion may be responsible for 
the reduced coupling of the magnetic field to the neutral particles.  Here
Ohmic diffusion dominates near the torque cutoff, corresponding to
\be 
\eta_{\rm O} \gtrsim -\frac{B_R}{B_{\phi}}\Omega R^{2},
\ee
or, equivalently, 
\be
x_e < 5\times10^{-17}\left|{\frac{B_{\phi}}{B_R}}\right| \left(\frac{T/100\,\rm{K}}{R/\rm AU}\right).
\ee
We find that $B_\phi/B_R \gtrsim 10^2$ is generated by linear winding
in and below the active layer.  The cutoff in laminar torque then sits at
$\delta\Sigma_g \sim 10^2$ g cm$^{-2}$ below the disk surface, and $x_e \sim 10^{-14}$.

\begin{figure}[!]
\epsscale{1.2}
\plotone{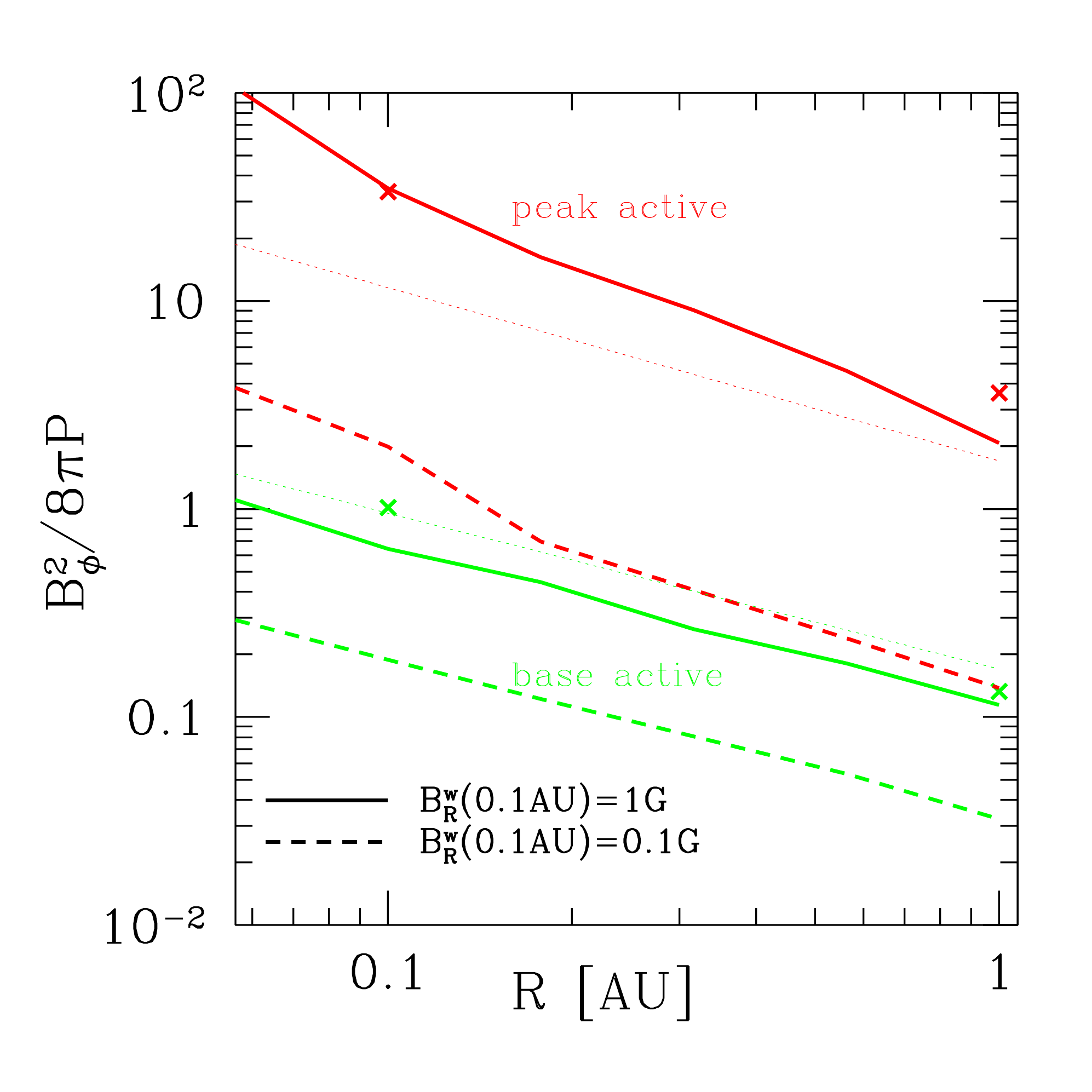}
\caption{Maximum and minimum toroidal magnetization in the active layer of the disk.
Curves refer to the more strongly magnetized hemisphere.  The vertical magnetization profile
has a sharp maximum outside $\sim 0.1$ AU from the protostar;  the magnetization continues
to drop below the plotted minimum below the base of the active layer.  The analytic estimate 
(\ref{e:betaeq}) is drawn as red (green) dotted lines for $\delta\Sigma_g=0.01$ (10) g cm$^{-2}$.
The peak magnetization exceeds the analytic estimate at small radii, where it sits below 
0.01 g cm$^{-2}$. `X' marks the magnetizations that result from raising $k_B T_X$ to $5$ keV from $1$ keV. }
\vskip .1in
\label{fig:betapeak}
\end{figure}

\begin{figure*}[ht]
\epsscale{1.1}
\plottwo{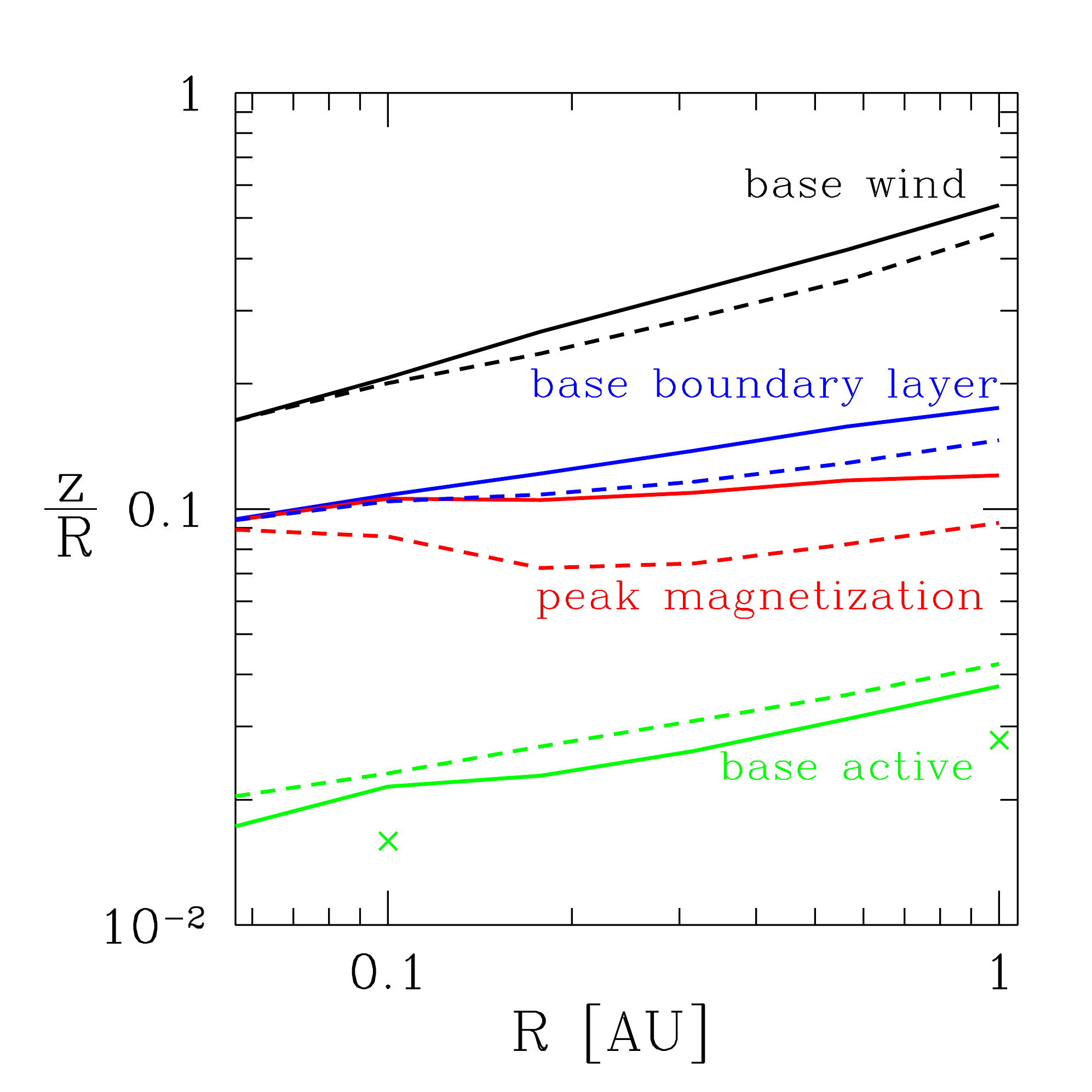}{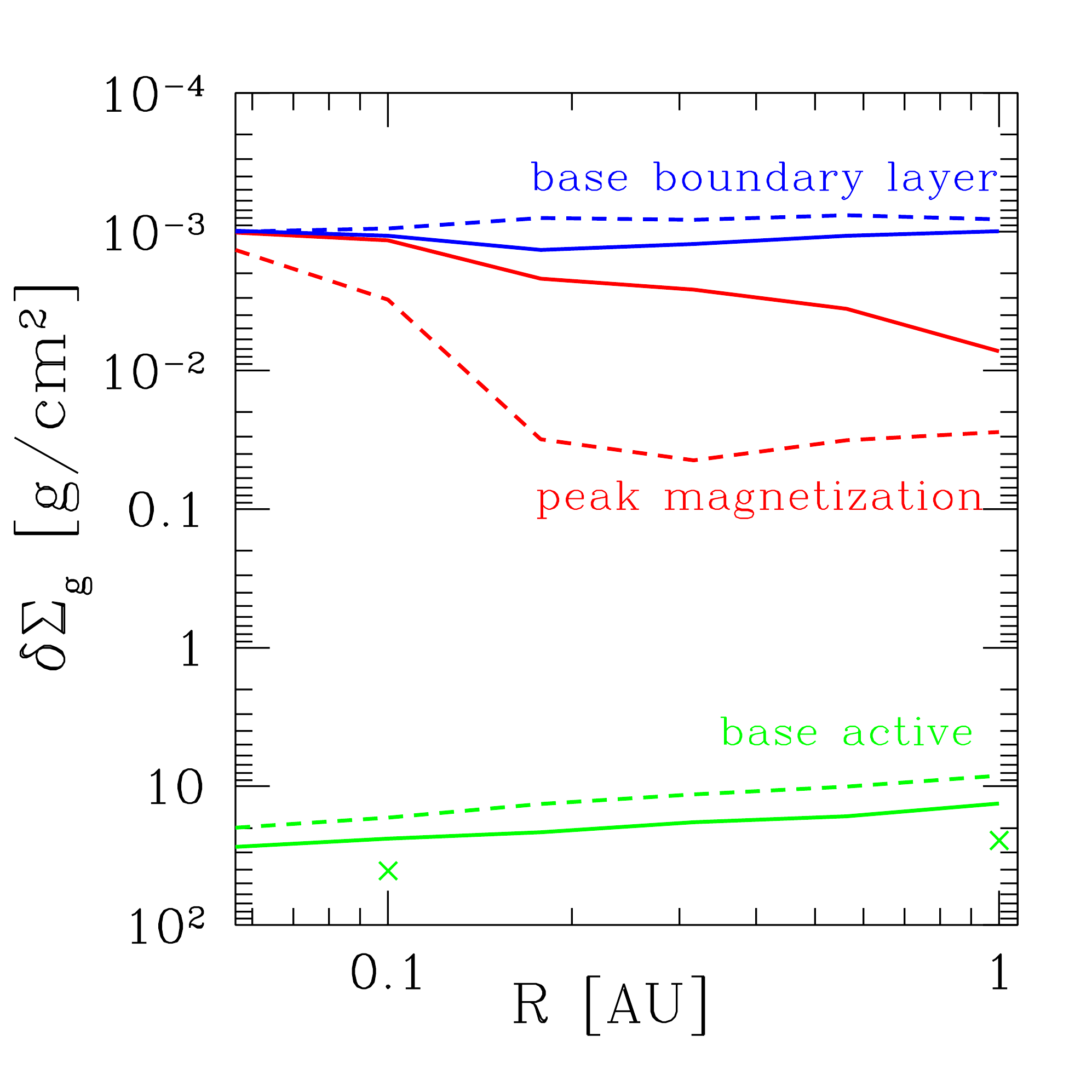}
\caption{\textit{Left panel:}  heights of important reference points in the vertical disk profile, 
as a function of orbital radius.  Solid lines: imposed magnetic field at full strength ($\epsilon_B = 1$);
dashed lines:  $\epsilon_B = 0.1$.  \textit{Right panel:}  corresponding gas column as measured from the
disk-wind interface.`X' marks the base of the active layer when $k_B T_X$ is raised to $5$ keV from $1$ keV. }
\vskip .1in
\label{fig:hR}
\end{figure*}

\section{Radial Profiles}\label{sec:glob}

We now present radial profiles of the disk magnetization, turbulence, and rate of mass transfer.  The radial
magnetic field that is imposed at the surface of the disk is take to be
\be\label{eq:BRapp}
B_R(R) = \epsilon_B\times 10^{-2}\left({R\over{\rm AU}}\right)^{-2}\;{\rm G}.
\ee
Vertical disk profiles are obtained at cylindrical radii $R=10^{-1.25}$--1 AU (Paper I), and are now
used to construct radial profiles of various disk properties.  Section \ref{sec:evo} treats
the evolution of the disk. 

We start with a total gas column $\Sigma_g = 200$ g cm$^{-2}$, not much larger than the MRI-active column.  
Part of the motivation here is to check whether the disk evolves self-consistently to a low column.  
Given that the solids initially present in the inner disk have mainly settled to the midplane,
the structure of the active layer would not be strongly modified by the presence
of a thicker, quiescent layer below it. In addition, $\Sigma_g$ grows with radius inside $\sim 2$--3 AU
in the aftermath of an ionization-driven instability (\citealt{zhu09}; Appendix A).  This means that the final
evolution of the inner, layered disk may start with relatively little gas compared with uniform-$\alpha$ accretion models.

\subsection{Equilibrium Magnetization}\label{sec:Beq}

The upper disk becomes magnetically dominated in response to the embedding of a radial field of strength 
$\epsilon_B \sim 0.1$--1.  The magnetization $B^2/8\pi P$ shows a pronounced peak at a relatively shallow
depth within the active layer (Figure \ref{fig:betapeak}), which disappears inside $\sim 0.1$ AU.
Below this peak, the magnetization drops monotonically toward the midplane, and in particular is lower
at the base of the active layer where most of the radial mass transfer is concentrated.

The vertical profiles compare well with the analytic estimate (\ref{eq:mageq}).  Substituting 
$P = \delta\Sigma_g c_g \Omega$ on the right-hand side of that equation and employing the temperature 
scaling $T = 180\,(R/{\rm AU})^{-3/7}$ K, gives the equilibrium magnetization
\be \label{e:betaeq}
\frac{B_{\phi}^{2}}{8\pi P}\approx 0.38 \left({\epsilon_{B}\over\widetilde\alpha_{{\rm MRI},-1}}\right)^{2/3}
\left(\frac{R}{\rm AU}\right)^{-0.76} \left(\frac{\delta\Sigma_{g}}{{\rm g~cm}^{-2}}\right)^{-1/3}.
\ee
This is also plotted in Figure \ref{fig:betapeak} at the column $\delta\Sigma_g \sim 10^{-2} (R/{\rm AU})$ g cm$^{-2}$ 
of peak magnetization.

\subsection{Vertical Structure}

The disk is comprised of several layers, the transitions between which are shown in Figure \ref{fig:hR}.
From the top: (i) the nominal boundary between hydrostatic disk and T-Tauri wind, where thermal pressure balances
the normal component of the wind ram pressure;  (ii) the base of the disk-wind boundary layer, where radiative
cooling balances the input of turbulent kinetic energy from the damping of velocity shear; (iii) the 
surface of peak magnetization; and (iv) the base of the active layer, where $\Lambda_{\rm O} \sim \Lambda_{\rm O, crit}$.
Our treatment of each of these transitions is discussed in more detail in Paper I.

The heights of these transitions all flare with radius. In spite of the high magnetization that is reached in
the uppermost parts of the inner disk, this flaring is similar to that obtained in a thermally supported, Keplerian
disk.  This result is partly the result of our imposition of marginal stability to undular Newcomb-Parker modes:  the 
zone of marginal stability expands to fill the active layer when the imposed magnetic field approaches the 
full strength considered, $B_R(0.1~{\rm AU}) \sim 1$ G ($\epsilon_B = 1$).  

The column density $\delta\Sigma_{\rm act}$ through the active layer is shown in the right panel of Figure \ref{fig:hR}.
The analytic fit $\delta\Sigma_{\rm act}\approx 13.4\,(R/{\rm AU})^{-0.27}$ g cm$^{-2}$ applies at $\epsilon_B = 1$.
The negative radial slope arises from the increased ionization rate in the inner disk, combined with the 
decrease of the toroidal field with radius, $B_\phi \propto R^{-1.33}$ near the midplane.  By comparison, the base
of the disk-wind boundary layer sits at a relatively uniform depth $\delta\Sigma_g \sim 10^{-3}$ g cm$^{-2}$.

An optical absorption layer is present in the disk as long as the mass fraction of $\sim \mu$m sized solid
grains exceeds $X_d \sim 10^{-6}$, as we show in Section \ref{sec:opt}.  Its height varies from $\sim 2h_g = 
2c_g/\Omega$ -- within the lower part of the active layer -- up to $\sim 6h_g$ as $X_d$ is raised to solar abundance.

\begin{figure}[!]
\epsscale{1.2}
\plotone{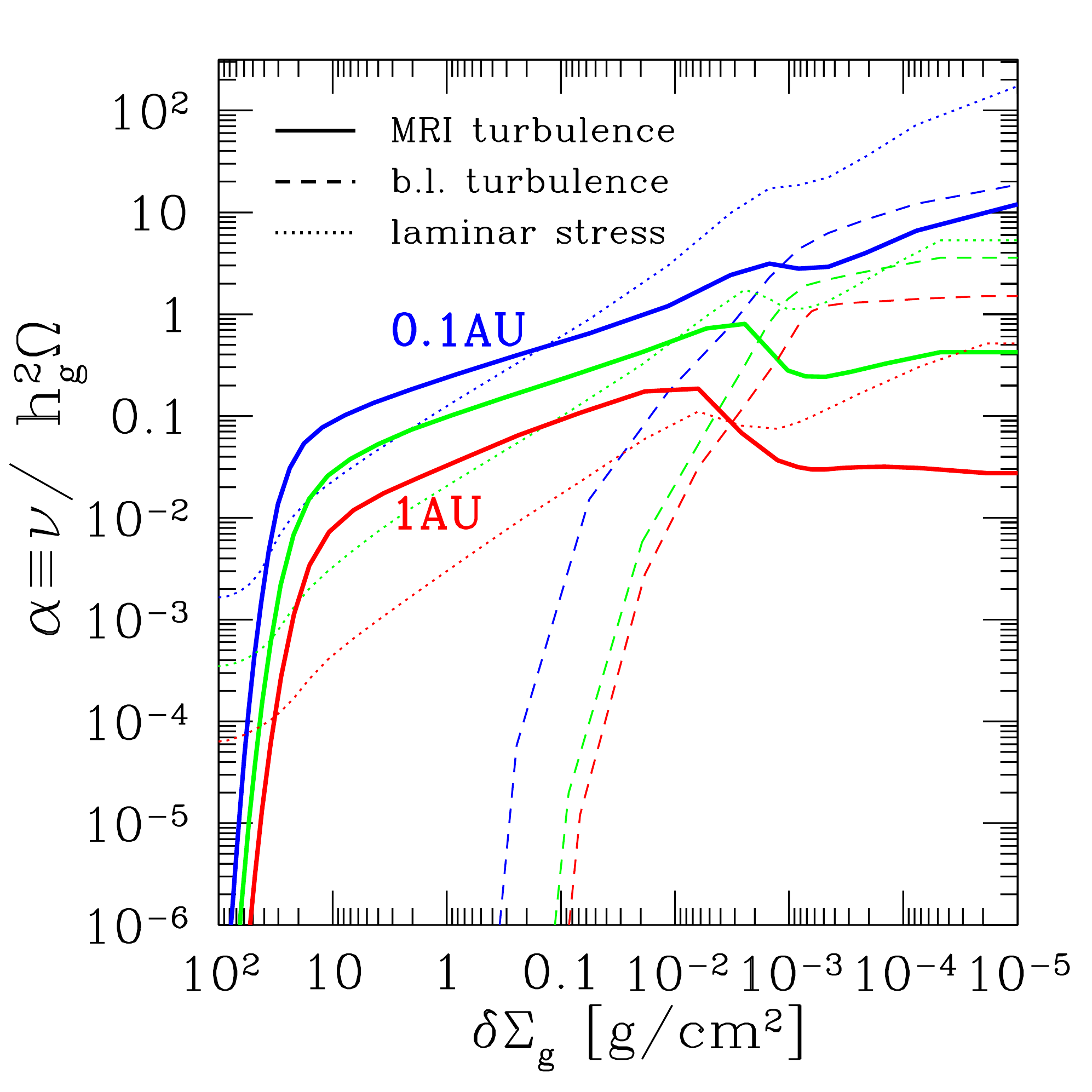}
\caption{Solid lines: amplitude $\alpha_{\rm MRI}$ of MRI-driven turbulence, Equation (\ref{e:alpha}), as
a function of column measured below disk surface.
Dashed lines: amplitude $\alpha_{\rm mix}$ of turbulence driven by disk-wind mixing (Equation (\ref{eq:numix}) 
combined with Equation (48) of Paper I).  Dotted lines:  corresponding parameter (\ref{eq:alphalam}) measuring the laminar stress
$B_RB_\phi/4\pi$.  Blue, green and red lines show the profile at $0.1$, $0.32$ and $1$ AU
orbital radius.}
\vskip .1in
\label{fig:alpha}
\end{figure}

\begin{figure*}[!]
\epsscale{1.1}
\plottwo{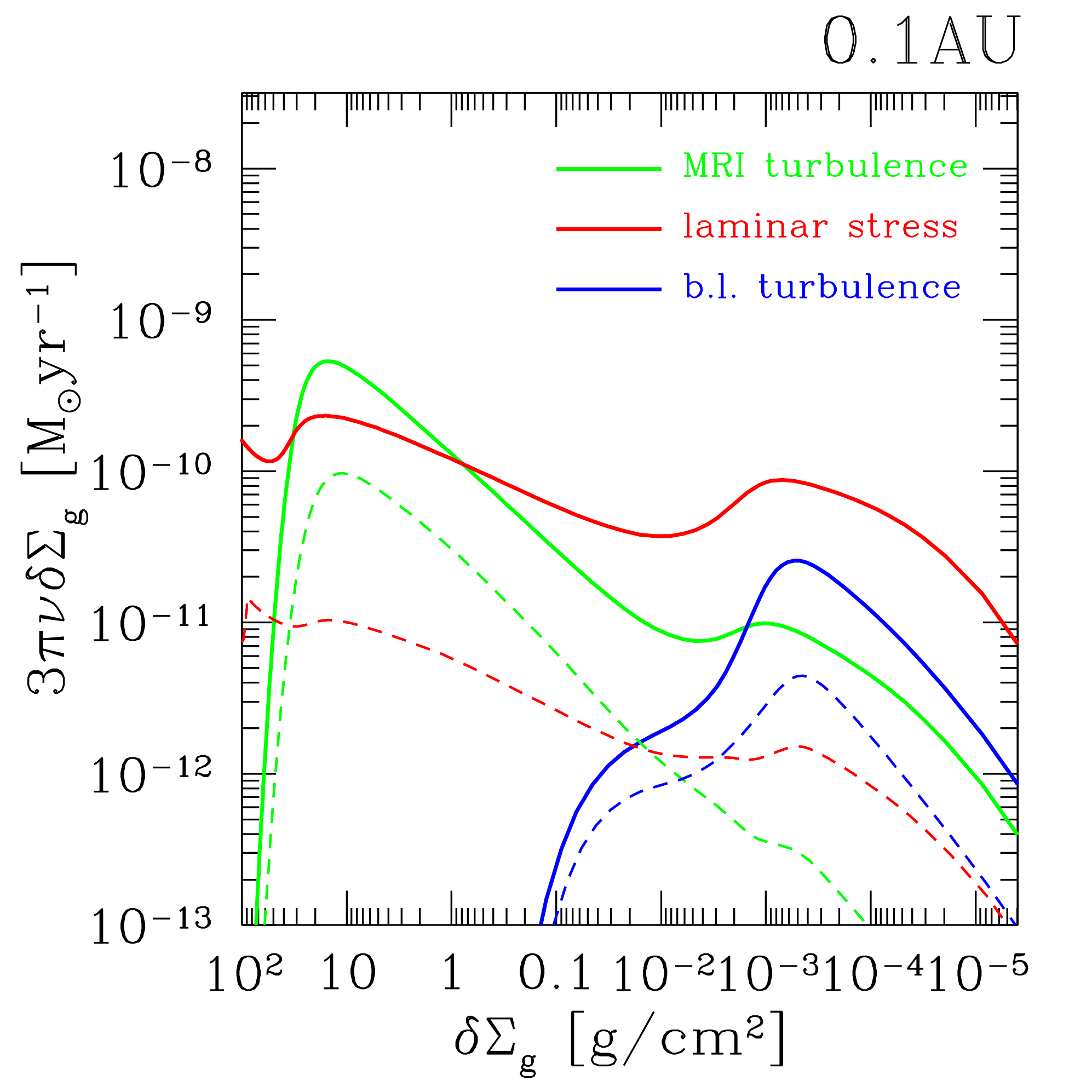}{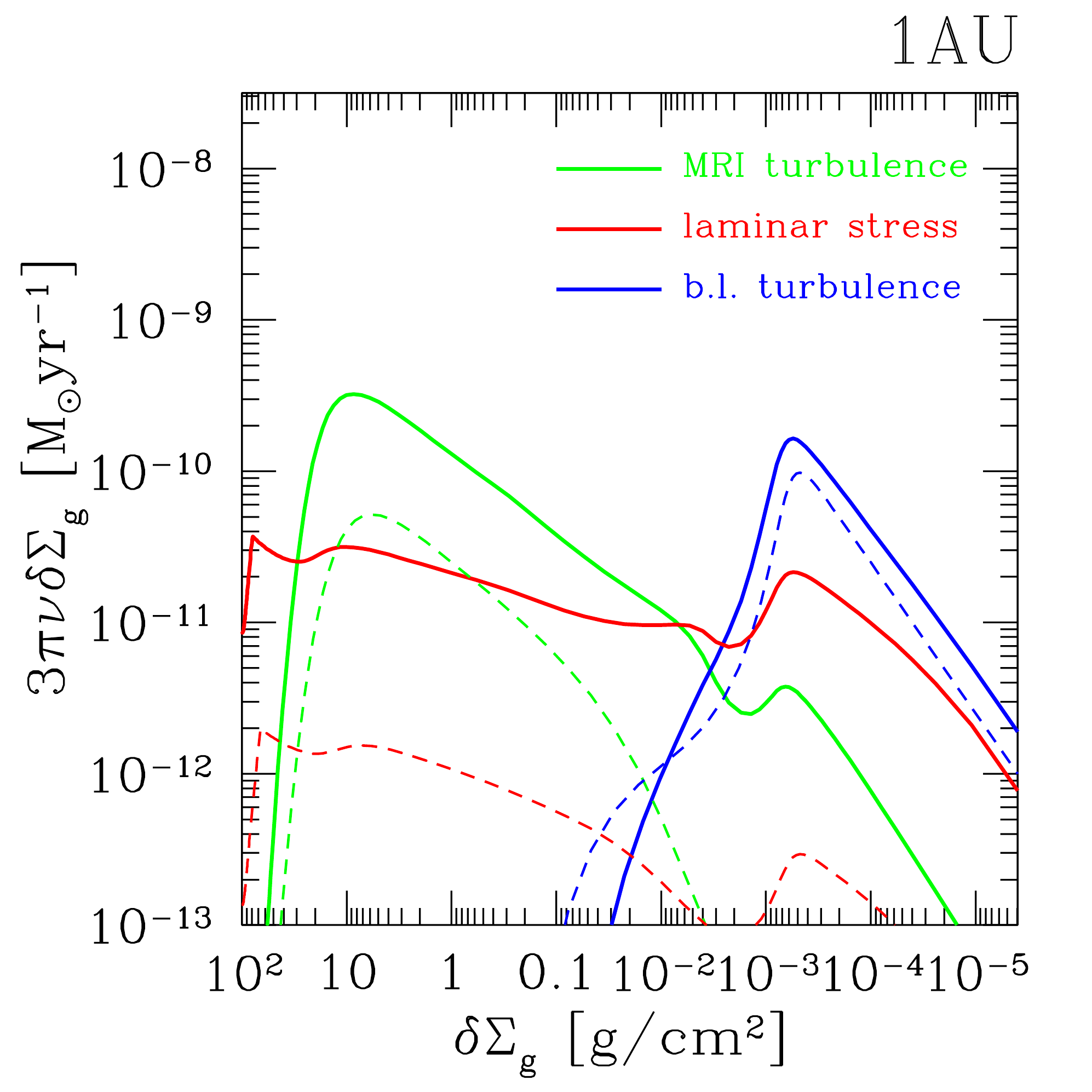}
\caption{Differential contribution to radial mass transfer, versus column $\delta\Sigma_g$ below disk surface.
The quantity $3\pi\nu\delta\Sigma_{g}$ is plotted separately for MRI, disk-wind turbulence, and laminar MHD stresses
(green, blue, and red curves).  For ease of comparison we define $\nu_{\rm lam} = \alpha_{\rm lam} c_g^2/\Omega$ 
using Equation (\ref{eq:alphalam}).  When estimating accretion rates in this paper, we generally ignore the contribution
from disk-wind turbulence.  Solid and dashed lines correspond to $\epsilon_B = 1$, 0.1 where the imposed radial field
is $B_R= \epsilon_B\left(R/0.1AU\right)^{-2}$ G.}
\vskip .1in
\label{fig:MdotSig}
\end{figure*}

\begin{figure*}[!]
\epsscale{1.1}
\plottwo{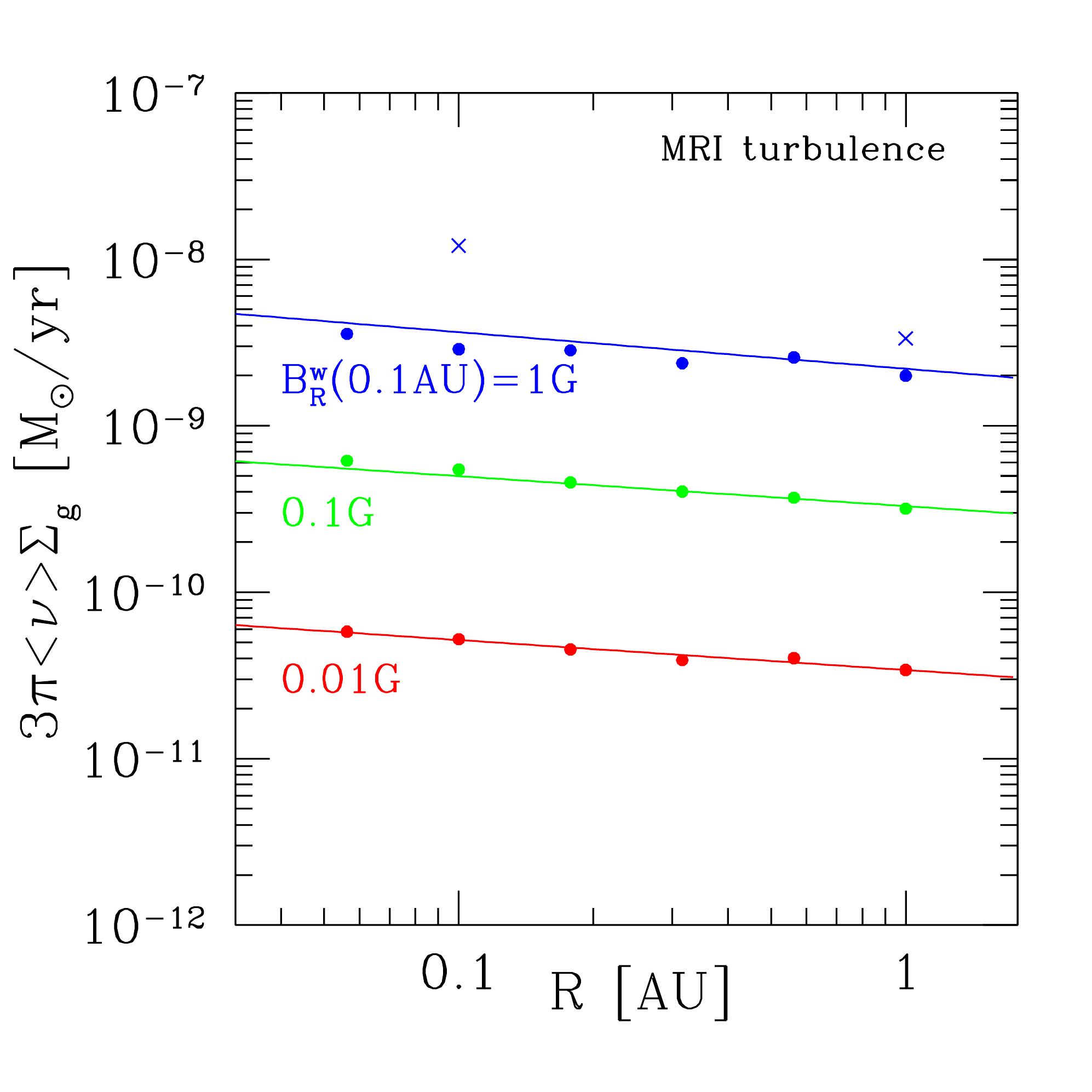}{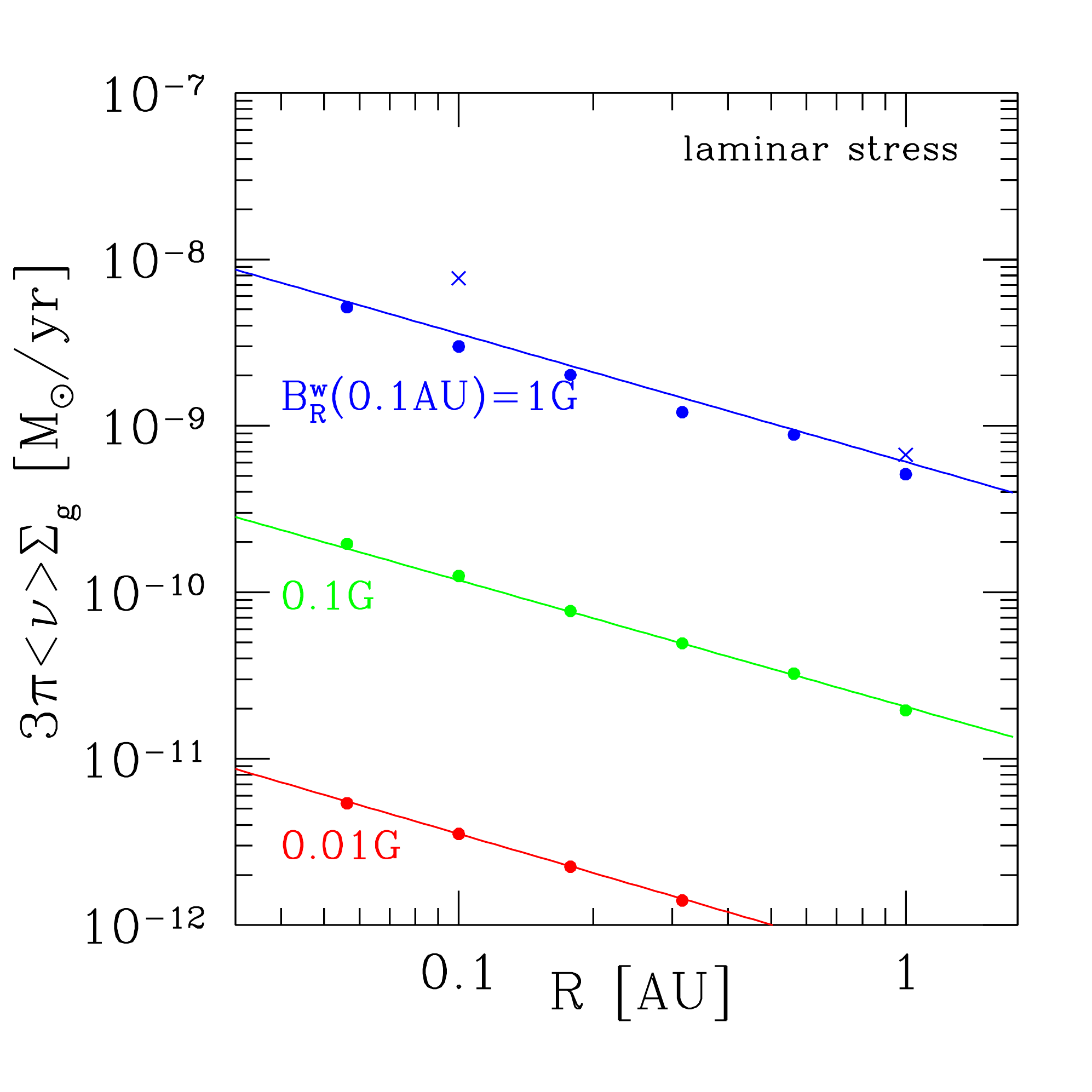}
\epsscale{0.55}
\plotone{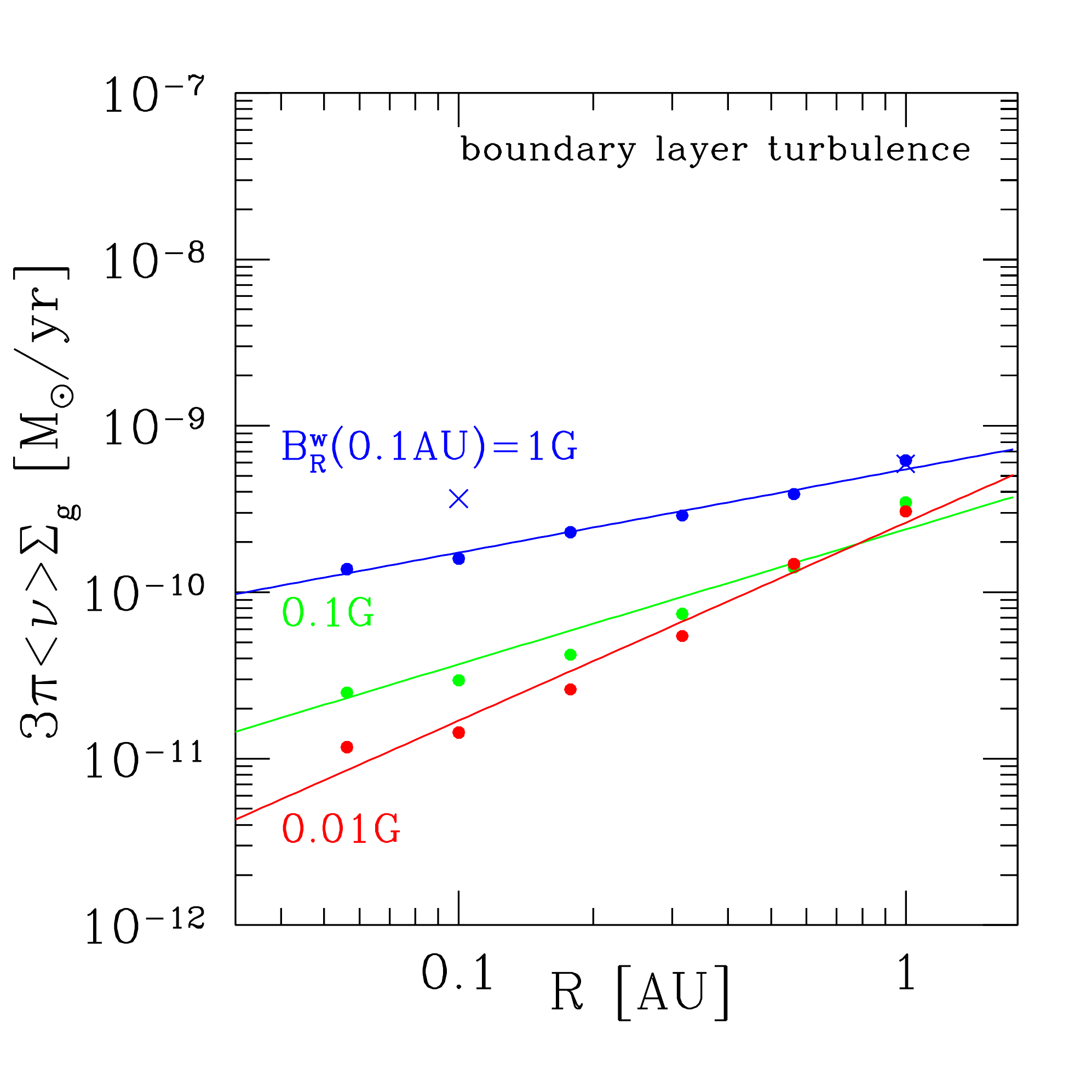}
\caption{Mass transfer rate that is driven separately by turbulent, laminar and boundary layer
stresses.  The quantities plotted in Figure \ref{fig:MdotSig} are integrated over height in the 
more highly magnetized hemisphere, and doubled to represent a disk that is seeded by a 
reflection-symmetric wind magnetic field.   Power-law fits (solid lines) for the strongest
applied field ($B_R = 1$ G at 0.1 AU, $\epsilon_B = 0.1$) are, top to bottom,
$2~\times~10^{-9}(R/{\rm AU})^{-0.22}$, $6.1~\times~10^{-10}(R/{\rm AU})^{-0.77}$, and 
$5.5~\times~10^{-10}(R/{\rm AU})^{0.5}\,M_\odot$ yr$^{-1}$.}  
\label{fig:MdotRfit}
\end{figure*}

\begin{figure}[!]
\epsscale{1.2}
\plotone{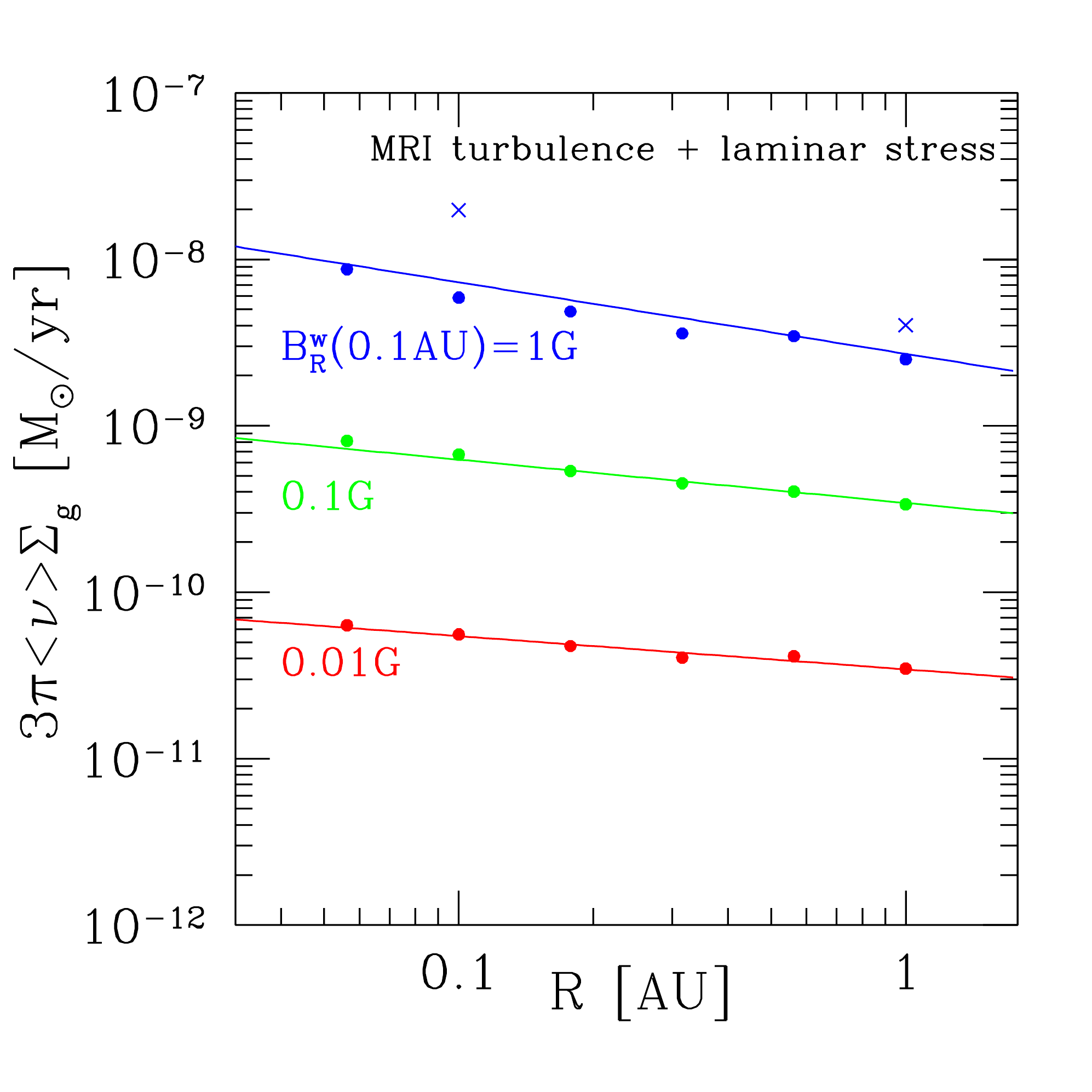}
\caption{Net mass transfer rate through both hemispheres that is driven by
turbulent and laminar stresses.  Power-law fit for the strongest applied
radial magnetic field: $3\pi\langle\nu\rangle\Sigma_{g,0}\approx
2.5\times 10^{-9}(R/{\rm AU})^{-0.4}\,M_\odot$ yr$^{-1}$.}
\vskip .1in
\label{fig:MdotRtotfit}
\end{figure}

\newpage
\subsection{Mass Transfer Rate}\label{sec:massflux}

Mass transfer in the disk is driven by a combination of MRI turbulence, the laminar stress $B_RB_\phi/4\pi$, and 
turbulence in the disk-wind boundary layer.   
Figure \ref{fig:alpha} shows the vertical profile of the turbulent amplitudes $\alpha_{\rm MRI}$ and 
$\alpha_{\rm mix}$.   Turbulence driven by disk-wind mixing is cut off sharply below the boundary layer.
The drop in $\alpha_{\rm MRI}$ above $\delta\Sigma_g \sim 10$--30 g cm$^{-2}$ reflects the onset of rapid Ohmic
diffusion.  The effective $\alpha$ parameter (\ref{eq:alphalam}) representing the laminar torque is also shown.

The mass transfer rate that is contributed by these stresses is shown versus column depth $\delta\Sigma_g$ 
in Figure \ref{fig:MdotSig}.  Although $\alpha_{\rm lam}$ only exceeds $\alpha_{\rm MRI}$ in
the diffuse upper disk and in the dead zone (where it is $\lesssim 10^{-5}$), the laminar stress still
contributes significantly to $\dot M$ closer than $\sim 0.1$ AU to the star.  

Figure \ref{fig:MdotRfit} shows the 
vertically summed contribution of each type of stress to the mass transfer rate.   Each varies with distance from the
protostar in a different way.  Importantly, the two dominant contributions (driven by the MRI and by the laminar
stress) both decrease with radius.  

The MRI-generated flow depends on the strength of the linearly amplified (seed)
toroidal field and the depth of the X-ray ionized layer.
Combining Equations (\ref{e:nut}), (\ref{eq:pran}), (\ref{e:Mdot}) and (\ref{e:betaeq}) gives
\ba \label{e:alphat}
\dot M_{\rm MRI}(R) &\,\approx\,& 3\pi (1+2\gamma_\nu) \nu_{\rm MRI} 
      \cdot 2\delta\Sigma_{\rm act} \nn
     &\approx& 1.8\times 10^{-8}\,\widetilde\alpha_{{\rm MRI},-1}^{1/3}\epsilon_B^{2/3}(1+2\gamma_\nu) \nn
     &&\times \left({2\delta\Sigma_{\rm act}\over 30~{\rm g~cm^{-2}}}\right)^{2/3}
     \left({R\over{\rm AU}}\right)^{0.31}\;M_\odot~{\rm yr}^{-1}.\nn
\ea
One finds $\dot M_{\rm MRI}(R) \propto R^{0.13}$ using the power-law fit to $\delta\Sigma_{\rm act}(R)$.
This is to be compared with the slightly negative gradient that is derived from the full vertical profiles
(Figure \ref{fig:MdotRfit}).  The difference arises from a somewhat larger $\alpha_{\rm MRI}$ near 
the base of the active layer in the inner disk.

We find that the laminar stress is competitive with the MRI stress at small radii, causing a non-negligible mass flux below 
the active layer. It also provides a strong contribution near the base of the boundary layer.
One finds (Equation (8) of Paper I)
\ba
{\dot M_{\rm lam}\over \dot M_{\rm MRI}} &\sim& 2 \left({9\widetilde\alpha_{\rm MRI}\over 4}\right)^{-2/3} 
     \left({B_R^2\over 8\pi P}\right)^{1/3} \nn
   &=& 0.1\,\left({\epsilon_B\over \widetilde\alpha_{{\rm MRI},-1}}\right)^{2/3}
    \left({\delta\Sigma_g\over 30~{\rm g~cm^{-2}}}\right)^{-1/3}\left({R\over {\rm AU}}\right)^{-0.65}.\nn
\ea
Note that this expression does not include the additional contribution to $\dot M_{\rm lam}$ from the laminar zone below the
MRI-active layer.  In agreement with the numerical results, it shows a stronger decrease with radius than $\dot M_{\rm MRI}$,
and comes close to reproducing the amplitude plotted in Figure \ref{fig:MdotSig}.

The sum $\dot M_{\rm MRI} + \dot M_{\rm lam}$ decreases with radius in both the numerical results and this analytic
approximation (Figure \ref{fig:MdotRtotfit}).
The analytic scalings $\dot M_{\rm MRI} \propto \epsilon_B^{2/3}$ and $\dot M_{\rm lam} \propto \epsilon_B^{4/3}$ are
confirmed from the vertical profiles.

The mass flux driven by boundary layer turbulence increases with radius (recall that $h_g^2\Omega\propto R^{3/2}$ for 
an atomic layer of uniform temperature $5000$ K).  It is negligible compared with $\dot M_{\rm MRI}$,
$\dot M_{\rm lam}$ when $\epsilon_B \sim 1$.   Because the structure of the boundary layer is not yet calibrated
by hydrodynamical simulation, we neglect its contribution to the disk evolution in Section \ref{sec:evo}. 

The active column $\delta\Sigma_{\rm act}$ is only logarithmically dependent on the inclination angle between disk 
surface and the stellar X-ray source.  The height $z_{\rm act}$ of the ionized column is pushed to a larger number
of thermal scale heights $c_g/\Omega$.  The pressure $P \sim \delta\Sigma_{\rm act} g(z_{\rm act})
= \delta\Sigma_{\rm act} \Omega^2 z_{\rm act}$ also increases.
On the other hand, the scale height in the active layer decreases, $h_g \sim c_g^2/g(z_{\rm act})$.  The net effect is
to {\it decrease} the turbulent viscosity as deduced from Equations (\ref{e:nut}), (\ref{eq:mageq}), and (\ref{eq:pran}):
\be
\nu_{\rm MRI} \propto \left({B_R^2\over 8\pi P}\right)^{1/3} h_g c_g \propto z_{\rm act}^{-4/3}.
\ee
This effect is only partly compensated by a mild increase in the equilibrium gas temperature resulting from the
increased irradiation.  We conclude that a negative scaling of $\dot M_{\rm MRI} + \dot M_{\rm lam}$ with radius
is actually enhanced if the inner disk starts with a higher column than we are using to construct our vertical
profiles (as it almost certainly does).

\subsection{Inner Disk without Settled Particles}\label{s:nopart}

We now construct the surface density profile corresponding to steady accretion, $d\dot M/dr = 0$, through an inner
disk without a significant mass of settled particles or embedded dust grains.  Strong depletions of gas, to
a surface density below below $\delta\Sigma_{g,\rm ion}$, are then obtained at accretion rates below 
$\sim 10^{-8}\,M_\odot$ yr$^{-1}$.   If the disk is able to reach this state,
then macroscopic (mm-cm sized) particles are self-consistently removed 
on a short timescale by inward drift.  The case where the decrease of
$\Sigma_g$ is halted by a continuous supply of particles from the outer 
disk is explored in detail in Section \ref{sec:solid}.

\begin{figure}[!]
\epsscale{1.2}
\plotone{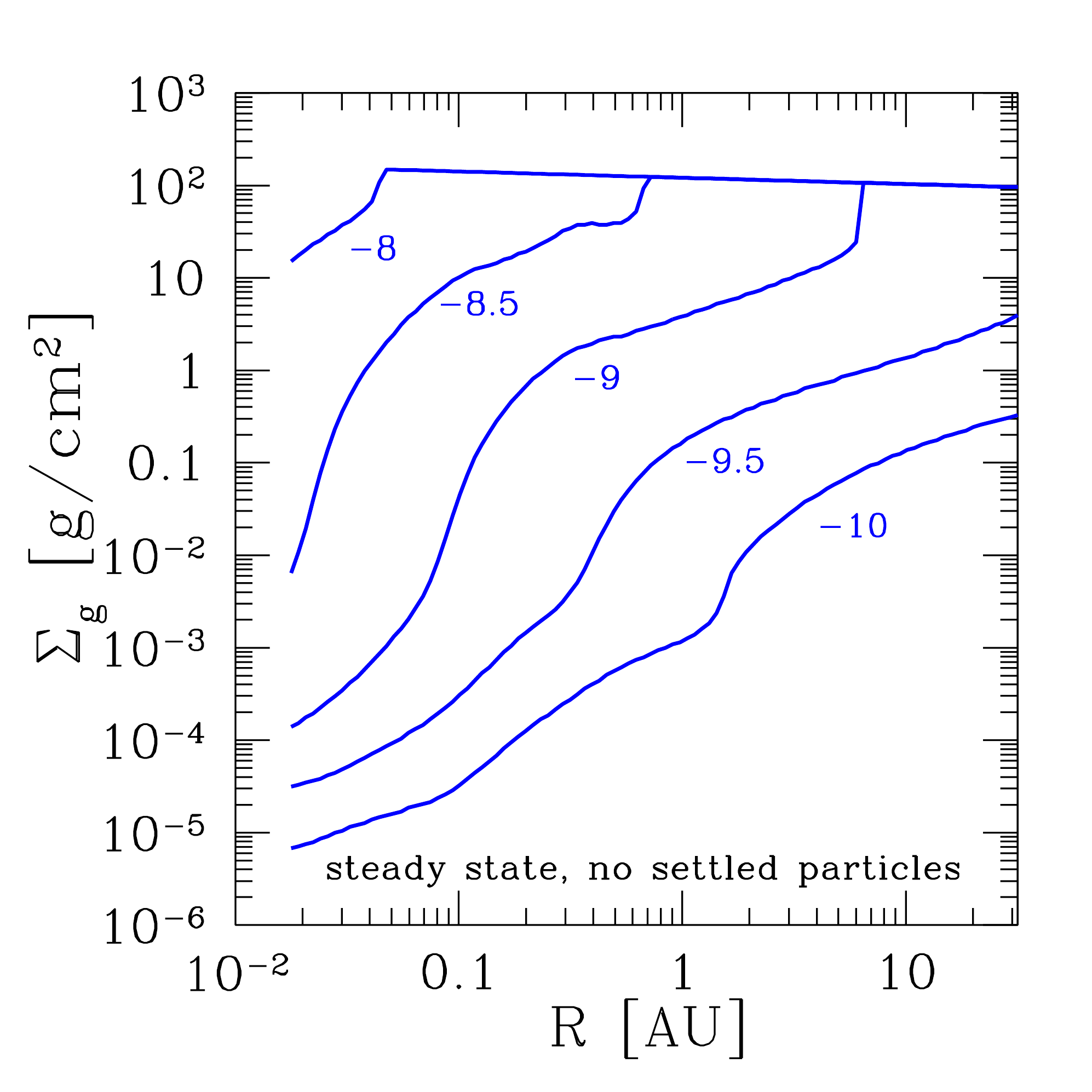}
\caption{Steady state column density corresponding to accretion at a uniform rate in a dust-free disk.
Curves obtained by integrating downward the vertical profiles of MRI and laminar stress. 
Labels correspond to $\log_{10}(\dot M_{\rm SS}/M_\odot~{\rm yr^{-1}}$).}
\vskip .1in
\label{fig:SigSS}
\end{figure}

The total gas column is obtained for a given $\dot M$ by inverting $\dot M = \dot M_{\rm MRI} + \dot M_{\rm lam}$
using the profiles of $\nu_{\rm MRI}$ and $B_RB_\phi/4\pi$ plotted in Figure \ref{fig:MdotSig}.  
The result is shown in Figure \ref{fig:SigSS}.  

One finds that the laminar torque dominates at small $\dot M$, 
resulting in very small total disk columns $\Sigma_g$.  An accretion rate $\sim 10^{-9}\,M_\odot$ yr$^{-1}$
can be supported by a total column of {\it atomic} gas as small as $\Sigma_g \sim 10^{-4}$ g~cm$^{-2}$ in the inner disk.  
This is primarily due the highly efficient laminar stress in the boundary layer, where the mass transfer rate is
\be
\dot{M}_{\rm lam} = 8.5\times10^{-11}\left|{\frac{B_{\phi}}{B_R}}\right|
\left(\frac{R}{0.1~{\rm AU}}\right)^{-1}T_{5000}^{1/2}\epsilon_{B}^{2}\;M_{\odot}{\rm yr}^{-1}.
\ee

This disk solution applies inside the sublimation radius $R_{\rm sub}$ of silicates (Equation (\ref{eq:rsub})), even if
settled particles are present further from the protostar.  The resulting outward gradient in $\Sigma_g$
has interesting implications for planetary migration (Section \ref{s:disc}).

\section{Optical Absorption Layer in a Dust-Depleted Disk}\label{sec:opt}

The disk profiles examined here assume that embedded dust has a minor impact on the ionization level
sustained by stellar X-rays.  This implies a mass fraction of $\mu$m-sized grains smaller 
than $X_d \sim 10^{-4}$.  Such a level of dust depletion still allows the formation of an optical
absorption layer.  In other words, the disk can remain optically thick to stellar light even while the
dust abundance is too small to reduce appreciably the equilibrium ambipolar Elsasser number ${\rm Am}$. 
Here we show how the height of the optical absorption layer depends on the level of dust depletion.

The inner part of a PPD will not appear as a transition disk if its column is reduced to $\sim \delta\Sigma_{g,\rm ion}$
and then sustained there by the fragmentation of settled particles.  The brightness of the disk in 
the near-IR is more ambiguous if the column falls further to the low level shown in Figure \ref{fig:SigSS}.

Because the number density of dust grains is much smaller than that of free electrons, electron adsorption on
grains is suppressed by the build-up of electric charge (see \citealt{ilgner2006,baigood2009}, and Sections
3.3 and 3.4 of Paper I).  Depletion of free electrons is
then mediated by the adsorption of positively charged molecular ions and metal atoms, followed by recombination
on grain surfaces.  An estimate of the critical value of $X_d$ is obtained by balancing the adsorption rate
\be\label{eq:gamads}
\Gamma_{\rm ads} = \pi a_d^2 n_d \left({8\mu_g\over\pi m_i}\right)^{1/2} c_g,
\ee 
where $a_d$ and $n_d$ are the grain radius and number density, against the recombination rate $\alpha_{\rm eff} n_e$
with a free electron in the gas phase.  Our choice of recombination coefficient $\alpha_{\rm eff}$ is given by
Equation (23) in Paper I, corresponding to an abundance $x_M = 10^{-8}$ by number of free metal atoms.  The 
geometric optical depth of spherical grains of mass density $\rho_s$ through a disk of scale height $h_g$ 
at a column $\delta\Sigma_{\rm act}$ below its surface is
\be
\tau_{{\rm opt},r} \sim {r\over h_g} {3X_d \delta\Sigma_g\over 4\rho_s a_d}.
\ee
Combining the above two equations with the relation between ${\rm Am}$ and $x_e$ gives
\be
\tau_{{\rm opt},r} \lesssim 1\times 10^2 {({\rm Am/10})\over (R/{\rm AU})^{0.2}(T/200\,\rm{K})^{0.4}}
    \left({\delta\Sigma_{\rm act}\over 10~{\rm g~cm^{-2}}}\right)^{-0.2}.
\ee
The vertical models constructed in Paper I reach ${\rm Am} \sim 100-300$ in the active layer, and so 
allow a substantial optical depth.

\begin{figure*}[!]
\epsscale{1.1}
\plottwo{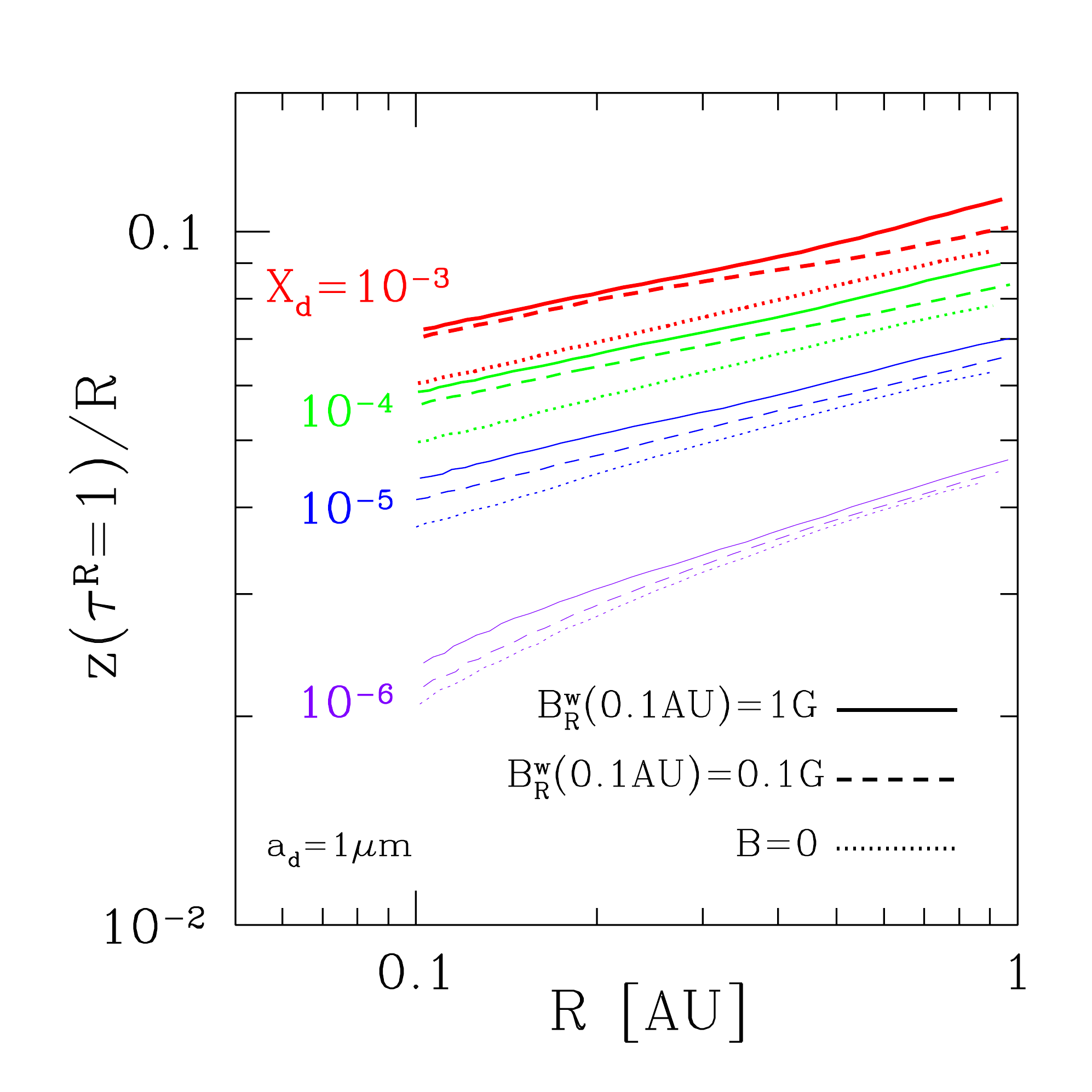}{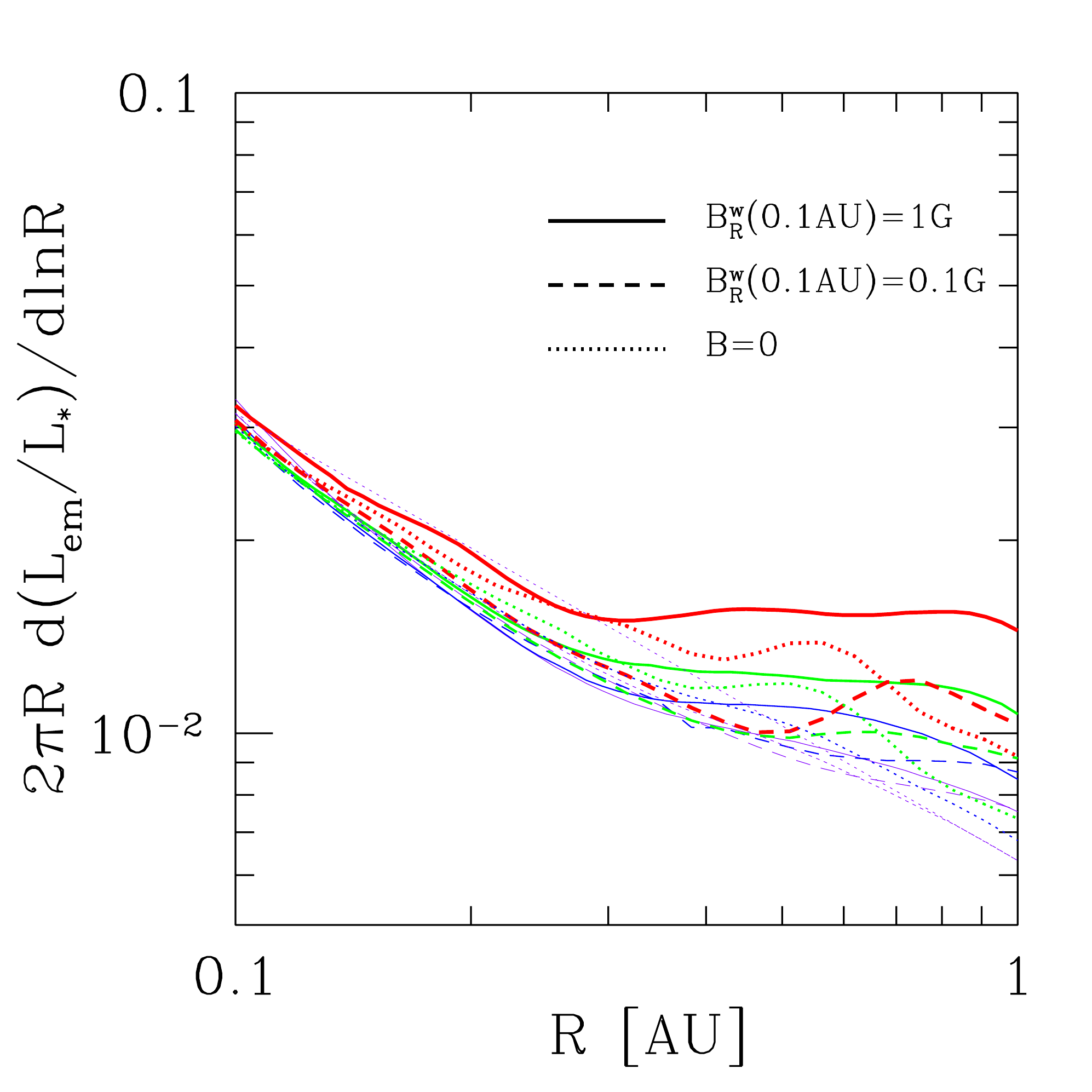}
\caption{\textit{Left panel:} height of the optical absorption layer, as a function of radius.  Stellar photon source is taken to sit at $(R,z)=(0.9,0.4)R_\star$.  
Colors show different dust loadings of the MRI-active layer, as measured by
$X_d/a_d$, where $X_d$ is mass fraction of small grains of radius $a_d$.  Disk ionization fraction
is only weakly perturbed for $X_d/a_d \lesssim 1$ cm$^{-1}$.  Solid (dashed) curves: applied radial 
magnetic field $\epsilon_B = 1$ (0.1).  Dotted curves:  unmagnetized disk.
\textit{Right panel:}  corresponding fraction of stellar luminosity that is re-radiated 
in various radial bins.}  
\vskip .1in
\label{fig:phot}
\end{figure*}

At millimeter wavelengths, where the outer parts of PPDs
are spatially resolved, depletion factors range from $0.1$ to $10^{-3}$ compared with the interstellar dust/gas
ratio $X_d\approx 10^{-2}$, and about $50\%$ of PPDs in the Taurus region are depleted by more than a factor
$\sim 10^{-3}$ in large grains \citep{furlan2006,dalessio2006}.  This depletion is frequently interpreted as
evidence of grain growth and settling.  

Previous modelling of the IR spectral energy distributions of T-Tauri and
Herbig Ae stars is consistent with an optical absorption layer sitting $1-5$ scale heights above
the disk midplane \citep{dalessio1999,chiang01}.  There is some evidence that the height of this layer 
decreases with age, reaching $z_{\rm opt} \sim (2-3) h_g$ at a few Myr and before the formation
of a transitional disk.  The radial dependence of $z_{\rm opt}/h_g$ inside $\sim 1$ AU is only weakly
constrained by mid-IR spectra, except in cases where a dust cavity is present \citep{williams11}.

We perform the exercise of varying $X_d$ while fixing the disk mass profile.
Here $X_d$ is, for simplicity, assumed to be independent of radius.
We focus on the radial distribution of the reprocessed IR emission from the flared disk surface.
We only consider the reradiated fraction of the stellar luminosity $L_\star$.  The spectrum of the re-emitted
IR also depends on the size distribution of grains, which is a complicated matter, and for that
reason we do not attempt to model it here.

Our procedure is to calculate the optical depth along stellar rays over
a range of polar angles, using a two-dimensional spline (in $R$, $z$) of the density profiles obtained
in Paper I. The optical absorption layer is defined by the position on each ray where $\tau = 1$ in the
visual band assuming $1\,\mu$m sized grains.  Rays are taken to originate from a point $(R,z)=(0.9,0.4)R_\star$
on the star's photosphere (determined by its average height), with $R_\star = 2R_\odot$.

The height of the resulting absorption layer is shown in Figure \ref{fig:phot} for $X_d/a_d = 10^{-1}$, 
$1$, $10~{\rm cm}^{-1}$.  It increases from 
$\sim 2h_g$ to $\sim 5h_g$  as $X_d/a_d$ is raised from $10^{-2}$ to 10.  It should be kept in mind that
fragmentation of grains in the upper disk will cause $a_d$ to decrease inward, so that $X_d/a_d$ need
not be constant even for a uniform mass fraction of grains advected by accreting gas.  The absorption
layer begins to disappear from the inner disk when $X_d/a_d$ drops below $10^{-2}$ cm$^{-1}$.  

Figure \ref{fig:phot} also shows that the fraction of $L_\star$ that is intercepted
by each radial annulus of the disk varies weakly with the dust loading when 
$X_d/a_d\lesssim 1$, as is considered in this work.  That is because the absorption layer flares more
strongly outward for lower values of $X_d/a_d$, thereby compensating the smaller total angle $z_{\rm opt}/R$
that is subtended by the layer.  In the upper range of $X_d/a_d$ considered, the absorption layer 
maintains a simple profile, $z_{\rm opt} \propto R^{9/7}$.  

The stronger flaring of the absorption layer that is seen for
smaller $X_d/a_d$ is partly explained by the larger angle of incidence of the stellar photons 
reaching the inner disk:  emission from a finite height above the disk tends to reduce the path length
through the gas and push the absorption layer to a higher vertical gas column.

The absorption layer also expands vertically with increasing strength of the imposed radial magnetic field.
In our disk model, this effect is strongly curtailed by imposition of (marginal) Newcomb-Parker stability. 
A mild flattening of the photosphere in the inner disk is caused by the slightly stronger
magnetic support there.  This also has the effect of introducing some mild shadowing of the disk at
intermediate radii ($\sim 0.4$ AU).  The net effect is that the IR emission from our model disk is
only weakly sensitive to the large radial variation in its internal magnetization.

\section{Radial Gas Flow Limited by \\ the Stirring of Solids}\label{sec:solid}

Our focus in this paper is on the flow of gas in a PPD after its self-gravity stops playing
a significant role in angular momentum transport.  Then solids condensed from the vapor phase will
settle to the disk midplane over a broad range of radius \citep{chiang10}.  Small grains
rapidly stick to form larger particles at gas columns $\gtrsim \delta\Sigma_{g,\rm ion}$, where MRI turbulence
is strongly suppressed. Particles may grow to mm-cm sizes during settling \citep{gw73}.

Here we consider the constraints on radial spreading that are imposed by the continuing interaction of the gas
with a settled particle layer.  We are interested here in the behavior of the disk at a stage where $\Sigma_g$ 
has {\it dropped} to the ionization threshold $2\delta\Sigma_{g,\rm ion}$.  
We have been considering disks whose upper layers are depleted in dust.
But even a relatively small surface density
$\Sigma_p$ in settled particles can suppress the MRI at $\delta\Sigma_g
\sim 1-10$ g cm$^{-2}$ if the particles are lofted high in the disk, 
followed by 
catastrophic fragmentation.  This corresponds to $\Sigma_p \gtrsim 10^{-4}(a_d/\mu{\rm m})
\delta\Sigma_{g,\rm ion}$,
where $a_d$ is the size of the `dust' fragments.  

\begin{figure}[!]
\epsscale{1.15}
\plotone{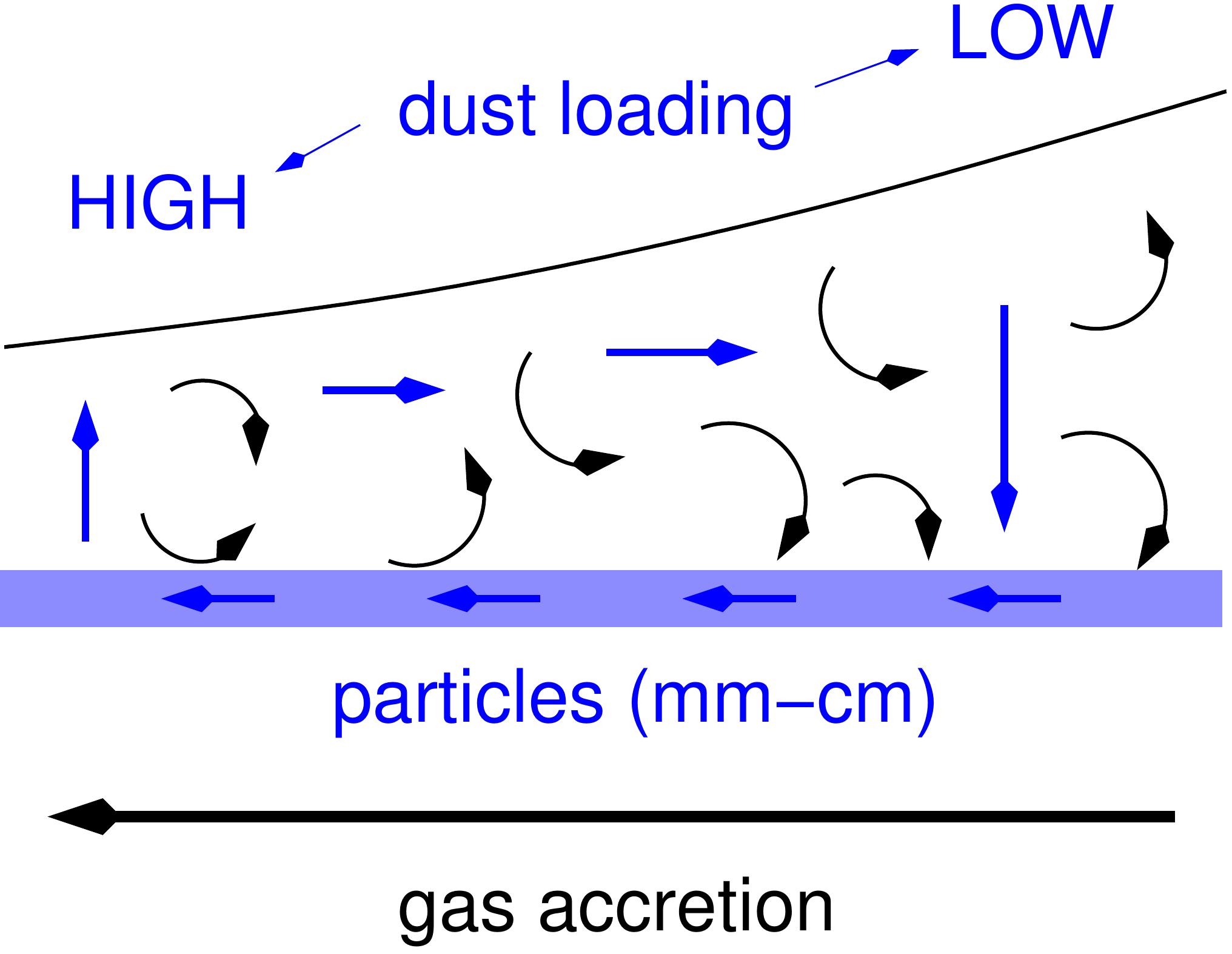}
\caption{Cycling of macroscopic settled particles and lofted dust grains in a disk with gas column
near the threshold for X-ray ionization of the midplane ($\Sigma_g \sim 10-30$ g cm$^{-2}$).
The settled particles drift inward to supply a higher equilibrium dust abundance in the inner disk.
The higher dust concentration there leads to outward diffusion of dust grains through the accreting
gas.  Re-adhesion of grains into larger particles allows excess dust to be removed from the outer
disk.}
\vskip .1in
\label{fig:dustcycle}
\end{figure}

The settled particles orbit more slowly than the gas 
where the gas pressure decreases outward, and therefore drift inward
toward the star.  A surface density of settled particles at least
comparable to the interstellar abundance $\sim 10^{-2}\delta\Sigma_{g,\rm ion}$
can be sustained by inflow from the heaviest parts of the disk outside $\sim 1$ AU.  We focus 
here on values of $\Sigma_p$ below the critical value $(c_g/\Omega R) (\ell_P/h_g) \Sigma_g$ where vertical
shear in the orbital velocity induces a strong Kelvin-Helmholtz instability in the particle layer \cite{sekiya98}.
Here $\ell_P = P/|dP/dr|$ at the midplane.  Then the mean density $\bar\rho_p$ of particles in
the layer remains well below $\rho_g$, and the inward drift speed of particles (of stopping time $t_{\rm stop}$
and Stokes parameter ${\rm St}_p = t_{\rm stop}\Omega$) is $v_r = - {\rm St}_p\, c_g^2/l_P \Omega$ \citep{gw73,weiden77}.

The mass flow in particles that will sustain a given $\Sigma_p$ is
\ba\label{eq:dmdtp}
\dot M_p &=& -2\pi r v_r \Sigma_p = 2\pi\epsilon_{\rm dr} \left({\Sigma_p \over \Sigma_g}\right) 
    {\rho_s a_p R c_g^2\over \Omega l_P} \nn
    &=& 1.4\times 10^{-12}\,\left({\Sigma_p/\Sigma_g\over 10^{-3}}\right) 
   \left({\epsilon_{\rm dr}\rho_s a_p\over {\rm g~cm^{-2}}}\right)
 \left({R\over {\rm AU}}\right)^{3/2}\nn
  &&\times \left({l_P\over r}\right)^{-1}\left({T\over 200\,\rm{K}}\right)\;M_\odot~{\rm yr}^{-1}.
\ea
The disk column is low enough that even macroscopic particles are in the Epstein drag regime;  here
$\epsilon_{\rm dr} \sim 0.4$.
The radial drift speed of the particles is high when $\Sigma_g \lesssim 2\delta\Sigma_{g,\rm ion}$.  
It nonetheless remains below the fragmentation speed unless the particles grow beyond $\sim$ cm radius.

We find that $X_d$ is a strong decreasing function of radius in the MRI-active layer, when
the dust abundance adjusts so as to sustain a steady inflow of {\it gas} through the inner disk.
Then the inward drift of macroscopic particles is compensated by an {\it outward} diffusive flow of lofted dust grains,
as depicted in Figure \ref{fig:dustcycle}.   The cycle can close within the disk when settled particles in the
inner disk are lofted by turbulent gas eddies, and excess dust in the outer disk re-adheres into larger particles.
We describe these processes in Section \ref{s:loft}.

\subsection{Equilibrium Surface Density Resulting from Particle Lofting and Midplane X-ray Ionization}\label{s:accdust}

The gas column that is maintained in the inner $\sim$ AU of an actively accreting PPD is the result of 
a competition between two effects:  increased X-ray penetration to the midplane (which accelerates 
MRI-driven turbulence), and the lofting of solids from the midplane (which suppresses the turbulence).  
We note that the feedback of settled dust on the MRI has been considered by \cite{jacquet12}, 
but without taking into account long-term secular changes in the gas column, or the effects of particle fragmentation.

\begin{figure*}[!]
\epsscale{1.1}
\plottwo{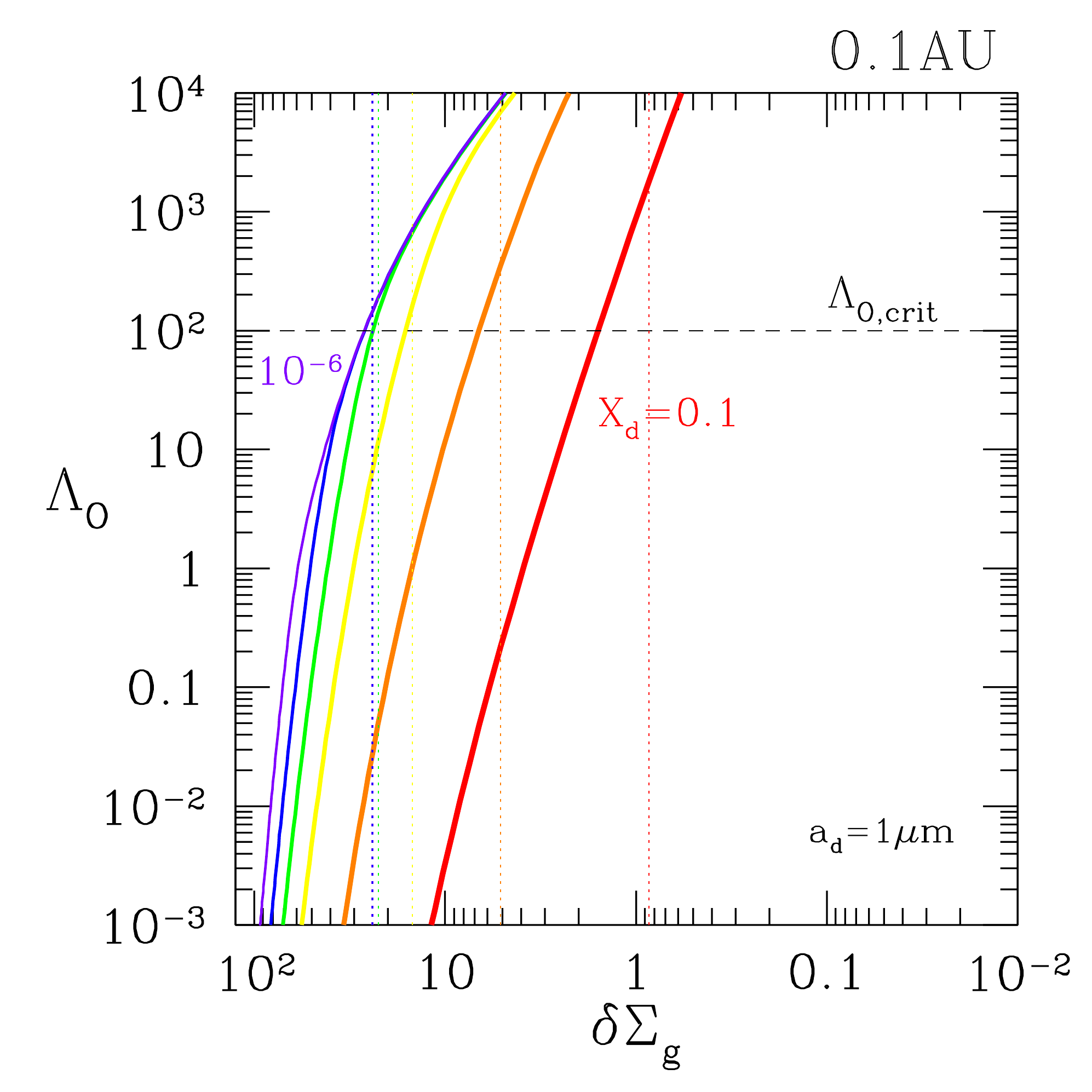}{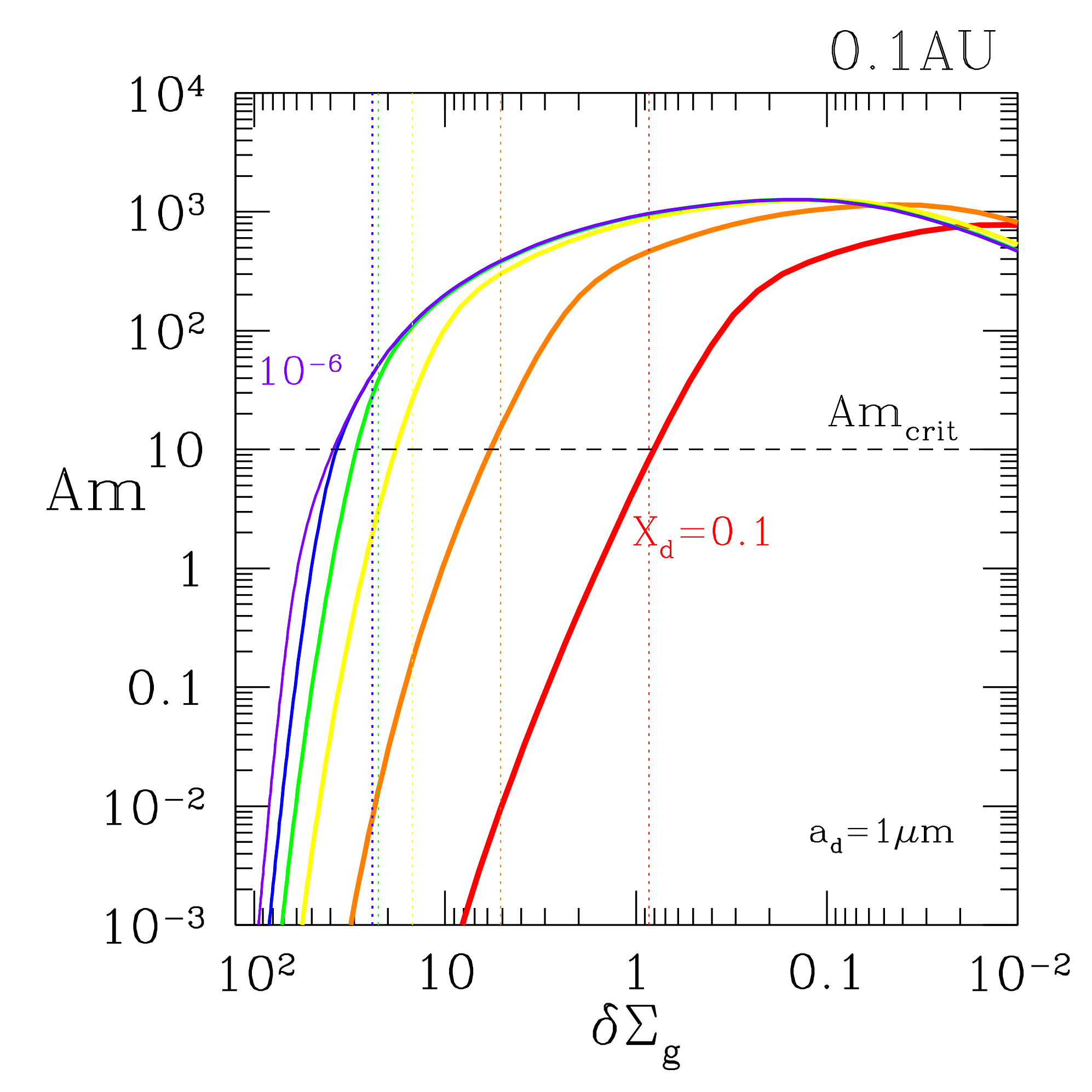}
\caption{\textit{Left panel:} Ohmic Elassser number as a function of column 
$\delta\Sigma_g$ below the disk surface. {\it Right panel:}  corresponding ambipolar Elsasser number.  
Dust with concentration parameter $X_d/a_d$ is added to the vertical profile that was constructed in Paper I
assuming $X_d = 0$.  The ionization fraction is reduced according to Equation (\ref{eq:xi1}).
Vertical dotted lines show the active column $\delta\Sigma_{\rm act}$, at
which non-ideal effects suppress $\alpha_{\rm MRI}$ by a factor $1/2$ according to Equation (\ref{e:alpha}).  The drop
in $\delta\Sigma_{\rm act}$ due to rising dust abundance is mainly caused by enhanced Ohmic diffusion 
at low $X_d$, and ambipolar drift at high $X_d$.}
\vskip .1in
\label{fig:sigact}
\end{figure*}

At any depth in the disk, there is a critical dust loading above which the ionization level begins to
be suppressed.  Two effects are important here.  First, an increase in the mass density of metals
increases the X-ray opacity.  This can reduce the ionization rate $\Gamma_i$ by an
order of magnitude at $\delta\Sigma_g \sim 10$ g cm$^{-2}$ when the dust abundance is restored to the solar value.  

Second, as is reviewed in Section \ref{sec:opt}, the abundance of free electrons is suppressed by the 
charging up of grains, followed by the adsorption of positive metallic or molecular ions on grain surfaces (Equation (\ref{eq:gamads})).
In equilibrium we have $x_e \Gamma_{\rm ads} = \Gamma_i$, which gives
\be\label{eq:xi1}
x_e = {4\rho_s a_d\over 3\delta\Sigma_g\Omega}\left({\pi m_i\over 8\mu_g}\right)^{1/2} {\Gamma_i\over X_d}.
\ee
This expression holds when $x_e$ is much smaller than $(\Gamma_i/\alpha_{\rm eff} n_H)^{1/2}$; otherwise
we revert to the expression corresponding to negligible adsorption.

Thus an enhancement in the dust abundance forces a reduction
in the active column $\delta\Sigma_{\rm act}$, which is defined by
setting the cutoff factors in Equation (\ref{e:alpha}) to $1/2$.
The result is shown in Figure \ref{fig:sigact}.  The reduction 
in $\delta\Sigma_{\rm act}$ is driven mainly by enhanced Ohmic 
diffusion at low $X_d$, and by ambipolar drift at high $X_d$.  

\begin{figure*}[!]
\epsscale{1.1}
\plottwo{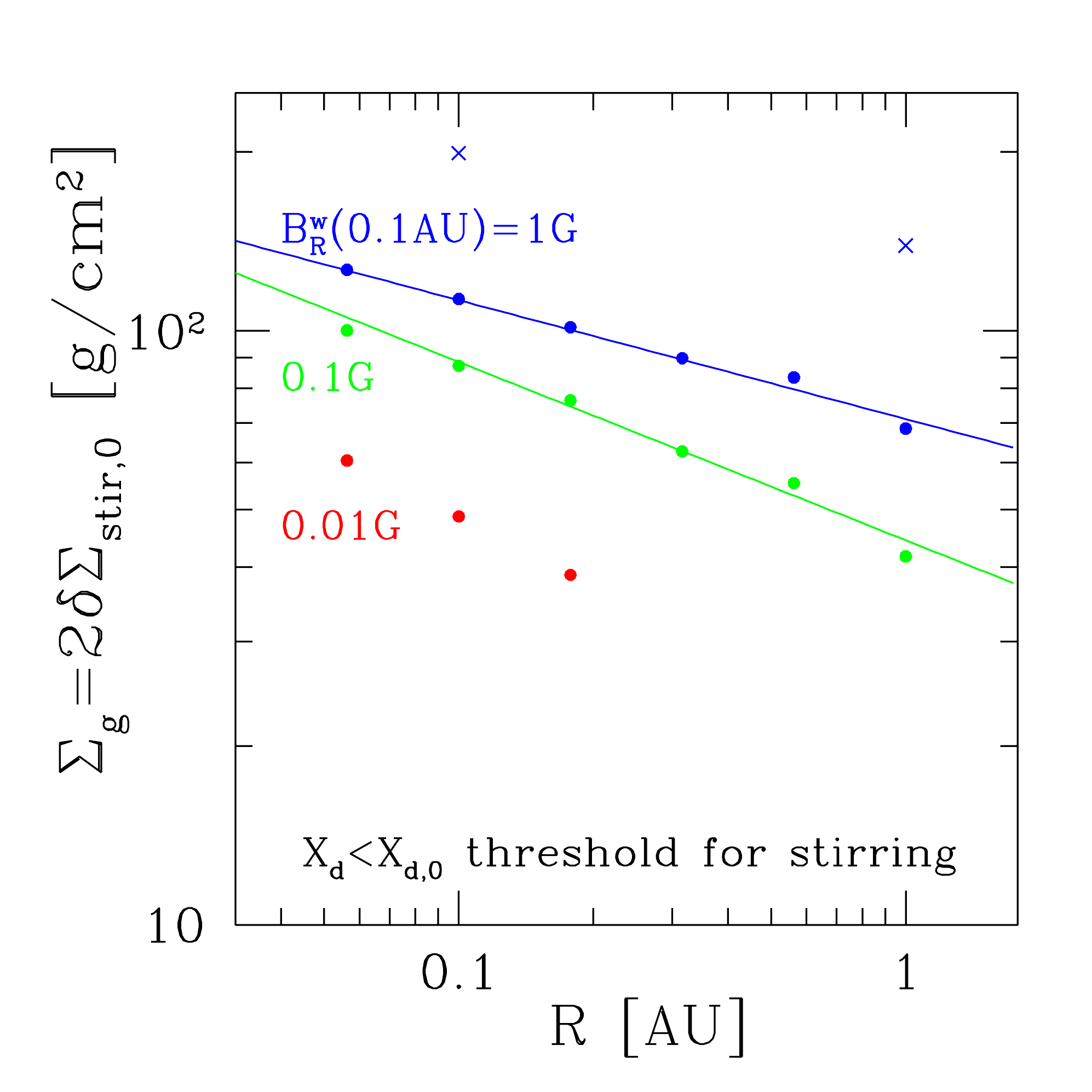}{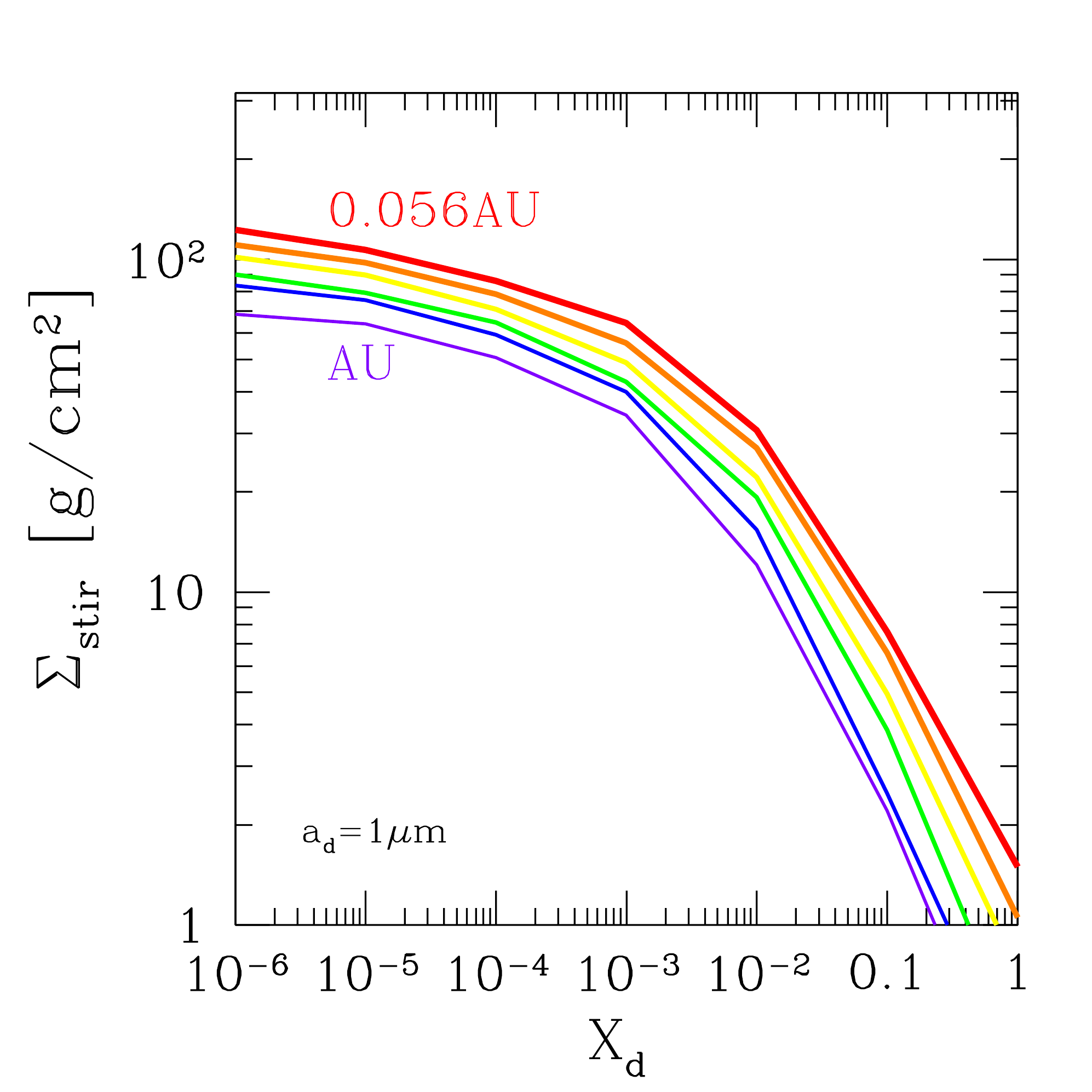}
\caption{\textit{Left panel:} threshold total gas column below which macroscopic (mm-cm sized) particles
will be stirred up from the disk midplane, leading to catastrophic fragmentation and loading of
the upper disk with small grains.  Critical $\alpha_{\rm MRI}$ for stirring is
taken to be $10^{-4}$, as described in the text. Colors correspond to different imposed radial magnetic fields.
Power law fits are $\Sigma_g= 70$ $\left(R/\rm{AU}\right)^{-0.2}$ [44 $\left(R/\rm{AU}\right)^{-0.3}$] g cm$^{-2}$ for $B_R =1$ [0.1]
G at 0.1 AU ($\epsilon_B = 1$, 0.1).  \textit{Right panel:} suppression of stirring column caused by
rising dust abundance with uniformly mixed micron-sized grains.  Colors correspond to different values of $R$, separated
by $0.25$ in $\log_{10}R$.  Maintaining uniform $\dot M$ in the inner disk requires progressively higher
dust loadings toward smaller $R$.}
\vskip .1in
\label{fig:stir}
\end{figure*}

Below the layer of strong MRI activity lies a deeper and more weakly ionized zone, within which the
turbulent motions are still fast enough to loft macroscopic particles from the midplane.  We take
for illustration a critical turbulent amplitude $\alpha_{\rm MRI} \sim 10^{-4}$ for particle lofting.
The corresponding stirring column $\delta\Sigma_{\rm stir,0}$ is plotted in the left panel of 
Figure \ref{fig:stir} for a few values
of the applied $B_R$.  This column is obtained from our dust-free vertical disk profiles.
For the maximum field strength considered ($\epsilon_B = 1$), we find $\delta\Sigma_{\rm stir,0}
\approx 35(R/{\rm AU})^{-0.2}$ g cm$^{-2}$.  By way of comparison, the critical column for stirring is roughly three times
larger than the total active column across the disk (Figure \ref{fig:hR}).  

Adding a uniformly mixed population of small grains (we choose $a_d=1$ $\mu$m) to the gas reduces the ionization level and
both the active column $\delta\Sigma_{\rm act}$ and the stirring column $\delta\Sigma_{\rm stir}$.  For each value of 
the grain mass fraction $X_d$, one can work out the adjusted ionization profile $\Gamma_i$ and ionization level $x_e$ from 
Equation (\ref{eq:xi1}) and calculate the resulting ambipolar number ${\rm Am}$.  (Note that the form of the ambipolar 
number given by Equations (\ref{eq:etas}) and (\ref{eq:lam0}) is modified when grains are the main charge carriers, but 
this occurs at columns greater than $\delta \Sigma_{\rm act}$ and so has very little effect for our purposes.)
The adjustment in $x_e$ is greater at larger $\delta\Sigma_g$, and disappears below
the column where $X_{d,0} = X_d$.  Here we take a short cut by maintaining the magnetic field profile 
$B_\phi(\delta\Sigma_g$) of our zero-dust disk solutions, and then adjusting $\Lambda_{\rm O}$ through 
Equation (\ref{eq:lam0}).  This in turn gives the re-scaled ambipolar number and MRI diffusivity 
$\nu_{\rm MRI}$ using Equation (\ref{e:alpha}).

\begin{figure*}[!]
\epsscale{1.1}
\plottwo{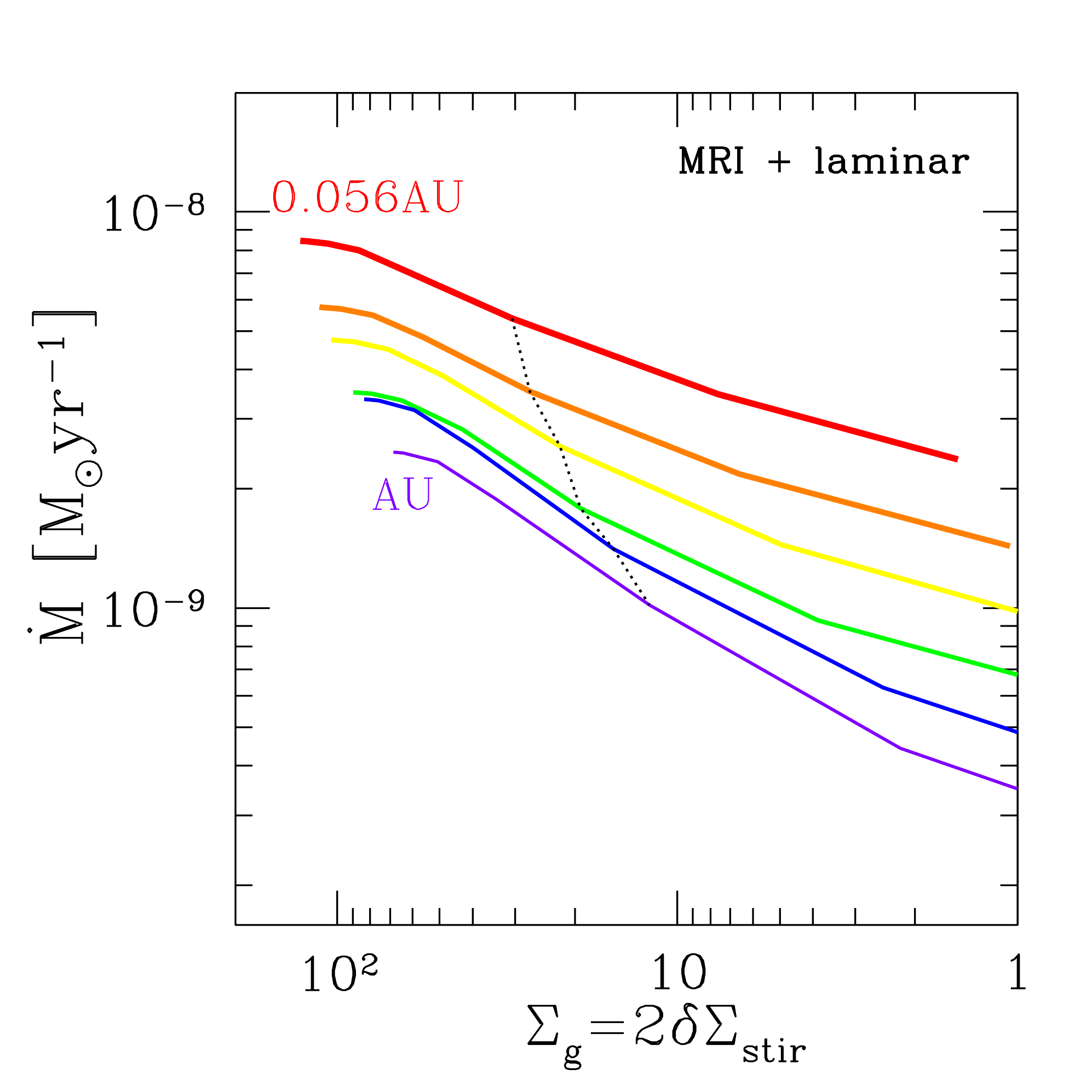}{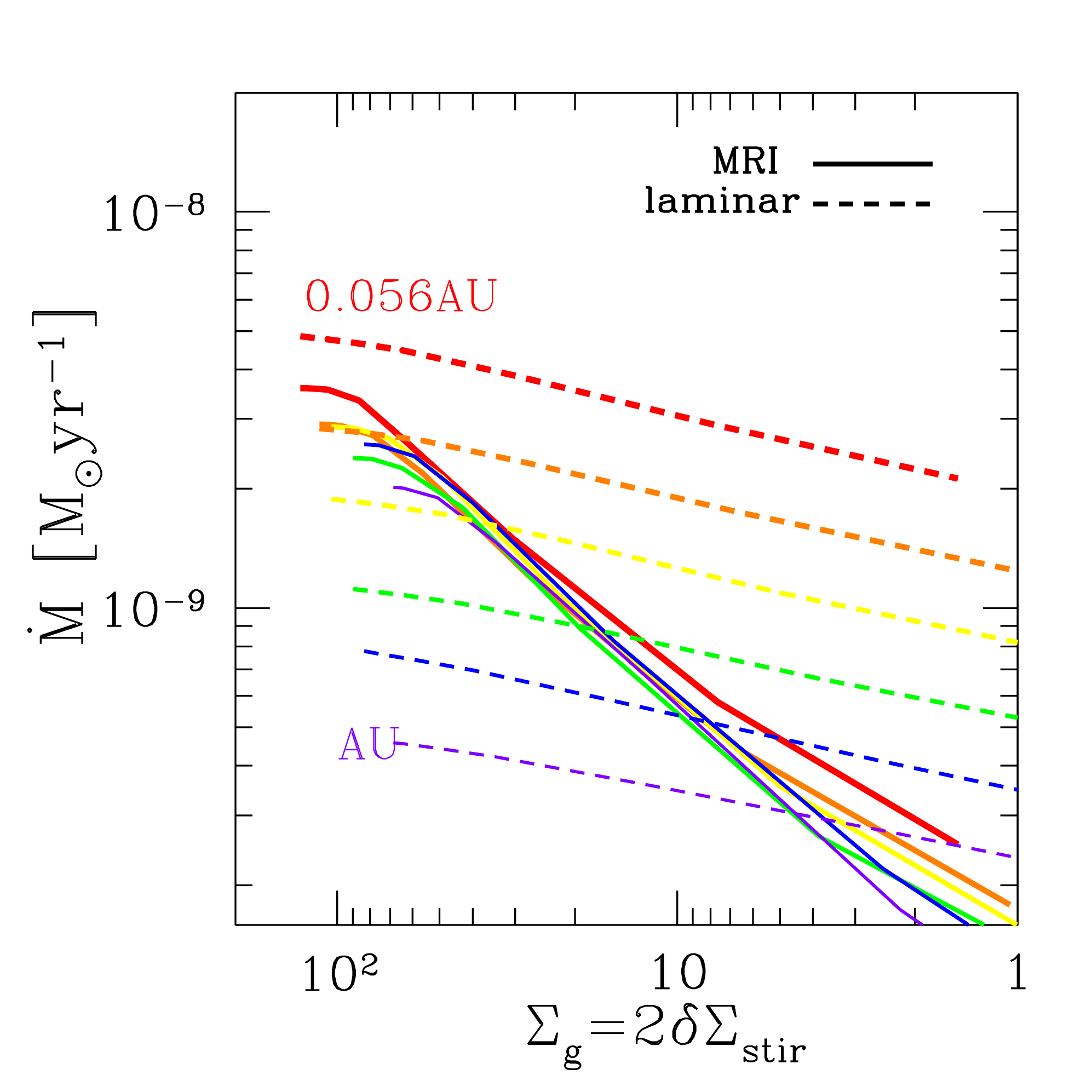}
\caption{\textit{Left panel:} total accretion rate in a disk whose active layer is limited in thickness by the lofting
of particles and fragmentation into dust (Equations (\ref{eq:nudust}) and (\ref{eq:mdotdust})).
Points to the right correspond to higher dust mass fraction (up to $X_d=1$ with micron-sized grains).  
Black dotted line:  accretion rate $X_d\dot M$ of the lofted dust
grains reaches a fraction $10^{-2}$ of the total accretion rate $\dot M$.
Small grains must be sourced by inward-drifting particles in the midplane.  When the 
corresponding solid particle accretion rate $\dot M_p$ drops below $X_d\dot M$, the gas
surface density is no longer buffered from below.  Interior to this point, the disk transitions 
to much lower $\Sigma_g$ as shown in Figure \ref{fig:SigSS}.   \textit{Right panel:} separate contributions 
to the accretion rate from MRI and laminar torques.  The MRI contribution 
is directly related to $\delta\Sigma_{\rm stir}$, hence the overlap of the solid curves.}
\vskip .1in
\label{fig:mdotdust}
\end{figure*}

The last step is to calculate a vertically summed accretion rate from the truncated profile
of MRI activity at finite $X_d$,
\be\label{eq:nudust}
\dot M_{\rm MRI} = 3\pi \langle \nu_{\rm MRI}\rangle \cdot \delta\Sigma_{\rm stir} = 
3\pi\int_0^{\delta\Sigma_{\rm stir}} \nu_{\rm MRI} d\Sigma_g
\ee
and
\be\label{eq:mdotdust}
\dot M_{\rm lam} = {1\over 2\Omega}\int^{\delta\Sigma_{\rm stir}} {B_RB_\phi\over\rho_g} d\Sigma_g.
\ee
Figure \ref{fig:mdotdust} shows these quantities, together with the summed $\dot M$.  
We also obtain a vertically averaged viscosity coefficient $\langle \alpha \rangle =
\dot{M}/3\pi h_g^2\Omega$, which is used in Section \ref{s:disc} to calculate the migration rates
of planets within the disk.

In summary, we have obtained the mass transfer rate in a low-column disk where the ionization
level is determined by a balance between (i) activation of the MRI near the midplane by X-ray ionization; and 
(ii) dust formation driven by the fragmentation of lofted particles.  Analytic expressions for the separate contributions
from turbulent and laminar Maxwell torques are
\be \label{e:MdotstirMRI}
\dot{M}_{{\rm MRI}}\approx4\times10^{-9}\left(\frac{2\delta\Sigma_{{\rm stir}}}{10^2~{\rm g~cm^{-2}}}\right)  \; 
{\rm M_\odot~yr^{-1}}
\ee
and
\be \label{e:Mdotstirlam}
\dot{M}_{{\rm lam}}\approx5\times10^{-10}\left(\frac{2\delta\Sigma_{{\rm stir}}}{10^2~{\rm g~cm^{-2}}}\right)^{0.2}
\left(\frac{R}{\rm AU}\right)^{-0.7} {\rm M_\odot~yr^{-1}}.
\ee

The surface density profile $\delta\Sigma_g(R)$ corresponding to steady accretion through the inner part 
of a PPD can be obtained by inverting $\dot M = \dot M_{\rm MRI}[\delta\Sigma_g] + \dot M_{\rm lam}[\delta\Sigma_g(R)]$.
The radial dependence of $\delta\Sigma_g$ is mainly determined by the radial variation of $\dot M_{\rm lam}$.

Maintaining a steady mass flow in the inner disk requires the dust loading to increase to sharply towards the star,
reaching $X_d\approx (0.1$--$1)\left(a_d/\mu m\right)$ inside $0.1$ AU.  Although small grains are tightly bound to
the gas, the strong negative $dX_d/dr$ implies outward transport of grains within the MRI-active layer. Therefore
a partly closed cycle of solids is possible in the inner disk, as sketched in Figure \ref{fig:dustcycle}.

The residency time of the solids depends on their rate of leakage through the inner boundary of the
dust-loaded disk.  This boundary cannot push inside the dust sublimation radius,
\be\label{eq:rsub}
R_{\rm sub} = 0.04\,\left({T_{\rm sub}\over 1800\,\rm{K}}\right)^{-2}\left({L_\star\over L_\odot}\right)^{1/4}\;{\rm AU}.
\ee
Closer to the star, where the ionization level returns to the dust-free case,
the surface density must drop sharply, to the level plotted in Figure \ref{fig:SigSS}.  
The inward transport of lofted grains could be suppressed outside $R_{\rm sub}$ by the strong outward force imparted 
by the absorption of stellar optical light.  The transport of settled particles depends on how how laminar the 
flow is near the inner boundary of the dust-loaded disk, an issue which is addressed in Section \ref{sec:dust}.

A higher loading of dust in the inner disk will raise the height of the optical absorption surface, thereby 
opening the possibility of disk self-shadowing.   The absorption surfaces shown in Figure \ref{fig:phot} are calculated
for a uniform dust loading $X_d$, and when $X_d/a_d\gtrsim 0.1$ cm$^{-1}$ 
they coincide with nearly fixed multiples of the scale height.
The finite size of the star reduces the shadowing effect of higher
$X_d$ in the inner disk.  We find that the value of $X_d$ 
that is implied by uniform accretion would produce enough absorption to block light from the stellar 
equator, but not from the poles.  This partial self-shadowing would be further compensated by a modest reduction in gas scale height
in response to the reduced irradiation of the inner disk. A self-consistent model of shadowing effects in a radially inhomogeneous
and dust-loaded disk is beyond the scope of this paper.

\subsection{Particle Lofting and Fragmentation, and Dust Settling}\label{s:loft}

The accretion solutions obtained in Section \ref{s:accdust} do not depend on the details of how macroscopic particles
are lofted and fragment.  Here we examine how an equilibrium concentration of small grains can be maintained
by a competition between fragmentation and mutual grain sticking.

Even massive particles with ${\rm St}_p > \alpha_{\rm MRI}$ can be advected upward from the upper edge of the 
particle layer by individual eddies of speed $\alpha_{\rm MRI}^{1/2}c_g$.   When ${\rm St}_p < \alpha_{\rm MRI}^{1/2}$,
this process only allows particles to reach a height $\sim c_g/\Omega$ above the midplane.   But $\alpha_{\rm MRI}$
increases rapidly upward to a value $> {\rm St}_p$, meaning that some particles will continue to diffuse away
from the midplane, into a thin column\footnote{The stopping time $t_{\rm stop} \sim \epsilon_{\rm dr} 
\rho_s a_p/\rho_g c_g$ when the particle size is smaller than the mean free path of H$_2$ molecules.  
Hence ${\rm St}_p \sim \epsilon_{\rm dr} \rho_s a_p/\delta\Sigma_g$.}
\be
\Sigma_{g,\rm min} \;\sim\;  {\epsilon_{\rm dr}\rho_s a_p\over\alpha_{\rm MRI}}
 \;\sim\; 10^{-1}\left({a_p\over{\rm mm}}\right)\alpha_{\rm MRI,-1}^{-1}
     \; {\rm g~cm^{-2}}
\ee
below the disk surface.  Here the vertical diffusion speed of a particle of radius $a_p$
drops below the vertical drift that is imposed by the disk gravity, $\alpha_{\rm MRI} \sim {\rm St}_p$.

A key point is that collisions between diffusing particles are delayed
until their relative speed $v_{\rm col}$ rises well above the fragmentation speed ($v_{\rm frag} \sim 1$ m 
s$^{-1}$ for conglomerates of $\mu$m-sized silicate grains).  That is because the vertical diffusion time
$\sim \alpha_{\rm MRI}^{-1}\Omega^{-1}$ is shorter than the mean time between collisions.  
The relative velocity of particles of somewhat different sizes is comparable
to their drift speed through the gas,\footnote{Here the velocity field is assumed to have a Kolmogorov
spectrum on a scale $< \alpha_{\rm MRI}^{1/2} h_g$.} 
\ba
v_{\rm col} &\sim& (\alpha_{\rm MRI}{\rm St}_p)^{1/2} c_g \nn
&\sim& 20\alpha_{{\rm MRI},-1}^{1/2}{ (a_p/{\rm mm})^{1/2}
(T/200\,\rm{K})^{1/2}\over (\delta\Sigma_g/30~{\rm g~cm^{-2}})^{1/2} }
\;\;{\rm m~s^{-1}}.\nn
\ea
 
The net abundance of small grains results from a balance between destructive collisions
involving a large particle with a high drift speed, and low-velocity collisions between grains
which lead to coagulation.  The production of small grains is dominated by a `sandblasting'
effect:  a high speed collision between a large particle and a grain will eject a multiple
$Y$ of the grain mass \citep{jacquet14}.  Then the mass density in grains $\bar\rho_d$ is determined by 
\be
  Y {\bar\rho_p\over m_s}{\bar\rho_d\over m_d} \pi a_p^2 v_{\rm col}(a_p)
 = \left({\bar\rho_d\over m_d}\right)^2 4\pi a_d^2 v_{\rm col}(a_d).
\ee
giving
\be\label{eq:sbal}
{\bar\rho_d\over\bar\rho_p} = {Y\over 4} \left({a_d\over a_p}\right)^{1/2}.
\ee
Experiments suggest $Y(v_{\rm col}) \sim (v_{\rm col}/v_{\rm frag})^\beta$ with $1 < \beta < 2$ 
\citep{housen83,holsapple93}.

This channel for dust production by stirred particles also allows
for the removal of dust by the re-growth of particles, as we find is
necessary to sustain a steady cycle of solids in the inner disk (Figure \ref{fig:dustcycle}).

The upward flux of particles of mass $m_s$ can be written as
\be 
{dN_s\over dA dt} \sim \nu_{\rm MRI}\rho_g {\partial\over\partial z}\left({\bar\rho_p\over m_s\rho_g}\right).
\ee
Given that collisions are slow compared with vertical diffusion, the gradient of the particle concentration
$\bar\rho_p/\rho_g$ is small.  Then $\bar\rho_p \propto \rho_g$.  The dust mass density varies
with depth in the disk since it also depends on the particle drift speed through the yield $Y$
(Equation (\ref{eq:sbal})).

\section{Evolution of the Inner Disk}\label{sec:evo}

We now consider the clearing of mass from a PPD.
This process is largely completed over the first $\sim 3$--10 Myr \citep{hernandez2007}, it appears 
by internal torques inside 1--2~AU, combined with a wind that is driven from the outer disk
by the intense UV and X-ray flux from the protostar.  
A direct magnetorotational outflow
from the disk surface is assumed to be suppressed by the pressure of the stellar wind, for the reasons outlined in Paper I. 
We also neglect turbulence that is driven by a purely hydrodynamical instability, e.g. vortices \citep{marcus14}.

We first prescribe the initial surface density profile.
Three initial conditions are considered: the immediate aftermath of a FU Ori-like outburst following the recondensation
of silicates; the `minimum-mass solar nebula' (MMSN, e.g. \citealt{hayashi81}); and a flatter density profile 
$\Sigma(R) \propto R^{-1}$ with the same normalization at 2 AU as the MMSN.
The first profile is derived in Appendix \ref{s:appendixa}:
\ba\label{eq:sigfuori}
\Sigma_g(R) &=& 3\times 10^3\left({R\over {\rm AU}}\right)\;{\rm g~cm^{-2}} \quad (R < R_Q \sim 2~{\rm AU});\nn
\Sigma_g(R) &=& {c_g\Omega\over \pi Q(R)} = {c_g\Omega\over 2\pi} \quad (R > R_Q).
\ea
Here $Q = c_g\Omega/\pi\Sigma_g$ is the \cite{toomre64} parameter for (axisymmetric) gravitational stability of
a thin Keplerian disk.  The inner zone is stable by this criterion ($Q > 2$), and the outer zone is marginally stable.  
In the second case,
\be
\Sigma_{g,\rm MMSN}(R)=1700\left({R\over {\rm AU}}\right)^{-3/2}\;{\rm g~cm^{-2}}.
\ee
\begin{figure*}[!]
\epsscale{1.1}
\plottwo{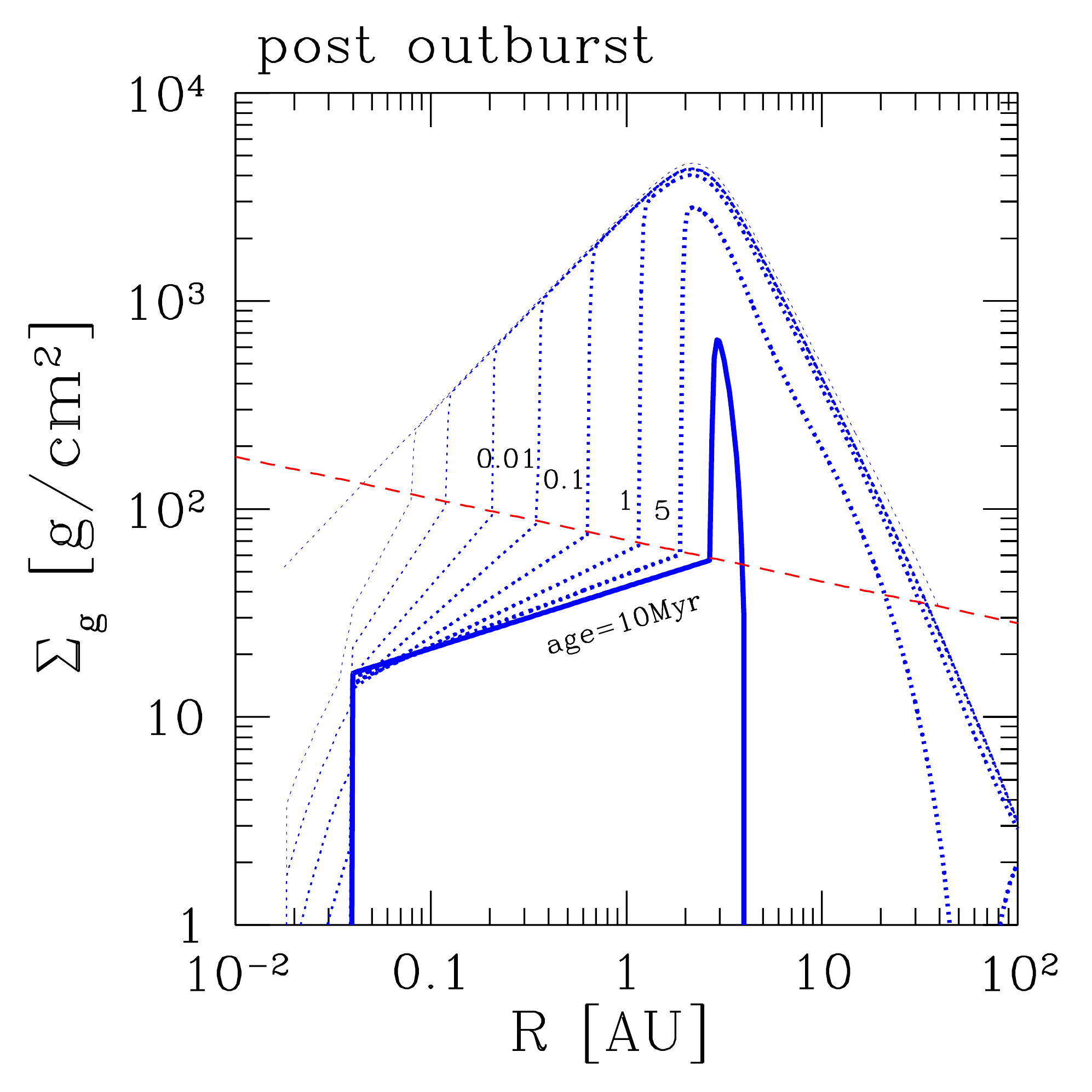}{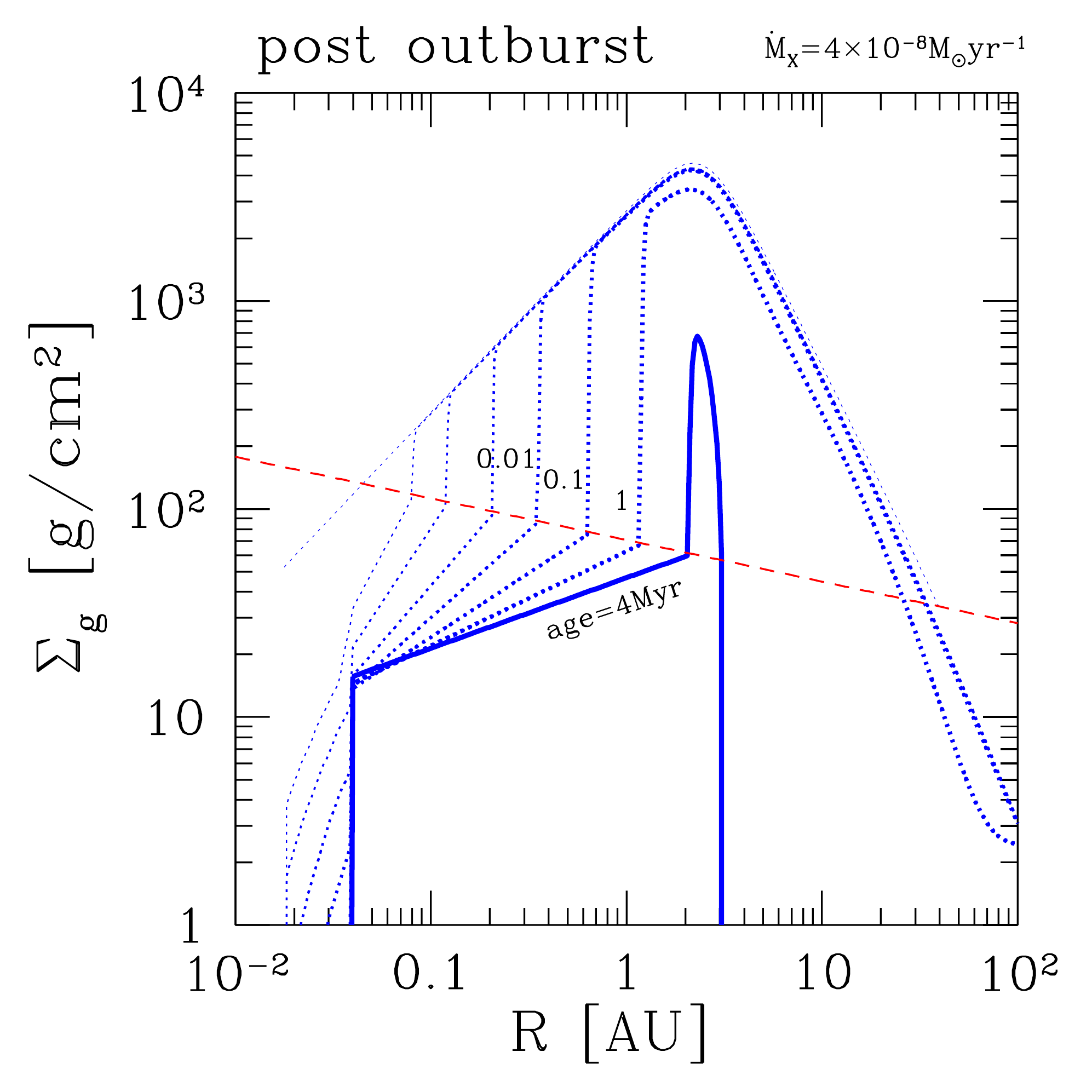}
\plottwo{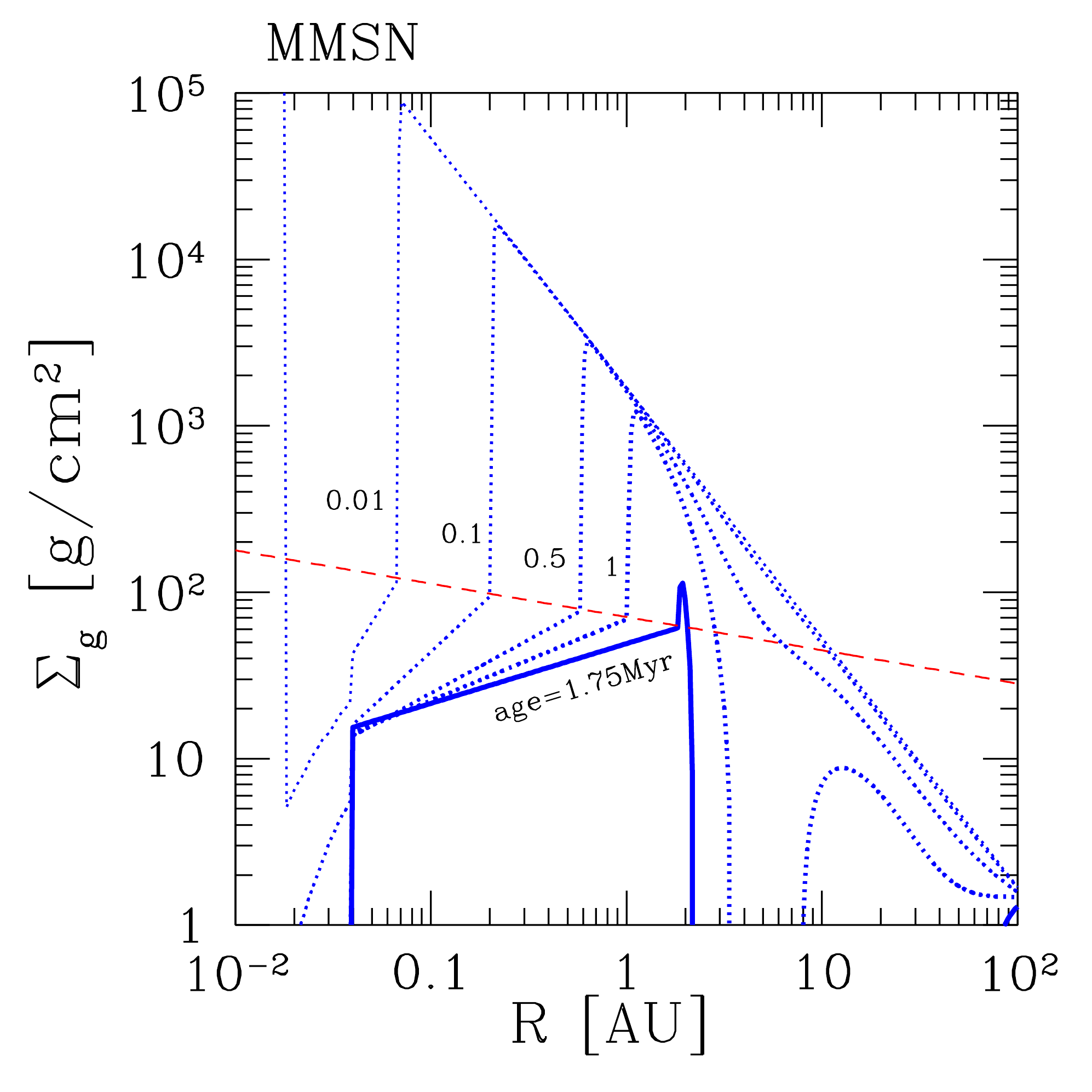}{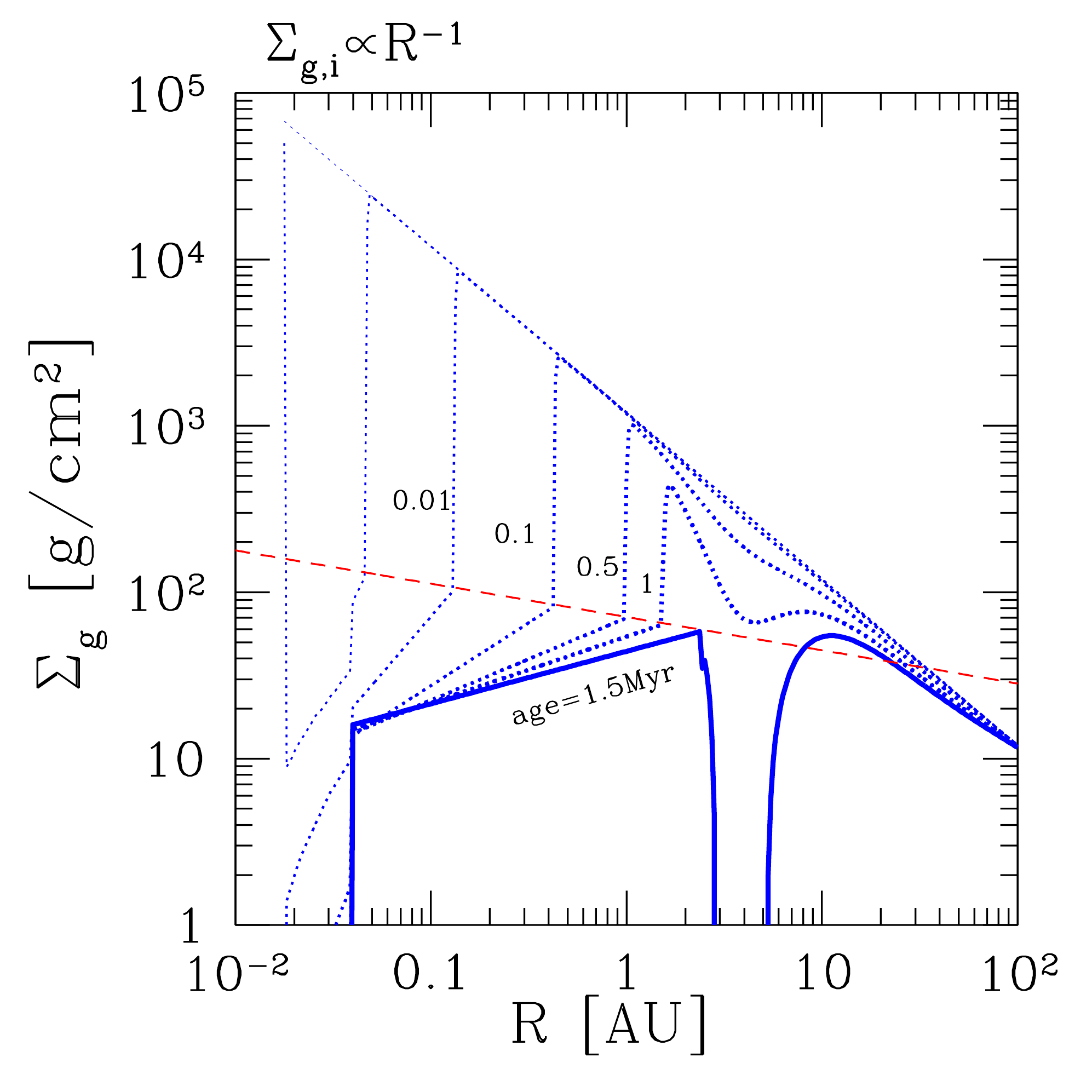}
\caption{Long-term evolution of gas column, as driven by a combination of MRI and laminar MHD stresses,
combined with the photoionization-driven mass loss rate of \cite{owen2012}.  Buffering of the
column at $\delta\Sigma_{\rm stir}\approx 40(R/{\rm AU})^{-0.2}$ g cm$^{-2}$ is due to the stirring up 
of solid particles and loading of the MRI-active layer by dust.  \textit{Top panels:} evolution following
an FU Ori like outburst, starting from a time when silicates recondense in the disk and the ionization
level near the midplane is suppressed by the formation of grains.  Initial disk is marginally
self gravitating ($Q=2$) outside $\sim 2$ AU.  Lifetime of the gas around
$\sim 1$ AU varies inversely with the normalization of the X-ray and FUV driven mass
loss rate ($\dot M_{\rm X} = 1.3\times 10^{-8}$ and $4\times 10^{-8}\,M_\odot$ 
yr$^{-1}$ at left and right).
\textit{Bottom left panel:} evolution starting from the minimum-mass solar nebula;  {\it Bottom right panel:}
initial gas profile $\Sigma_g \propto R^{-1}$ with same normalization as MMSN at 2 AU.}
\vskip .1in
\label{fig:Sigt}
\end{figure*}

The gas surface density is evolved according to
\be \label{e:dSig}
\frac{\partial\Sigma_{g}}{\partial t}=\frac{1}{2\pi R}\frac{\partial\dot{M}}{\partial R}
+ \left(\frac{\partial\Sigma_{g}}{\partial t}\right)_{\rm X},
\ee
where $\dot{M}$ includes contributions from MRI and the laminar Maxwell stress.  The last term on the right is the 
ionization-driven wind from \cite{owen2012}, specifically for a PPD with 
$\dot{M} \approx 1.3\times 10^{-8}M_\odot$yr$^{-1}$. 
In the parts of the disk with $\Sigma_g > 2\delta\Sigma_{g,\rm ion}$, we use the power-law fit 
\be\label{eq:mdotin}
\dot{M} \simeq 2.5\times10^{-9}(1+2\gamma)\left({R\over {\rm AU}}\right)^{\gamma}\;M_{\odot}~{\rm yr}^{-1}
\ee
from Figure \ref{fig:MdotRtotfit}.  Here $\gamma=-0.4$ measures the gradient in
the combined accretion rate $\dot M = \dot M_{\rm MRI} + \dot M_{\rm lam}$.  This fit is used 
beyond the maximum radius of 1 AU for which vertical disk profiles were constructed in Paper I, 
but close to this radius the ionization-driven wind begins to dominate.  

Key outputs of the calculation include:
\vskip .05in
\noindent 1. The surface density inside $\sim 0.3$ AU, where Kepler has discovered many planets.  This decreases
with time given the radial profile (\ref{eq:mdotin}) of $\dot M$.  This behavior is in sharp contrast with a 
uniform-$\alpha$ disk model, where $\langle\nu\rangle$ increases with radius due to the disk flaring and
$\Sigma_g$ increases with time from the initial state (\ref{eq:sigfuori}); see \cite{zhu09}.
\vskip .05in
\noindent 2. The outer radius $R_{\rm stir}$ of the `stirring' region where the column has been reduced to 
$2\delta\Sigma_{\rm stir,0}$ (Figure \ref{fig:stir}) and the depletion of gas slows.  This zone expands with time as
gas is removed from the inner PPD.    We recall from Section 
\ref{sec:opt} that this depleted inner disk can maintain optical absorption surface, and at first will not appear
as a transition disk.
\vskip .05in
\noindent 3. The relative timing of the removal of the outer and inner disks, due to an ionization-driven wind and
internal viscous stresses, respectively.  We find that the outer disk is only modestly depleted after the inner disk 
reaches a surface density $\sim \delta\Sigma_{g,\rm ion}$.   
\vskip .05in
Because the accretion time is very short inside radius $R_{\rm stir}$ (compared with the time to excavate
the intermediate parts of the disk), we take $\dot M$ to be uniform in the inner, depleted disk.  
Then a significant drop in accretion rate is sustained at $R_{\rm stir}$, and a sharp outward radial
gradient in $\Sigma_g$ forms there, associated with a local pressure maximum (Figure \ref{fig:Sigt}).   
We set the gradient scale of $\Sigma_g$ to $4c_g/\Omega$ near the pressure maximum.
No significant changes in the disk evolution are noticeable for order unity adjustments of this value.

The outward progression of the radius $R_{\rm stir}$ can be accurately estimated from
\be
{dR_{\rm stir}^2\over dt} = {\dot M_- - \dot M_+\over \pi \Sigma_g(R_{\rm stir})} = 
     {1.27(R/{\rm AU})^{-0.4}\over (\Sigma_g(R)/10^3~{\rm g~cm^{-2}})}\;{{\rm AU}^2\over{\rm Myr}}.
\ee
Here $\dot M_-$ is the accretion rate everywhere inside radius $R_{\rm stir}$, and is given by Equation 
(\ref{eq:mdotin}) at $R_{\rm stir}$ with $\gamma = 0$.  The accretion rate $\dot M_+$ just outside 
$R_{\rm stir}$ is smaller by a factor $1+2\gamma \simeq 0.2$. 

The surface density profile interior to $R_{\rm stir}$ (and outside the sublimation radius (\ref{eq:rsub}))
is obtained as follows.  
The dust-loaded disk developed in Section \ref{sec:solid} gives $\Sigma_g$ as a function of the accretion 
rate $\dot M_-$ through the inner disk.  Expressing $\dot M_-$ in terms of $R_{\rm stir}$, we find the 
steady-state column to be
\be \label{eq:siginner}
\Sigma_g = 2\delta\Sigma_{{\rm stir},0}(R_{\rm stir})\left(\frac{R}{R_{\rm stir}}\right)^{\gamma_s} ; \quad  
\gamma_{s}=0.5\left(\frac{R_{{\rm stir}}}{\rm AU}\right)^{-0.5}
\ee
at $R_{\rm sub} < R < R_{\rm stir}$.  The mild drop in $\Sigma_g$ with time seen in the inner disk in Figure \ref{fig:Sigt} is
thus a result of the decreasing mass flux sourced by material near $R_{\rm stir}$.

The dust loading $X_d/a_d$ that is implied by Equation (\ref{eq:siginner}) varies strongly with radius.
Figure \ref{fig:mdotdust} shows that the disk will sustain a given accretion rate at smaller $R$ with 
a lower $\Sigma_g$, as a consequence of a higher dust loading.  The accretion rate at $R_{\rm stir}$ is
determined by the disk profile with vanishing dust loading.   Moving inward toward the star, $X_d/a_d$ rapidly increases.  

The changing value of the index $\gamma_s$ in Equation (\ref{eq:siginner}) also deserves comment.  At early times,
only a small inner part of the disk has a column reduced to $\sim \delta\Sigma_{\rm stir}$.  We find that the mass
flux close to the star is mainly driven by the laminar torque (see Figure \ref{fig:mdotdust}).  Since the laminar
Maxwell stress decreases strongly with radius, a variation in $\Sigma_g$ is required to maintain a constant mass flux.
As $R_{\rm stir}$ approaches 1~AU, MRI stresses begin to dominate. Since $\dot{M}_{\rm MRI}$
is nearly independent of radius at fixed $\delta\Sigma_{\rm stir}$, the surface density profile flattens out.  
Additional flattening of $\Sigma_g(R)$ is caused by a weakening of the imposed radial field ($\epsilon_B \lesssim 0.1$
in Equation (\ref{eq:BRapp})), because the laminar stress scales more strongly with the applied field.

\begin{figure}[!]
\epsscale{1.2}
\plotone{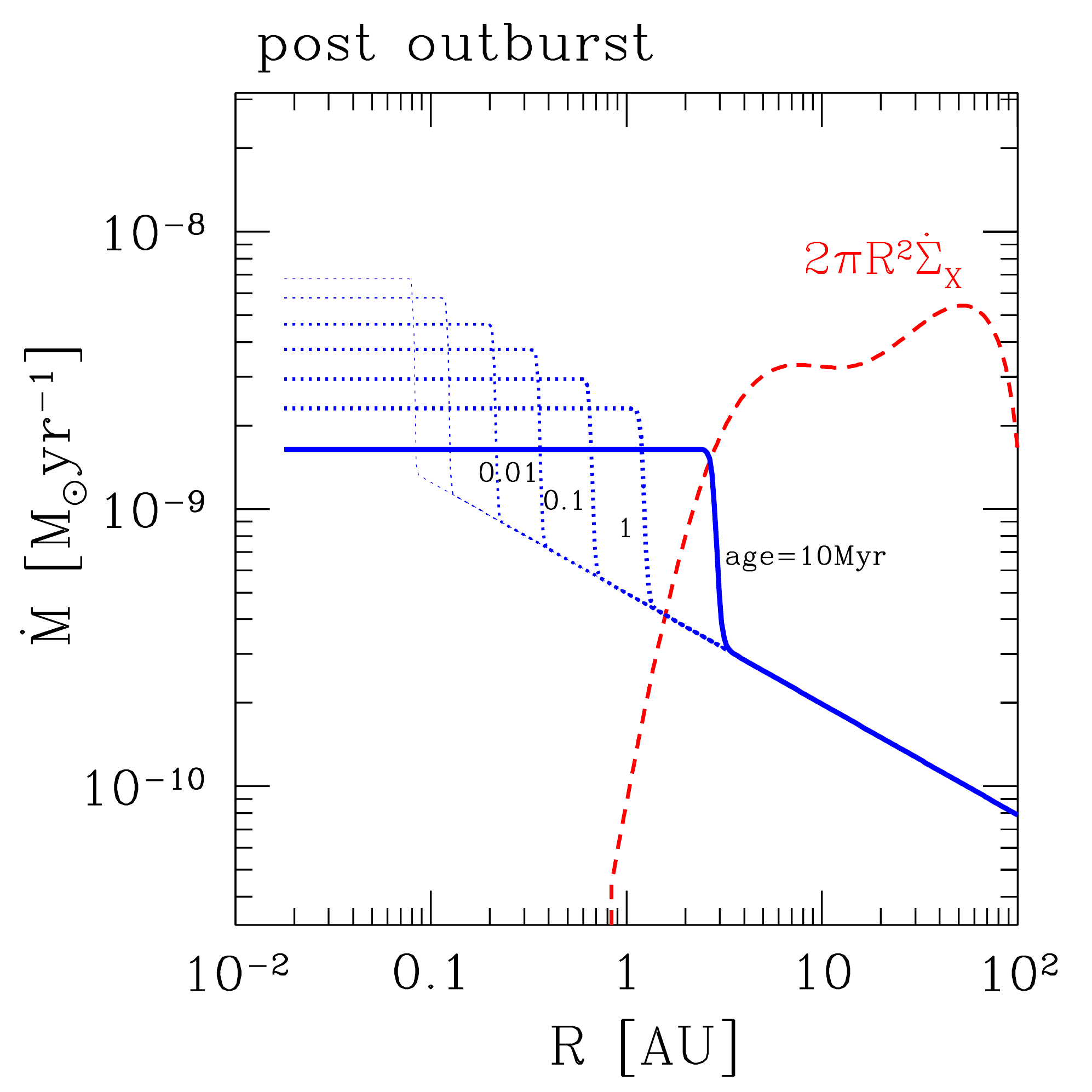}
\caption{Blue lines: accretion rate versus radius and time corresponding to the top left panel of Figure \ref{fig:Sigt}.  
 Red dashed line:  mass loss rate in photoionization-driven wind according to the model of \cite{owen2012}.}
\vskip .1in
\label{fig:Mdot}
\end{figure}

\subsection{Details of Disk Evolution}\label{sec:results}

Consider first the evolution of the disk from a post-FU Ori outburst configuration.  Figure \ref{fig:Sigt} shows
that within $\sim 1$ Myr the inner disk is reduced to the stirring column out to $\sim 1$ AU.  Over the same interval,
the wind model of \cite{owen2012} produces little clearing;  this becomes significant only 
after $\sim 5$ Myr.  Shortly after 9 Myr, a gap opens up outside 4 AU.  The density bump
inside it disappears in the next Myr, causing the inner disk to drain rapidly onto the star
in $\sim 0.1$ My. The disk lifetime scales nearly inversely with the rate of mass loss due
to photoevaporation, as is shown in the top right panel of Figure \ref{fig:Sigt}.

There is a slow decrease in the accretion rate onto the protostar
as the dust-loaded inner disk expands (Figure \ref{fig:Mdot}):
from $\sim 10^{-8}M_\odot$ yr$^{-1}$ when $R_{\rm stir}$ sits at 0.06 AU, 
down to $2\times 10^{-9}\,M_\odot$ yr$^{-1}$ as $R_{\rm stir}$ expands to $2$ AU.

The dust-free solution with very low $\Sigma_g$ is obtained inside $R_{\rm sub} \sim 0.04$ AU, explaining the sharp
upward rise in $\Sigma_g$ at that radius in Figure \ref{fig:Sigt}.   This inner pressure bump may expand outward,
depending on the particle flux through the midplane region into the inner disk.

A lower total mass must be dispersed when starting the evolution with the MMSN model, because the Toomre 
parameter $Q \simeq 20$ is about 10 times larger than we choose in the `post-outburst' disk model.  
(How the disk would reach such a state is not clear in the present context:  angular momentum transport
by spiral density waves would freeze out at lower $Q$, and so an additional source of torque would need to
be invoked before the photoionized wind considered here would start.)  

Clearing of the inner disk is initially a bit slower in this case, due to the inward-peaked surface density profile,
but then picks up speed.  Dispersal of the disk is completed within 2 Myr; this time is less sensitive to
photoevaporative losses than in the post-outburst case since the density peaks at a smaller radius, where viscous
evolution dominates photoevaporation.

The evolution of the flattest ($\Sigma_g \propto R^{-1}$) initial profile is distinguished from the other two by
the persistence of gas at 10--100 AU after gas at $\sim 1-2$ AU is largely removed.  This is the same profile
assumed in the models of \cite{owen2012}; not surprisingly, we also find that a distinct cavity develops around 
2-3 AU.  This case shows the greatest resemblance to transition disks, and the behavior of the `UV switch' 
advanced by \cite{clarke01} to explain their relatively brief appearance.

\section{Global Transport of Solid Material}\label{sec:dust}

Grains are supplied to the MRI-active layer of the inner disk through two channels:  inward advection from a
gas reservoir sitting outside $\sim 1$ AU; and a second, more indirect, channel involving the inward drift
of macroscopic particles through the disk midplane, followed by lofting and catastrophic fragmentation.

We first consider the residual dust mass fraction $X_d$ when the turbulent motions at the disk midplane are
too weak to loft particles.  Then we consider the effect of a hydrodynamic instability, such as a baroclinic instability
\citep{klahr03}, which may be excited near the outward-propagating density peak.  Here a narrow annulus of the disk
will be directly exposed to a higher stellar radiation energy flux, and will develop a strong radial
temperature gradient.  

We must first develop a working criterion for the lofting of particles.   This depends on
establishing a connection between the turbulent amplitude and particle size.  Here we employ a simple model of a
`bouncing barrier' \citep{zsom10}, representing the maximum size of compact
solid particles that have been compressed to a low porosity by repeated mutual collisions.  Surface van der
Waals forces facilitate the sticking of $a_1 \sim \mu$m-sized silicate monomers at relatively high collision speeds,
up to $v_{\rm frag} \sim 1$ m s$^{-1}$.  The critical speed for the sticking of larger, compact conglomerates
(radius $a > a_1$) scales roughly as $v_{\rm stick} \sim v_{\rm frag}(a_1/a) \equiv k_{\rm stick}/a$ \citep{chokshi93}.

Particles with Stokes parameter ${\rm St}_p = t_{\rm stop}\Omega \lesssim \alpha$ move through the gas, and with 
respect to each other, with a small velocity $v_{p-g} \sim (\alpha {\rm St}_p)^{1/2}c_g \ll$ 
m s$^{-1}$.  Balancing this with $v_{\rm stick}$ at a column $\delta\Sigma_g \sim \rho_g c_g/\Omega$, one obtains
\be
a \sim \left({k_{\rm stick}^2\delta\Sigma_g\over \alpha \epsilon_{\rm dr} \rho_s c_g^2}\right)^{1/3}.
\ee
(Here stopping is self-consistently in the Epstein regime.) 

At the threshold for setting, $\alpha \sim \epsilon_{\rm dr} \rho_s a_p/\delta\Sigma_g$, and one finds
\ba
a_p &\sim& 0.01\left({k_{\rm stick}\over 10^{-2}~{\rm cm^2~s^{-1}}}\right)^{1/2}
\left({\delta\Sigma_g\over 10^3~{\rm g~cm^{-2}}}\right)^{1/2}\nn
&&\times\left({T\over 200\,\rm{K}}\right)^{-1/2}\,{\rm cm}.
\ea
The $\alpha$ parameter below which the particles settle corresponds to
\ba\label{eq:aset}
\alpha < \alpha_{\rm set} &=& 3\times 10^{-5}\,\left({k_{\rm stick}\over 10^{-2}~{\rm cm^2~s^{-1}}}\right)^{1/2}
\left({\delta\Sigma_g\over 10^3~{\rm g~cm^{-2}}}\right)^{-1/2}\nn
&&\times\left({T\over 200\,\rm{K}}\right)^{-1/2}.
\ea

\subsection{Residual Grain Abundance in a Layered Disk}

First consider a layered disk in which $\alpha$ attains a large value $\alpha_{\rm max} \sim 10^{-2}-10^{-1}$ at the top
of the disk, and drops rapidly at columns $\delta\Sigma_g \gtrsim 10$ g cm$^{-2}$.  Then $\alpha_{\rm set}
\sim 3\times 10^{-4}$ for silicate particles from Equation (\ref{eq:aset}).   Embedded particles diffuse vertically
over a timescale $\sim \alpha^{-1}\Omega^{-1}$.  Even in the relatively quiescent settling layer ($\alpha \sim \alpha_{\rm set}$), this is faster than radial spreading through the turbulent upper disk, which occurs over the timescale $(r/h_g)^2 
\alpha_{\rm max}^{-1} \Omega^{-1} \sim 10^{3-4}\alpha_{\rm max}^{-1}\Omega^{-1}$.  Solid particles are therefore
nearly uniformly mixed down to the depth at which mutual sticking allows them to settle out.

We are interested here in the residual density of small grains, which are easily suspended by MRI turbulence.
The preceding considerations show that their depletion from the gas is limited not by mixing, but by the rate
of mutual collisions, which is highest for the smallest particles.   It is easy to see that the largest
contribution to the net collision rate (per unit area of disk) comes from the base of the layer where
$\alpha$ begins to drop sharply (Figure \ref{fig:alpha}).  The collision rate between grains of mass 
$m_d \sim (4\pi/3)\rho_s a_d^3$ and space density $n_d = X_d\rho_g/m_d$ is
\be
h_g\cdot n_d^2 4\pi a_d^2 v_{d-g} \sim X_d^2{\delta\Sigma_g\Omega\over (4\pi/3)\rho_s a_d^3}
\left(\alpha_{\rm max}\,\epsilon_{\rm dr}{\delta\Sigma_g\over \rho_sa_d}\right)^{1/2}.
\ee
Averaging vertically over the disk gives a characteristic collision time 
$\bar t_{\rm col}$ between small grains.  

The formation of larger particles is suppressed if $\bar t_{\rm col}$ is longer than the time for radial inflow,
\be
\bar t_{\rm col} \sim {1\over 3X_d\Omega} \left(\alpha_{\rm max}\,\epsilon_{\rm dr}{\delta\Sigma_{\rm act}\over \rho_s a_d}\right )^{-1/2}  \gtrsim  {1\over \alpha_{\rm max}\Omega} \left({h_g\over R}\right)^{-2},
\ee
corresponding to a dust loading
\ba\label{eq:xdmax}
X_d &\;\lesssim\;& {\sqrt{\alpha_{\rm max} {\rm St}_d}\over 3\epsilon_{\rm dr}} \left(\frac{h_{g}}{R}\right)^{2} 
\;\approx\; 9\times10^{-7}\,\alpha_{{\rm max},-1}^{1/2} \nn
&&\times \left(\frac{R}{\rm AU}\right)^{4/7}\left(\frac{a_{d}}{\mu{\rm m}}\right)^{1/2}
\left(\frac{\delta\Sigma_{\rm act}}{20~{\rm g~cm^{-2}}}\right)^{-1/2}.\nn
\ea
Larger particles easily form when this condition is violated. 

The main conclusion here is that the dust population in the MRI-active parts of the inner disk is
regulated by settling in the outer disk that feeds it.  After the lapse of $\sim$ Myr, a small fraction of
the solids deposited initially at $1$--2 AU will have swept through the inner disk, sustaining a population
of $\mu$m or sub-$\mu$m sized grains.

\subsection{Turbulent Transport Across Density Peak}

Macroscopic particles near the outer edge of the depleted inner disk (radius $R_{\rm stir}$)
experience a combination of turbulent diffusion and secular drift toward the pressure maximum.  
In the absence of turbulence, the particles settle toward the midplane and the pressure maximum 
presents a barrier to their inward radial migration (e.g. \citealt{kretke07}).  A narrow density peak
may, however, be susceptible to a hydrodynamic (e.g. baroclinic) instability.
Without understanding details of such an instability, we can still formulate the
following question:  if the turbulence excited is strong enough to smear the density peak to a
radial lengthscale $\Delta r \gtrsim h_g$, will it also transport particles across the peak?  

Here it is essential to remember that the peak moves outward at a speed $dR_{\rm stir}/dt \gtrsim 
1~{\rm AU}/{\rm Myr}$.  Hence turbulence of amplitude $\nu_{\rm t} = \alpha c_g^2/\Omega$ will smear
the peak over a scale given by $(\Delta r)^2/\nu_t \sim (dR_{\rm stir}/dt)^{-1}\Delta r$.  The
value of $\alpha$ corresponding to a given value of $\Delta r$ is
\be
\alpha \sim {dR_{\rm stir}/dt\over c_g}{\Delta r\over h_g} \sim 3\times 10^{-5}\left({dR_{\rm stir}/dt\over
1~{\rm AU}/{\rm Myr}}\right)\left({\Delta r\over 3h_g}\right).
\ee
Equation (\ref{eq:aset}) indicates that his level of turbulence will suspend macroscopic
particles at a gas column $\gtrsim 10^3$ g cm$^{-2}$, which is comparable to the column that is
attained by our post-FU Ori disk model at $R \sim 1$ AU.  

We conclude that a hydrodynamic instability that smears the density peak will also facilitate the inward
flow of settled particles into the inner disk.  Indeed, the dust loading of the inner disk may easily 
exceed the level (\ref{eq:xdmax}) that is left behind in a layered disk with a quiescent midplane layer.

\section{Discussion}\label{s:disc}

We have demonstrated that the inner $\sim 1$--2 AU of a PPD will quickly evolve to a mass profile very different 
from the one that is obtained by assuming a uniform viscosity coefficient $\alpha$ (or, indeed,
as manifested in the popular MMSN disk model and its variant obtained from the {\it Kepler}
planetary systems:  \citealt{chiang13}).   Mass is removed from the inner disk as soon as a stellar
wind flowing across its surface deposits a radial magnetic field into the upper layers of the disk.
This radial field is wound up by the disk shear and then transported downward by a combination of
turbulence and non-ideal MHD effects.  

The vertical disk model so constructed in Paper I forms the basis for a constrained calculation of
PPD evolution.  The radial profile of the seed magnetic field is obtained directly
from the profile of the imposed T-Tauri wind.  We find that a depletion of mass from
the inner disk is not sensitive to the details of how the MRI-generated stress depends on
the seed (linearly wound) toroidal magnetic field.   The rate of mass transfer through the disk
depends only weakly on the normalization of this relation.

More standard MRI-based models (e.g. \citealt{gammie96,zhu09}), which assume a vertical seed field,
have the drawback that the flux distribution across the disk surface cannot yet be constrained 
in a useful way.  The same limitation applies to models based on magnetorotational outflows from
a vertically magnetized disk \citep{pudritz86,suzuki10,bs13,lesur14,gressel15}:  these models
also depend on an uncertain parameterization of the flux-to-mass ratio of the disk, and cannot
yet be used to predict the sign of the change in mass in the inner part of a PPD.

Solid material in the inner disk is maintained by two sources:  (i) small ($\mu$m sized) grains which
are transported by MRI turbulence faster than they can stick together to form macroscopic
particles; and (ii) macroscopic (mm-cm sized) particles which settle to the quiescent midplane
and drift inward by the usual headwind effect \citep{gw73,weiden77}.  

We showed that suspended grains can have a small influence on the ionization level
even while forming an optical absorption layer above a height $\sim 2 h_g$.  The radial 
distribution of reprocessed stellar light is found to depend weakly on the mass fraction 
$X_d$ of dust grains in the MRI-active layer. 

We also showed that the depletion of gas from the inner PPD is buffered by X-ray ionization of the midplane
layer.  Although this tends to activate the MRI, the lofting of macroscopic (mm-cm sized) particles from
the disk midplane leads to catastrophic fragmentation higher in the disk,
which loads the MRI-active layer with 
small grains.  We found that steady accretion through the inner PPD can be sustained with a dust
loading that varies with radius, but only over a finite range of $\dot M$.  In this situation
there is a strong inward gradient in $X_d$, meaning that suspended dust diffuses {\it outward} from
the protostar.  In this way, it may be possible to sustain a closed cycle of solids inside $\sim 1$ 
AU even while gas continues to accrete through the inner disk.

\begin{figure}[!]
\epsscale{1.2}
\plotone{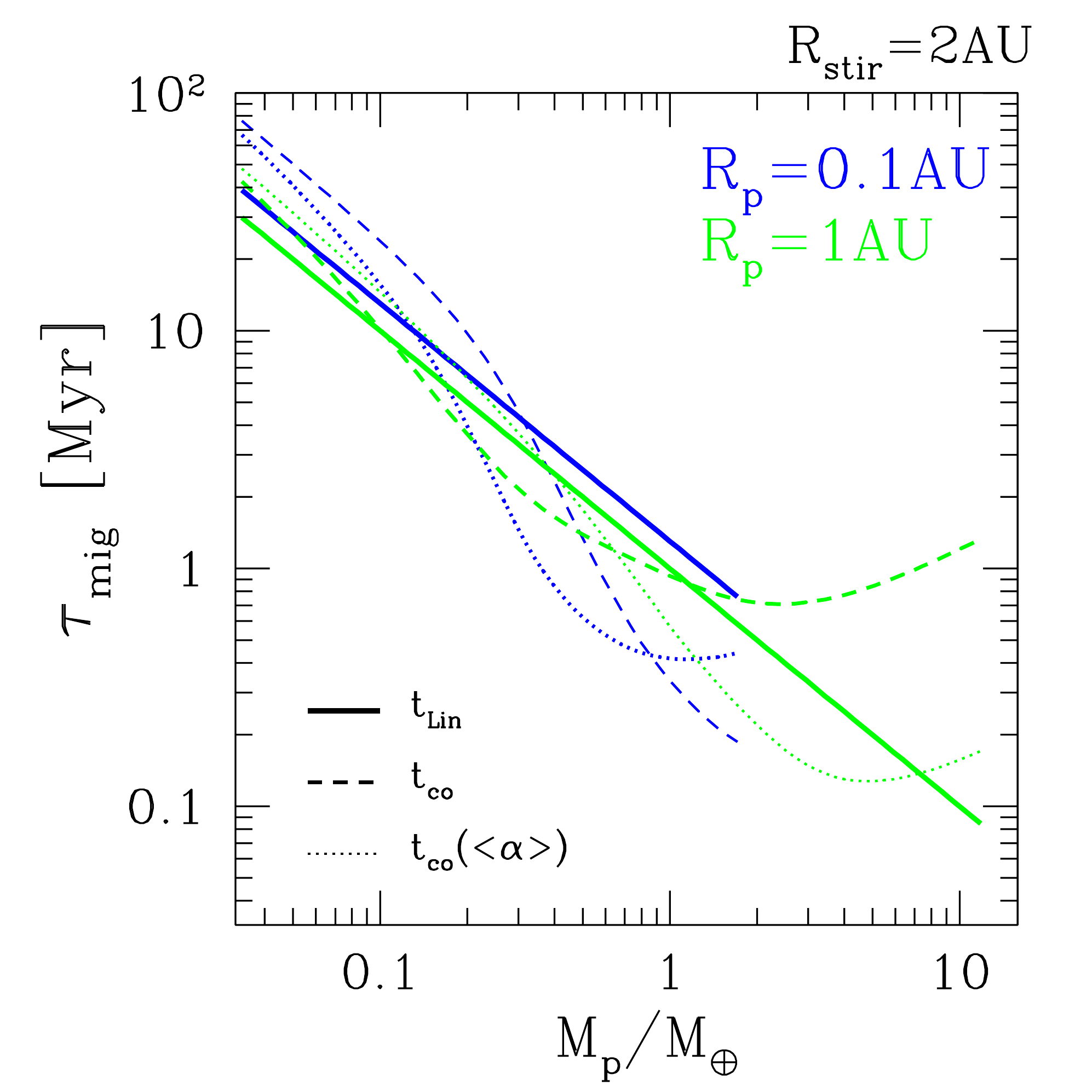}
\caption{Migration time $|a/\dot a|$ of a planet of mass $M_p$ in response to:
linear Lindblad torques (negative torque, solid lines);  or a combination of linear corotation
torques and the non-linear horseshoe torque (positive torque, dashed and dotted lines), based on
formulae tabulated in \cite{paard4}. In the second case, the transition from linear to non-linear torques and the saturation of the horseshoe torque is sensitive to the
disk viscosity.   Dashed curves assume $\alpha$ is normalized at the disk half-column;
dotted curves make use of the vertically averaged viscosity $\langle\alpha\rangle$.
Planet mass is constrained to lie below the `Hill' mass given by 
$(M_p/3M_\star)^{1/3} = h_g/R$.}
\vskip .1in
\label{fig:tmigM}
\end{figure}

A conspicuous feature of our disk solution is a strong maximum in
surface density which moves outward as the inner disk clears.  This may create a visible excess of 
infrared radiation at a particular radius, although not as strong a spectral feature as would a
rim bounding a transparent inner zone.  

This maximum in $\Sigma_g$ is also associated with a maximum in pressure, at which
solid particles may collect.  In contrast with the pressure bump associated with an ice sublimation
surface (e.g. \citealt{kretke07}), this feature scans through a wide range of radius.  
As such it is a possible site for the formation of planetesimals -- although it should be kept in mind that
settled particles will lag inside the pressure maximum as it moves outward.  The gas near the bump
is also susceptible to a hydrodynamic instability, especially a baroclinic instability, as a result
of the strong radial temperature gradient that is generated by the raised disk profile \citep{klahr03}.  As a result, the
particles still experience differential orbital motion with respect to the gas, along with vertical stirring.

\subsection{Planets below the Gap Opening Mass}

The equilibrium gas column of the dust-loaded inner disk is in a
range that allows the inward migration of both Earth-mass and Jupiter-mass planets
on a $\sim$ Myr timescale.  Figure \ref{fig:tmigM} shows the migration time for planets smaller
than the `Hill mass', corresponding to $M_p = 3M_\star (c_g/\Omega R)^3$, based on the formulae 
of \citep{paard4} and the surface density profile shown in Figure \ref{fig:Sigt}.  The solid
lines in this figure are obtained from the sum of the inner and outer Lindblad torques, and
represent inward migration \citep{gt80,tanaka2002}. The dashed and dotted lines show the 
summed effect of the linear and non-linear corotation torques, the latter imparted by gas 
executing horseshoe orbits in the corotation zone \citep{ward1991}.  The corotation torque is positive, meaning
that the net torque is negative (inward migration) where the dashed (or dotted) line lies
{\it above} the solid line.  Migration is outward where the opposite inequality holds. 
For example, a $M_\odot$ planet would experience very slow migration at $1$ AU since 
positive and negative torques nearly cancel (with $\alpha$ normalized to the disk half column) 
but would migrate outwards at $0.1$ AU due to a strong positive corotation torque.

The calculation includes the feedback of viscous and thermal diffusion on corotation torque 
saturation, again using formulae from \cite{paard4}.
The disk viscosity has a strong vertical gradient (Figure \ref{fig:alpha}),
meaning that the upper disk can maintain an unsaturated corotation torque much more
easily than the lower, quiescent disk.   We evaluate the effect of the viscosity in 
two ways:  first by adopting the value of the viscosity
coefficient $\alpha$ at the half-column point in the disk (dashed lines);  alternatively by using
the vertically (mass) averaged  $\langle\alpha\rangle$ (dotted lines).  With the first prescription,
planets of mass $\sim 1$--$10\,M_\oplus$ migrate inward at 1 AU, but migration stalls before the
planet reaches 0.1 AU.  

It should be emphasized that the magnitude of the corotation torque experienced by a planet
below the gap-opening mass is sensitive to the dust loading of the gas in the co-orbital region.
Repulsion of settled particles from the planet's orbit (which requires a lower planet mass than
does the repulsion of gas) would lead to a reduction in dust loading.  Our flow solution with
low $X_d$ has a much lower gas column (Figure \ref{fig:SigSS}) than does the solution in which 
$\nu_{\rm MRI}$ is regulated by the lofting of dust (Figure \ref{fig:stir}).  Therefore the 
formulae of \cite{paard4} may overestimate the horseshoe torque in a PPD with settled particles.
Figure \ref{fig:tmigM} provides a reasonable estimate of the magnitude of the orbital torque,
but not necessarily of its sign.

\subsection{Jupiter-mass Planets}

\begin{figure}[!]
\epsscale{1.2}
\plotone{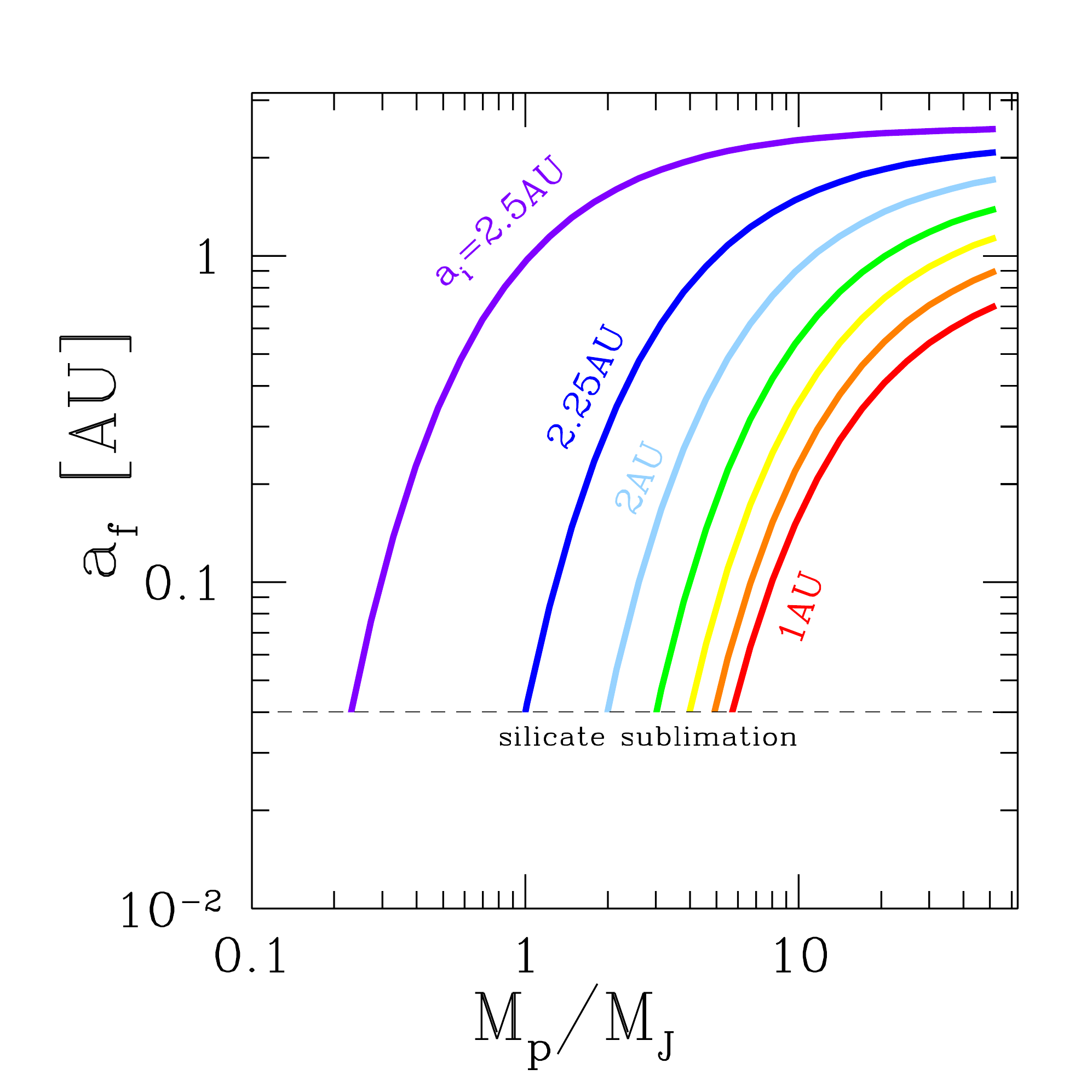}
\caption{Final orbital semi-major axis $a_f$ of a gap-opening planet, as a function of mass.
Planet migrates inward according to the rate determined by \citep{duffell2014} at various 
times in the evolution of the disk shown in Figure \ref{fig:Sigt}.  Planet is assumed
not to accrete from the disk; partial accretion of the gas leads to a modest increase
in $a_f$.}
\vskip .1in
\label{fig:Mmig}
\end{figure}

The inward migration of Jupiter-mass planets through our model disk is buffered by the
cumulative transport of mass, because the planet mass $M_p \gg \Sigma_g R^2$.  This migration
regime was first studied by \cite{syer1995}; we adopt a migration rate $\dot a/a \simeq
-\dot M/M_p$ as reported by \cite{duffell2014} in a numerical for a Jupiter-mass planet in 
a $\alpha = 10^{-2}$ PPD.  The planet is started at the outer boundary $R_{\rm stir}$ of
the dust-loaded inner disk and allowed to migrate inward without accreting.  Figure 
\ref{fig:Mmig} shows that over the final few Myr of disk dispersal, planets of mass $\sim 1$--$3\,M_J$ 
will migrate partly toward the protostar.  

This result depends mainly on the accretion rate through the inner
disk.  Realistically, the planet will accrete part of the mass incident on its orbit \citep{lubow99}, 
but it is straightforward to show that this does not substantially change the result.
% (Russo \& Thompson, in preparation).}

As the accretion rate drops,
as it inevitably must as the outer parts of the PPD are removed, 
$\Sigma_g$ must transition to the much lower steady-state value shown in Section \ref{s:nopart}, corresponding
to a radially magnetized disk in which the dust abundance is everywhere too small to perturb the ionization fraction.
This represents a second stage of inside-out partial clearing, during which the inner disk may become transparent 
to optical photons.   Once this happens, gas-mediated planetary migration must freeze out. 

A transition to a much lower gas column will also occur at the radius $\sim 0.04$ AU where 
solid silicate material that is directly irradiated by the disk will be sublimated.  As a result,
our disk model predicts the formation of a planet trap at this radius, where even the migration
of a Jupiter-mass planet will be stalled.  The existence of a `pile-up' of hot Jupiters at this
radius is presently controversial, depending on an evaluation of how the orbits are sampled
observationally \citep{winn2015}.

\subsection{Transition Disks as a Probe of the Disk Mass Profile During Planet Formation}

We have combined our mass transfer model for the disk inside $\sim 1-2$ AU with the 
photoionization-driven wind formulated by \cite{owen2012}.  Considering a range of initial mass profiles
(including a post-FU Ori outburst configuration, the MMSN profile, and a flatter $\Sigma_g \propto R^{-1}$
initial profile), we find that the clearing of gas from inside 1--2 AU is insensitive to the initial
profile.  In most cases, gas generally persists at 1--2 AU after the dispersal of gas at 10--100 AU.
The one exception to this is the flattest ($R^{-1}$) profile, which contains relatively more mass
at a large radius.  By comparison, some PPDs show large cavities extending out to $\sim 10$--30 AU \citep{williams11},
suggesting that -- at least in this subset of systems -- the raw material for planet formation is 
concentrated in the outer part of a PPD.

\begin{appendix}

\section{Post-FU Ori Disk Model}\label{s:appendixa}

Most PPDs may undergo one or many outbursts of accretion in the early stages of their evolution, while the disk is
still massive and self-gravity influences its dynamics.  These have been named `FU Orionis outbursts' after 
the first transient of this type to be discovered.  An interesting feature of theoretical models of such outbursts
(e.g. \citealt{zhu09}) is that the surface density profile inside $\sim 1-2$ AU departs significantly from the MMSN
model.  In fact, the surface density at $\sim 0.1$ AU can drop well below the corresponding value in the MMSN.

In time-dependent calculations which assume a layered disk with uniform $\alpha$, $\Sigma_g$ grows inside $\sim 1$ AU following an outburst.  We
show in Section \ref{sec:evo} that the opposite behavior is encountered if the seed magnetic field in the 
inner disk is dominated by the stellar wind immediately following the outburst.

Two significant feedbacks contribute to the growth of the MRI during a FU Ori type outburst.  First, the ionization
level shoots up deep in the disk (at $\Sigma_g \gtrsim \delta\Sigma_{g,\rm ion}$) when the temperature is high
enough ($T \gtrsim 1500\,^\circ$K) for silicate dust to sublimate.  Then free electrons are no longer absorbed on the surfaces
of grains, which are rapidly dispersed throughout the disk column following the catastrophic fragmentation of
macroscopic particles high in the disk (Section \ref{sec:solid}).  The second feedback occurs above several
$10^3\,^\circ$K as electronic transitions of H and scattering off H$^-$ contribute significantly to the opacity.

\begin{figure}[!]
\epsscale{.6}
\vskip 1in
\plotone{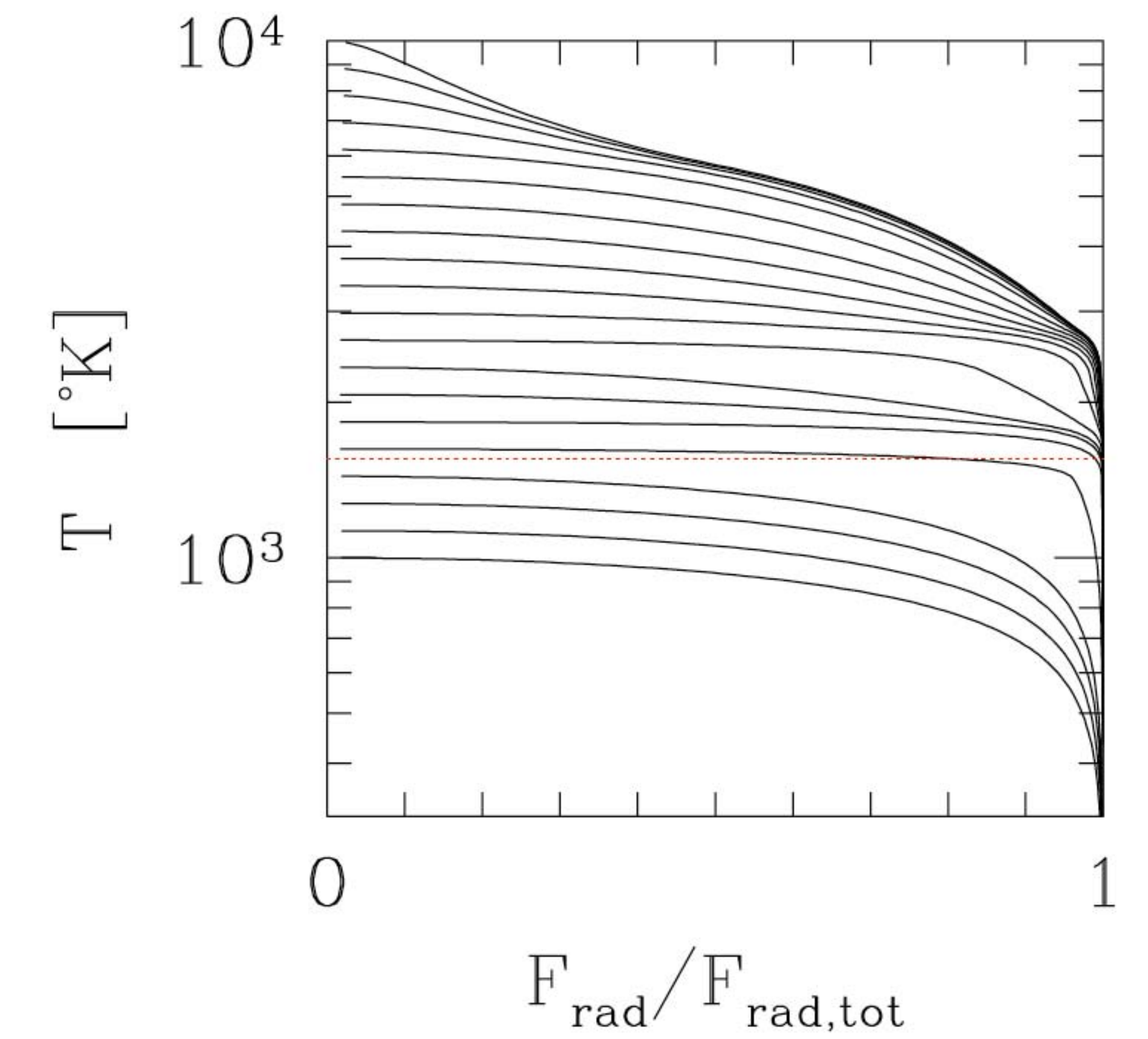}
\caption{Internal temperature of disk constructed from Equations (\ref{eq:verdisk}) versus cumulative radiative
flux (with $F_{\rm rad} = 0$ at the disk midplane).  Horizontal red dotted line:  sublimation temperature of 
silicate grains.  The dissipation in the disk is concentrated almost entirely at $T > T_{\rm sub}$ or $< T_{\rm sub}$
except for a narrow range of $T(0)$.}
\vskip .1in
\label{fig:tdisk}
\end{figure}

\begin{figure}[!]
\epsscale{.6}
\vskip .8in
\plotone{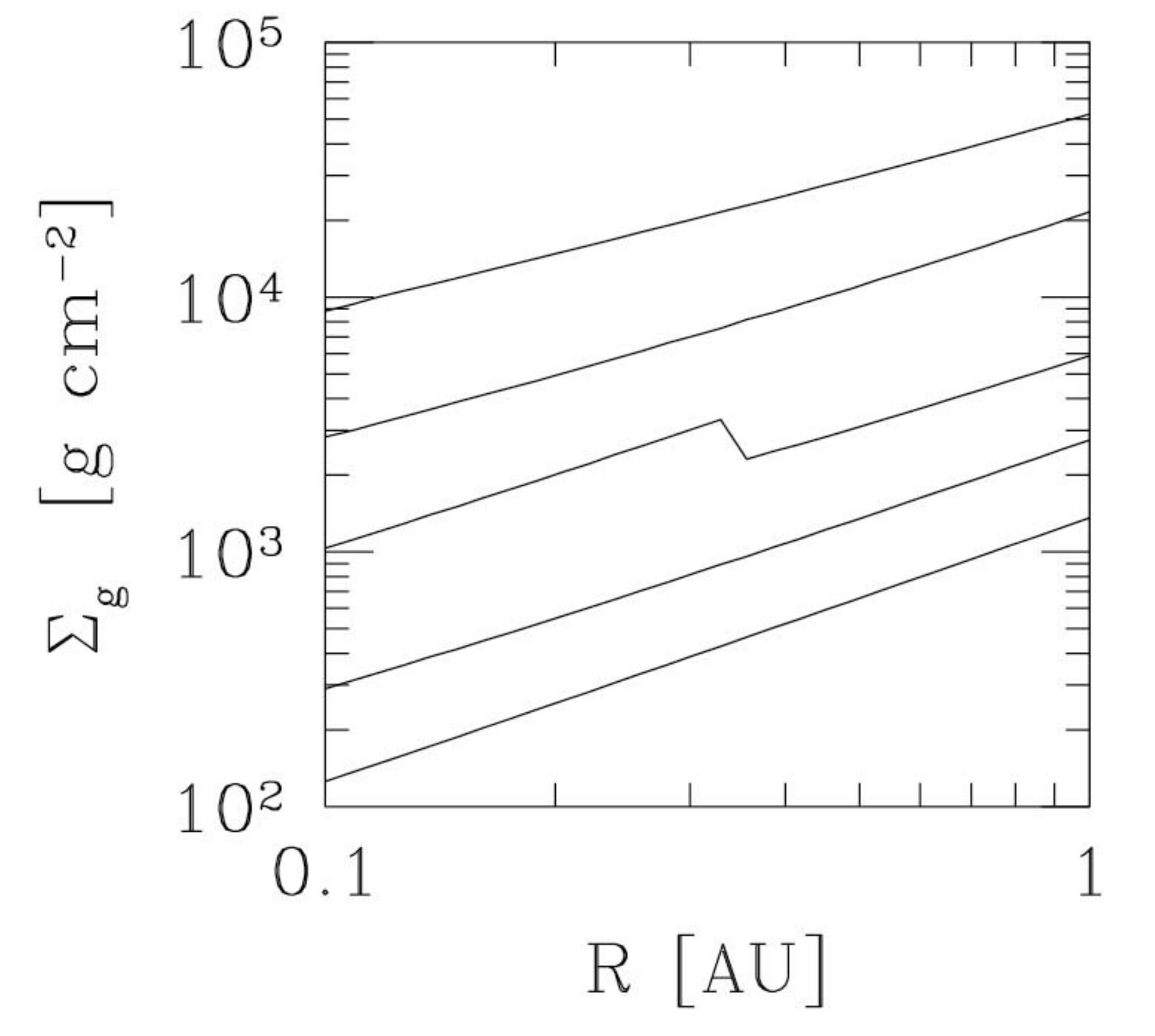}
\caption{Surface density versus radius of a disk with $T(0)$ adjusted to produce one-half of the outgoing radiative
flux at $T > T_{\rm sub}$, and one-half at lower temperatures.  Disks with surface density below this critical
value are dust loaded; those with higher surface densities are almost entirely dust free.  Curves correspond to different values of $\alpha$,
ranging from $0.1$ (bottom) to $10^{-5}$ (top).}
\vskip .1in
\label{fig:sigmacrit}
\end{figure}

Our focus here is on the aftermath of a FU Ori outburst, and so we consider the profile of the disk when it
transitions back to a dust-loaded state with strong MRI growth restricted to ionized surface layers \citep{gammie96}.
To this end, we have constructed uniform-$\alpha$
vertical disk models using the analytic fit to dust, molecular, atomic, free-free and bound-free opacity 
(collectively denoted by $\kappa$) provided by \cite{zhu09}.

Radiation transfer near the photosphere is handled using the Eddington approximation.  The entire disk is assumed
to be radiative, and convection is neglected.  Hence the equations solved are
\ba\label{eq:verdisk}
{dP\over dz} &=& -\Omega^2 z {P\mu_g\over k_B T};\nn
{dF_{\rm diss}\over dz} &=& {9\over 4}\alpha\Omega P;\nn
{dF_{\rm rad}\over dz} &=& \rho\kappa\left(aT^4-U_{\rm rad}\right)c;\nn
{dU_{\rm rad}\over dz} &=& - 3{\rho\kappa\over c}F_{\rm rad};\nn
F_{\rm rad} &=& F_{\rm diss}.
\ea
The lower boundary condition at $z = 0$ is $F_{\rm rad} = F_{\rm diss} = 0$, and the upper boundary
conditions at large $z$ are $P = 0$; $F_{\rm rad} = cU_{\rm rad}/\sqrt{3}$.   We do not 
explore here the origin of the seed field during the outburst, in particular the possibility that it
is also derived from the stellar wind.  Hence $\alpha$ is simply taken to be independent of height.

The solution to Equations (\ref{eq:verdisk}) has the property that the upward radiative flux is generated
in zones that are entirely above or below the sublimation temperature.  Figure \ref{fig:tdisk} shows a sequence
of models of increasing midplane temperature $T(0)$; the horizontal red dotted line denotes the sublimation temperature
used in the opacity fit.  Only for a narrow range of $T(0)$ is there a significant contribution to the outgoing
radiative flux from layers with $T < T_{\rm sub}$ {\it and} $T > T_{\rm sub}$.  

We identify the critical $T(0)$ for which one-half of the outgoing radiative flux is generated at $T > T_{\rm sub}$
(or $< T_{\rm sub}$),
a plot it in Figure \ref{fig:sigmacrit} as a function of $R$ for various values of $\alpha$.  To initialize our
post-FU Ori outburst calculation, we choose the curve corresponding to $\alpha = 10^{-2}$ (Equation (\ref{eq:sigfuori})).
This disk solution implies gravitational instability beyond $\sim 2$ AU. To avoid that, we smoothly interpolate 
to a marginally gravitationally unstable disk with $Q = c_g\Omega/\pi\Sigma_g = 2$ beyond $\sim 2$ AU.

The further evolution of $\Sigma_g$ in the inner disk is insensitive to the details of this initial profile.
The chosen value of $Q$ in the outer disk does influence whether this material can be entirely removed using
the prescription of \cite{owen2012} for a X-ray/FUV-driven wind:  that happens for $Q = 20$ but not for $Q=2$.  

\end{appendix}

\end{document}